\documentclass[preprint,12pt]{elsarticle}
\usepackage[letterpaper,margin=1in]{geometry}
\IfFileExists{lmodern.sty}{\usepackage[T1]{fontenc}\usepackage{lmodern}}{\usepackage[OT1]{fontenc}}
\usepackage[utf8]{inputenc}
\usepackage{microtype}
\usepackage{amsmath,amssymb,amsthm,mathtools,bm}
\usepackage{booktabs,array,enumitem}
\biboptions{sort&compress}
\usepackage{hyperref}
\input{xref/Supplement_labels.tex}
\hypersetup{colorlinks=true,linkcolor=blue,citecolor=blue,urlcolor=blue,pdftitle={The Quantum Pendulum of Turbulent Flow: Decaying turbulence as a thermal state of an integrable system},pdfauthor={Alexander Migdal}}
\newcommand{\ii}{\mathrm i}
\newcommand{\dd}{\,\mathrm d}

\newtheorem{theorem}{Theorem}[section]
\newtheorem{proposition}[theorem]{Proposition}

\theoremstyle{definition}
\newtheorem{definition}[theorem]{Definition}
\theoremstyle{remark}
\newtheorem{remark}[theorem]{Remark}
\numberwithin{equation}{section}
\makeatletter
\def\els@aparagraph[#1]#2{\elsparagraph[#1]{#2}}
\def\els@bparagraph#1{\elsparagraph*{#1}}
\makeatother
\newcommand{\ATWORE}{-0.08012}
\newcommand{\ATWOIM}{-0.25519}
\newcommand{\ABSD}{3.335\times10^{-3}}
\newcommand{\ARGD}{0.4812}

\newcommand{\OSCPER}{4.44288}

\newcommand{\ATWOACC}{10^{-3}}

\newcommand{\MCDEVFOUR}{1.9\%}

\newcommand{\MCERRFIVE}{1--4\%}

\newcommand{\QWALKN}{8000}% N of the exact zig-zag walk (mc/mc_check.py)
\newcommand{\QZHMAX}{1.28\times10^{4}}% largest h of the Floquet values of Zhat (12800)
\newcommand{\QGA}{0.0727643781}% a: end-layer resolvent quadrature, error 3e-12 (constants_closed.json); free fits of the 252-value data (h <= 0.02..0.07, a free, b = 0, terms to h^{7/2}) give 0.07276437809 +- 1e-10 (rel. 1.1e-9)
\newcommand{\QPIC}{0.0357837089}% c_{3/2}: end-layer resolvent quadrature, error 2e-15 (constants_closed.json pcoef); free fits of Pi (h <= 0.02..0.1, terms to h^{7/2} or h^4) give 0.03578371 (rel. <= 3.6e-7, 1.2e-7 for h <= 0.04)
\newcommand{\QSMALLAGREE}{10^{-7}}% relative agreement of the Floquet fits with the quadratures: a to 1.1e-9 (fits of g at h <= 0.02, 7 values), c_{3/2} to 1.2e-7 (fits of Pi at h <= 0.04, 10 values); both with the next terms (to h^{7/2}) included
\newcommand{\QGFITLO}{40}% lower end of the large-h fit of g: all 47 node values 40 <= h <= 3200 (dense 42..98 step 2 included); E', F change by < 3e-10 relative when the fit starts at 60, 100 or 200
\newcommand{\QGHMAX}{3200}% largest h at which Ghat is computed (Floquet); unchanged
\newcommand{\QZFITLO}{200}% lower end of the large-h fit of Pi (17 values to 12800); unchanged (c0 changes by 7e-8 for lower ends 100..500)
\newcommand{\QEPRIME}{5.510\times10^{-4}}% E' = 5.509664e-4: fit h^2 g_s = E' log h + F + F1/h + (E2 log h + F2)/h^2 + (E3 log h + F3)/h^3 on 40 <= h <= 3200 (rms 4.6e-11); fit-form spread 2e-10, statistical (err_Ghat) 3e-8, all values shifted by err_Ghat 1.5e-7
\newcommand{\QFTAIL}{7.729\times10^{-4}}% F = 7.728765e-4 (same fit); spread 2e-9, statistical 2.3e-7, shift by err_Ghat 1.3e-6. Provisional 7.730e-4 (7.72979e-4) used a fit with one correction term
\newcommand{\QGTAILRMS}{10^{-7}}% rms relative residuals of the fits with the next orders: h^2 g_s 4.6e-11, h^{5/2} G_osc 1.2e-7 (47 values, 40 <= h <= 3200)
\newcommand{\QDONE}{9.13\times10^{-3}}% |D1| = 9.1336e-3: fit h^{5/2} G_osc = D + D1/h + D2/h^2 + D3/h^3 on 40 <= h <= 3200 (rms 1.2e-7); D1 = 5.7079e-3 - 7.1304e-3 i, spread 3e-7, statistical 3e-5 (provisional 9.14e-3)
\newcommand{\QPICZERO}{0.05383}% c0 = 0.05383167 (fit 200 <= h <= 12800, residual 6e-14 relative; spread over fit ranges/forms 7e-8); no log(h)/h term: its free coefficient |c1| < 7e-9; free fit of kbar = 3.7360987483 = log 64 + gamma - 1 to 10 digits
\newcommand{\QAC}{5.524\times10^{-3}}% A_c = 4 sqrt(2pi) E' = 5.52427e-3
\newcommand{\QBC}{-1.289\times10^{-2}}% B_c = 4 sqrt(2pi)(F - kbar E') = -1.288997e-2
\newcommand{\QCLAWB}{1.4\times10^{-4}}% the same at h >= 400: 1.40e-4 (provisional 4e-7 not reproduced; with F1/h: 2.7e-6, see \QCLAWBONE)
\newcommand{\QCDIPLO}{1.424}% lower zero of g and C: 1.4237618 (+- 4e-7 from err_Ghat; interpolation error < 1e-10)
\newcommand{\QCDIPHI}{2.211}% upper zero: 2.2114911 (+- 6e-7)
\newcommand{\QCMIN}{-5.71\times10^{-4}}% minimum of C in the dip: -5.71291e-4
\newcommand{\QCMINH}{1.726}% its position: 1.725646
\newcommand{\QCMINREL}{0.46\%}% |C_min|/C(0) = 0.4570%
\newcommand{\QSPCUV}{0.0722219}% [Q] C_UV = 0.9925441785 a = 0.07222185988 (a = end-layer quadrature, error 3e-12); unchanged
\newcommand{\QSPUthree}{-0.0198}% [F] U_3 = -0.6394811 C_3 = -0.019777, C_3 = 0.030927 from the final c = 0.018177 (provisional -0.019830); same print
\newcommand{\QSPsumacc}{10^{-12}}% [N] |8 int_0^oo P dK - 1| = 4.7e-13 (Gauss panels in log K and log|K-sqrt2| to K = 1e4, closed-form UV tail; A0: 7.5e-13). Provisional 3e-4
\newcommand{\QSPIRacc}{3\times10^{-6}}% [N] max |law/computed - 1| of P/K for 1e-4 <= K <= 1e-2: maximum 3.3e-6 at K = 3.8e-3 (the decade points give -4.0e-07, -1.8e-06, 2.8e-06 at 1e-4, 1e-3, 1e-2) (smaller below: 4e-08 at 1e-6); M = 0.025386813. Provisional 1e-6
\newcommand{\QSPPKLO}{10^{-8}}% [N] lower end of the computed K, all with P > 0 (A0: 741 values 1e-8..1e4, 1201-point log grid; A1: 1401-point log grid from 1e-10). Provisional 1e-6
\newcommand{\QSPPKHI}{10^{4}}% [N] upper end (same). Provisional 1e3
\newcommand{\QSPPMAX}{0.0227}% [N] maximum of P: 0.022703919 (zero of K dP/dK); unchanged
\newcommand{\QSPKMAX}{1.524}% [N] its position K = 1.5242715; unchanged
\newcommand{\QSPnegwin}{10^{-3}}% [N] n < 0 for sqrt2 - K < 1.39e-03 and for K - sqrt2 < 4.75e-04; unchanged
\newcommand{\QSPTHDMIN}{10^{-12}}% [N] NEW: smallest |K - sqrt2| of the computed slopes (111 values per side, 1e-12..0.1; quadrature changes < 1e-16)
\newcommand{\QCOcone}{0.1207}% c_1 = 2^{3/4} Gamma(5/4)/Gamma(7/4) a_{1/2} = 0.12068890 (closed form x end-layer quadrature, 3e-12); computed c reproduces the 8-term series to 3e-16 at R = 1e-4
\newcommand{\QCOcthree}{0.0132}% gamma_3 C_3 = 0.42745107 x 0.0309273 = 0.0132199 [F: C_3 from the small-h fits of g and Pi with a_{1/2}, c_{3/2} fixed, spread 3e-7]; unchanged
\newcommand{\QCOZSMALL}{0.331}% (96/35) c_1 = 0.331032 (final); zeta_2 = 1.996668 at R = 1e-4 against the law 1.996690
\newcommand{\QCOATWO}{0.0360}% a_2 = 8 sqrt2 Im(D e^{i pi/4}) = 0.0359942 with the final Floquet D = 2.955991e-3 + 1.543284e-3 i (0.035994 from A_2); used by D5, D8, D9
\newcommand{\QCOKAPPA}{0.0178}% kappa_2 = 4 sqrt2 pi Re(D e^{i pi/4}) = 0.0177526 (Floquet D; 0.017751 from A_2); used by D5, D9
\newcommand{\QCOWAVEFITD}{0.3\%}% fit of R^3 Pi (c - c_removed)/sqrt(2pi) on 50 <= R <= 120 with Re[(D + E/R) e^{i(sqrt2 R - pi/4)}] + offsets {1, R Pi, R} (the provisional model): |D| within 0.31% (3.12e-3); 0.77% on [30,80]; with D + E/R + F/R^2 + G/R^3 and offsets {R Pi, Pi/R, Pi/R^3}: 1.3e-5
\newcommand{\QCOWAVEFITARG}{3\times10^{-4}}% the same fit: arg D within 3.3e-4 rad; with three orders in 1/R: 2.0e-5; the fitted E = 0.009104 - 0.012032 i agrees with sqrt2 [D_1 - i D (7/16 + eps/4)] = 0.009096 - 0.012045 i (eps = d log Pi/d log h = 0.126) to 1e-3
\newcommand{\QCOM}{0.0254}% M = 0.0253868 (direct regularized first moment of h C, m_direct.py; = the spectrum IR value 0.0253868 of spectrum_P_final.json to 9e-10); data errors +-3e-5; used by D5 eq:ev-IR
\newcommand{\QCOLAWACC}{2\times10^{-4}}% max |R^2 c/(law eq:osc-c-large + wave eq:osc-c) - 1| over all computed R >= 100 (0..125 step 0.01, 125..1100 step 0.1, log grid to 1e4): 1.8e-4 (7.3e-5 at R = 100 itself, 4.5e-5 at 316, 2.8e-5 at 1e3, 1.0e-5 at 1e4; <= 1e-4 for R >= 164, <= 3e-5 for R >= 910); without the wave 2.2e-3. CHANGED (was 10^{-4}: sampled at R = 100, 316, ... only)
\newcommand{\QCOLAWFROM}{100}% unchanged
\newcommand{\QCORMIN}{10^{-3}}% c(R) > 0 at all 22652 computed radii, 0 <= R <= 1e4 (log grid 1e-4..1e4, 0..125 step 0.01, 125..1100 step 0.1); min on 2..125 is 3.29e-6; kept at the range of Fig. fig:ev-corr(a)
\newcommand{\QCORMAX}{10^{4}}%
\newcommand{\QCONOMSIX}{0.8}% max |n_w^osc| on 6 <= R < 8: 0.810 (envelope eq:osc-n with c of the removed source 0.806; local amplitude at R = 6: 0.83); used by D8
\newcommand{\QCONOMTEN}{0.7}% max |n_w^osc| on 8 <= R < 12: 0.740; local amplitude at R = 10: 0.690 (middle form 0.659)
\newcommand{\QCONOMFIFTY}{0.4}% max |n_w^osc| on 45 <= R < 60: 0.384 (middle form 0.387); local amplitude at R = 50: 0.373; used by D8
\newcommand{\QCOZTEN}{0.486}% zeta_2(10) = 0.486429
\newcommand{\QCOZHUNDRED}{0.322}% zeta_2(100) = 0.322387
\newcommand{\QCOZTENFOUR}{0.197}% zeta_2(1e4) = 0.196734
\newcommand{\QCOZTWOTHIRDS}{3.39}% zeta_2 = 2/3 at R = 3.391931 (root finding; data errors 1e-6)
\newcommand{\QCOZOSCSIX}{6\times10^{-3}}% max |zeta_2^osc| near r = 6L: 6.4e-3 with the split of the oscillating part that needs no average over a period (anc/qsp/coordinate/verify/ana2.py); a local fit over one period gives 7.2e-3, 12% high, the plain running mean 8.4e-3. CORRECTED by the independent check

\journal{Annals of Physics}

\begin{document}

\begin{frontmatter}

\title{The Quantum Pendulum of Turbulent Flow\\[0.4em]{\large Decaying turbulence as a thermal state of an integrable system}}
\author[inst1]{Alexander Migdal}
\ead{amigdal@ias.edu}
\affiliation[inst1]{organization={Institute for Advanced Study},
            city={Princeton},
            postcode={08540},
            state={NJ},
            country={USA}}

\begin{abstract}
We present a theory of decaying turbulence in which the statistics of a classical flow becomes quantum mechanics in loop space.
The loop functional is the wave function, the Navier--Stokes equation makes it evolve linearly, and the viscosity is the Planck
constant. Turbulence at infinite Reynolds number is defined by a regularized loop equation on polygons with $N$ sides and a
renormalization. This equation has an exact manifold of decaying solutions, the Euler ensemble, each a singular point of the
deterministic evolution. We interpret this instability as spontaneous stochasticity: thermal noise only triggers a diffusion over
the Euler states, and its amplitude does not enter the statistics. The equal probability of these states is our only physical hypothesis; given this diffusion, it is proved as an H-theorem.
The equal-weight sum is a trace over fermions, which reduces to Gibbs averages of two quantum pendulums with the
Mathieu Hamiltonian (formally verified in Lean~4 in a forthcoming paper). The energy spectrum, a chain of explicit quadratures,
falls as $k^{-7/2}$ at large $k$. At $kL=\sqrt2$, with $L=\sqrt{\tilde\nu(t+t_0)}$, its slope diverges: the decay at a fixed wave
number is singular at a finite time, and the vorticity correlation oscillates in space. At small $k$ there are two solutions, for
a quenched and an annealed orientation of the loop: $k^{-1}\log(1/kL)$ and $k^{-2}/\log(1/kL)$, with energy decay indices of
$1.1$ to $1.2$ and about $1.45$. The comparison with simulations and experiment is left to a forthcoming paper.
\end{abstract}

%% Highlights are submitted as a separate file (Highlights.docx); uncomment to print them on a page before the title.
%\begin{highlights}
%\item Decaying turbulence becomes quantum mechanics in loop space
%\item Thermal noise only triggers the random motion over exact solutions
%\item The Euler ensemble reduces to two quantum pendulums; verified in Lean 4
%\item An explicit energy spectrum, singular at $kL=\sqrt2$: a finite-time singularity
%\item Two infrared solutions, quenched and annealed, with different decay indices
%\end{highlights}

\begin{keyword}
Decaying turbulence \sep Loop equation \sep Euler ensemble \sep Spontaneous stochasticity \sep Quantum pendulum \sep Integrable system \sep Finite-time singularity
\end{keyword}

\end{frontmatter}

\tableofcontents

\section{Introduction}
\label{sec:introduction}

Turbulence has resisted a theory from first principles for more than a century. The phenomenology that grew around it is rich and
useful, but it does not answer the question that a theoretical physicist would ask first: what is the probability distribution of
a turbulent flow, the analog of the Gibbs distribution of statistical mechanics? In this paper we address this question for freely
decaying turbulence. We do not look for a distinguished velocity field. We look for the statistical state: the probability law of
the observables, as in Hopf's functional formulation~\cite{Hopf1952}, which a whole class of initial distributions should approach
after the natural change of scale. Let us call it the turbulent attractor, as suggested by Hopf~\cite{Hopf1948}, though he had in
mind a finite-dimensional attractor, like those studied in nonlinear systems at the time. Our attractor will be
infinite-dimensional and will live in Hilbert space.

\begin{figure}[htb]
 \centering
 \includegraphics[width=\textwidth]{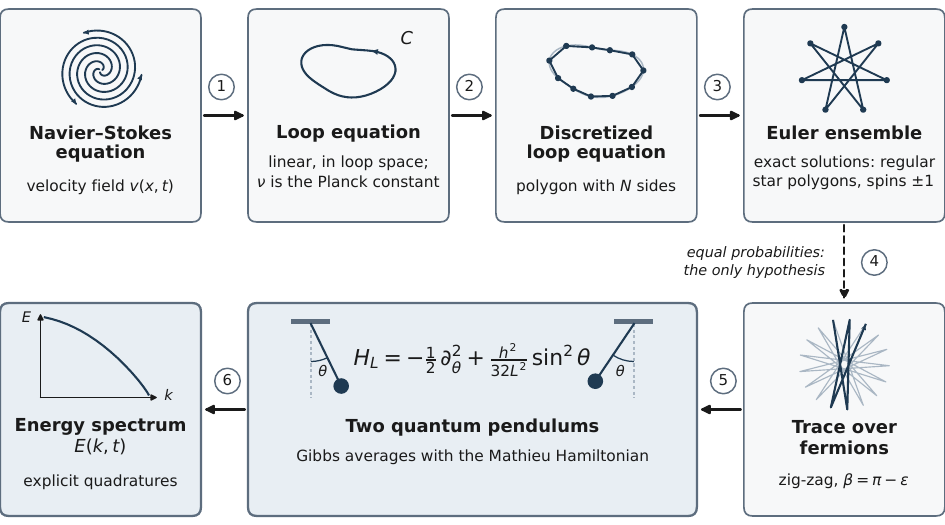}
 \caption{The path of the solution, from the Navier--Stokes equation to the energy spectrum. The pictures show a vortex of the
 flow; the loop $C$; its polygon; a regular star polygon, a state of the Euler ensemble; a nearly degenerate one, with the turning
 angle $\beta=\pi-\epsilon$, which dominates the statistical limit; the two pendulums; and the spectrum. Solid arrows are exact transformations, definitions or theorems; the dashed arrow is the only physical hypothesis
 (Table~\ref{tab:status}).
 Step~1: the circulation of the velocity around a loop; its average exponential obeys a linear equation (Sec.~\ref{sec:mle}).
 Step~2: the regularization, which approximates the loop by a polygon with $N$ sides (Sec.~\ref{sec:mle} and
 Remark~\ref{rem:intro-status}). Step~3: the exact solution (Sec.~\ref{sec:three-dimensional}). Step~4: equal probabilities of the solutions, the hypothesis; its mechanism is a diffusion over the solutions triggered by the
 thermal noise, for which it is an H-theorem (Secs.~\ref{sec:stoch-mechanism} and~\ref{sec:stoch}). Step~5: the limit
 $N\to\infty$, a theorem verified in Lean~4 (Secs.~\ref{sec:3d-fermion}, \ref{sec:cond-pendulum} and~\ref{sec:lean}). Step~6: the correlation of the vorticity, a conditional expectation by Bayes' formula, and the renormalization
 (Secs.~\ref{sec:cond-def} and~\ref{sec:3d-renorm}).}
 \label{fig:path}
\end{figure}

Our answer is unusual, and we state it at once: the turbulent attractor is a thermal state of an integrable quantum system, two
pendulums with the Mathieu Hamiltonian. This is not a model. Nobody would guess such a model, and we did not: the pendulum is
derived from the Navier--Stokes equation, by the chain of transformations shown in Fig.~\ref{fig:path}. The Navier--Stokes
equation is translated into loop space, where it becomes a linear equation. This equation, discretized on a polygon, is solved
exactly by a number-theoretic set of states, the Euler ensemble. In the statistical limit the sum over this set reduces to the
quantum pendulum. The temperature is not that of the fluid, and the quantum mechanics is not that of its molecules: both live in
loop space. The energy spectrum and the vorticity correlations then follow as explicit quadratures. Not every link of this chain is a theorem. One link, and only one, is a physical hypothesis: the equal probability of all the
states of the Euler ensemble. The other links are theorems, definitions and computations, and we say at every step which is
which.

\paragraph{Turbulence as quantum mechanics in loop space.} The story began with the loop equation~\cite{M93}. Instead of the
velocity field we follow the circulation $\Gamma_C$ of the velocity around a closed loop $C$ and its average exponential, the loop
functional $\Psi[C]=\langle\exp(\ii\Gamma_C/\nu)\rangle$. It carries the statistics of the flow on a contour, and incompressibility
and the elimination of the pressure are built into it. The Navier--Stokes equation makes it evolve by a linear functional equation,
the loop equation~\cite{M93,MigdalDuality}, much as a probability distribution evolves linearly under the Liouville equation while
the trajectories themselves are nonlinear.

The analogy with quantum mechanics is not a metaphor. By an exact operator identity (Sec.~\ref{sec:qsp-identity}) the loop
functional of any flow is the loop-ordered exponential of the covariant derivative $\hat D=\partial+\ii v/\nu$. The loop parameter
is the time of this quantum mechanics, the viscosity is its Planck constant, the vorticity is the commutator of two covariant
derivatives, and the loop functional is its wave function, or more precisely its
amplitude~\cite{migdal2024quantum,ReviewPaperAM}.

\paragraph{Statistical states and the ergodic hypothesis.} What is a statistical state of the fluid in this language? It is a
density matrix in loop space, and the loop functional is the ordered trace of the evolution along the loop. We call this the Quantum
Statistical Principle (Sec.~\ref{sec:qsp}). As a form it assumes nothing: a thermal distribution and a deterministic flow are
written in it on equal terms (Supplement~\ref{sec:S-qsp-examples}). The physics enters with the choice of the state. For decaying
turbulence we take the state reached at the end of the decay: the rescaled flow is carried to a fixed manifold of decaying
solutions, and every state on it has the same weight, as in the microcanonical ensemble. This is a quantum form of the ergodic
hypothesis of statistical mechanics, applied to the fixed decaying trajectory in loop space (Sec.~\ref{sec:qsp-principle} and
Supplement~\ref{sec:qsp-stochasticity}). It is the only physical hypothesis of the theory. The measure of the
observables that the vorticity correlation needs is not postulated, it is derived: the equal-weight sum over the states is a quantum
trace over fermions, an exact identity, and in the statistical limit this trace becomes the Gibbs trace of two quantum pendulums
(Secs.~\ref{sec:3d-fermion} and~\ref{sec:cond-pendulum}).

\paragraph{Regularization: a definition of turbulence.} Turbulence at infinite Reynolds number is singular at small scales. As in
quantum field theory, it has to be defined by a regularization and a renormalization. Our regularization approximates the loop by a
polygon with a large number $N$ of sides, on which the loop equation becomes a finite system of ordinary differential equations,
the discretized loop equation (Sec.~\ref{sec:mle}). The turbulent attractor is then defined as the limit $N\to\infty$ of its decaying
solutions, taken with a vanishing viscosity at a fixed turbulent viscosity $\widetilde\nu$, which survives as the single physical
scale. Is this a solution of the turbulence problem, or its definition? Remark~\ref{rem:intro-status} explains why it is a
definition rather than an incomplete solution of a problem posed in advance.

\begin{remark}
\label{rem:intro-status}
Let us note that there is no rigorous proof that our solution solves the Navier--Stokes loop equation. This solution, the Euler
ensemble defined below, is a discrete set of exact solutions of the discretized loop equation. The mathematical meaning of the loop
equations and a part of the construction of the solution were justified by Bru\'e and De~Lellis~\cite{DeLellisInprep}. The unproven part is the continuum limit of the discretized loop equation, as the number $N$ of
sides of the approximating polygon goes to infinity.

There is a deeper problem with this limit, and it changes the nature of the question. The Navier--Stokes equation is itself an
approximation to a physical fluid, which is made of moving molecules. Extreme turbulence, the limit of infinite Reynolds number, has
singularities at small scales: the circular Burgers vortex and the Burgers vortex sheet carry an infinite vorticity in an
infinitesimal region as $\nu\to0$~\cite{Burgers1948}, and turbulence has a hierarchy of such structures at smaller and smaller scales,
which conflict with the molecular structure of the fluid. In our solution the conflict is explicit: the enstrophy of the
regularized ensemble comes from wave numbers that grow with $N$ (Sec.~\ref{sec:3d-zero-separation}), and on the scale of the
viscosity the vorticity correlation is singular (Supplement~\ref{sec:hairpin-not-one}). A mathematical definition of extreme turbulence
therefore requires a regularization. This is the situation of the ultraviolet divergences of quantum field theory, which is
incomplete at small distances and is redefined there by a regularization; the theory at larger distances must come out
renormalizable, independent of the details of the regularization when expressed through physical scale parameters such as masses
and charges. Here the regularization is the polygon itself, the approximation of the loop by a polygon with a large number $N$ of
sides: its angular resolution $1/N$ plays the role of the lattice spacing, the vanishing viscosity $\nu_N=\widetilde\nu/N^2$ of an auxiliary flow the role of the bare coupling, and the
turbulent viscosity $\widetilde\nu$ the role of the physical scale (Sec.~\ref{sec:3d-renorm}). Renormalizability means that at fixed
$\widetilde\nu$ the whole dependence on $N$ reduces to a power of $N$ in front of a function that does not depend on $N$. This is
what we find for the energy spectrum: evidence for, though not a proof of, the renormalizability of the whole theory. Thus the loop
equations and their solution provide a \emph{definition} of turbulent flow at infinite Reynolds number, rather than a solution of a
problem defined in advance.

As a theoretical physicist, the author takes the usual position of treating this theory as a working hypothesis and investigating
its consequences. The statistics rests on one physical hypothesis, the equal probability of the Euler states, a counterpart of the
ergodic hypothesis of statistical mechanics; it is not derived from the Navier--Stokes equation. What we do is to compute the
statistical limit, starting from the Euler ensemble, as Onsager computed the statistical limit of the two-dimensional Ising
model~\cite{Onsager1944}. The underlying ergodicity, the equidistribution over the states, is not proven in either case: the ergodic
hypothesis behind the statistical mechanics of Boltzmann, Maxwell and Gibbs is still not proven. The limit is another matter. The
reduction of the Euler ensemble to the quantum pendulum, the statistical limit that we computed, is rigorously proved, with an
estimate of the rate of convergence (Supplement~\ref{sec:S-finiteN-proof}), and its convergence is formally verified in
Lean~4~\cite{JiangMigdalLean2026} (Sec.~\ref{sec:lean}). If the results are mathematically consistent, satisfy the basic physical
requirements and match simulations and experiments, the theory would describe reality while awaiting a full mathematical proof, as
quantum electrodynamics did when its observables were computed by renormalization before the procedure had any mathematical
justification. Quantum mechanics, relativity and Yang--Mills theory began in this way.
\end{remark}

\paragraph{The Euler ensemble.} The discretized loop equation has an exact manifold of decaying solutions, the Euler
ensemble~\cite{migdal2023exact} (Sec.~\ref{sec:three-dimensional}). Its states are regular star polygons in a complex
three-dimensional space, labeled by a rational turning angle $2\pi p/q$, a sequence of $N$ spins $\pm1$ and a winding number. On
each of them the viscous and the advective terms of the equation vanish separately. Counting these states brings in number theory,
through Euler's totient function, and the sector with zero winding, where every rational angle is allowed, outnumbers the others by
a factor of order $N$ (Sec.~\ref{sec:3d-ensemble} and Ref.~\cite{Zah23}). This even ensemble at zero winding is our regularized
turbulent attractor. Its alternatives, the odd ensemble and the planar solutions, remain exact solutions, but their vorticity
correlations are not positive (Supplement~\ref{sec:S-intro-details}).

\paragraph{Instability, spontaneous stochasticity and a model of ergodicity.} Here a warning is in order. The equal weight of the
Euler states is not what a direct integration of the discretized equation \eqref{eq:rescaled-full-mle} produces. A reader who solves
these ordinary differential equations near the Euler states will not see the trajectories settle on them with equal frequencies:
the perturbed trajectories leave the Euler states and grow (Fig.~\ref{fig:stoch}b). The reason is structural. In the even ensemble at
zero winding the time derivative term of the equation is not invertible, and at every state of this sector the matrix that should
determine the remaining part of the velocity vanishes identically (Proposition~\ref{prop:stoch-singular}). An arbitrarily small
displacement then produces a velocity of order one, in a direction that the displacement chooses. So, the deterministic evolution
is ambiguous exactly on the ensemble, and ergodicity cannot be a property of it.

We interpret this instability as the source of spontaneous
stochasticity~\cite{BernardGawedzkiKupiainen1998,FalkovichGawedzkiVergassola2001,BandakMailybaevEyinkGoldenfeld2024}. A vanishing
noise, such as the thermal noise that is present in every fluid, displaces the vertices of the polygon, acts in the undetermined
mode and results in a random motion over the manifold of the Euler states: a diffusion, by infinitesimal moves between close
states. The noise is a trigger and not a source: neither its amplitude nor the temperature of the fluid
enters the statistics of the motion that it starts (Sec.~\ref{sec:stoch-trigger}).

We cannot derive this random motion from the Navier--Stokes equation. What we can do is to study the simplest model of it
(Sec.~\ref{sec:stoch}). The logarithmic time of the rescaled equation is made discrete, and in every step the polygon passes from
one Euler state to another: by an exchange of two neighbouring spins, at a bond of the loop chosen at random, and by a symmetric
shift of the turning angle, reflected at both ends of its interval. The exchange is not an arbitrary choice. Every motion between
two Euler states violates the equation somewhere, and the exchange of two neighbours is the step that violates it the least
(Proposition~\ref{prop:stoch-least}). This violation vanishes as $N\to\infty$, while the random motion over the Euler states that
it produces does not, and this is how the model implements the spontaneous stochasticity. The model is a Markov chain with
symmetric transition probabilities, and for it ergodicity is a theorem, an H-theorem: the entropy grows in every step, and from every initial
distribution the system converges to equal weights of the spin sequences and to a uniform density of the angle
(Propositions~\ref{prop:stoch-H}, \ref{prop:stoch-exchange} and~\ref{prop:stoch-beta}). The choice of the moves is not unique and
is not derived from the Navier--Stokes equation, but the final state of maximal entropy is unique. The postulated state is thus the
attractor of a very general stochastic dynamics inside the Euler ensemble, and the quantum pendulum is in this sense a universal
attractor. The mechanism is the subject of Sec.~\ref{sec:stoch-mechanism} and the model of Sec.~\ref{sec:stoch}; what is proved
there and what is assumed is listed in Table~\ref{tab:status}.

\paragraph{From the Euler ensemble to the quantum pendulum.} How does one measure the correlation of the vorticity at two points
in loop space? On a narrow loop through both of them, a hairpin: the correlation is a second derivative of the loop functional with
respect to the area (Secs.~\ref{sec:contact} and~\ref{sec:3d-hairpin}). It is a conditional average, the vorticities inserted at
the two ends, normalized by the loop functional of the hairpin itself: this is the definition of a conditional probability, Bayes' formula
(Sec.~\ref{sec:cond-def}).
To compute it we write the equal-weight sum over the Euler states as a trace over fermions, one per side of the polygon, and
transform this trace exactly into a walk on a cyclic group (Sec.~\ref{sec:3d-fermion}).

Here is the most important point. Near the critical turning angle $\beta=\pi$ the polygon is a zig-zag whose direction, the slow
angle, wanders around the loop, while pairs of opposite spins carry a fast noise. The fast spins act as a thermostat for the slow
angle, the closed loop becomes its thermal circle, and the sum over the spins becomes the Euclidean evolution of two quantum
pendulums, one on each arc of the hairpin, with the Mathieu Hamiltonian $H_L=-\frac12\partial_\theta^2+h^2\sin^2\theta/(32L^2)$.
The partition function and the conditional correlations of the ensemble become Gibbs averages of these pendulums
(Sec.~\ref{sec:cond-pendulum}). This Gibbs measure is not postulated: it is what the equal weights of the Euler states become in
the limit. The temperature of the pendulums is a variable of the loop and not the temperature of the fluid: its inverse is
the variance of the walk of the slow angle around the loop (Sec.~\ref{sec:cond-thermal}). The pendulum is an integrable system. Let
us stress that this is a property of the reduced problem, not an integrability theorem for the Navier--Stokes evolution.

This spontaneous appearance of a thermodynamic ensemble from a nonlinear deterministic system resembles the appearance of the
Hawking temperature from the Schwarzschild solution of a black hole~\cite{Hawking1975}; in both cases the Gibbs distribution
appears by a dynamical mechanism, not as a result of an interaction with an external thermostat.

\paragraph{The energy spectrum and its singularity.} After the renormalization of Sec.~\ref{sec:3d-renorm} the energy spectrum is a
chain of quadratures of the resolvents of the two pendulums,
\begin{equation}
 \boxed{\begin{gathered}
 C(h)=\frac{\widehat G(h)}{\widehat Z(h)}
 =\frac1{16}\,\frac{\int_0^1\dd x\int_{-\infty}^{\infty}\dd\omega\;\mathcal T(h,x,\omega)}
 {\int_0^1\dd x\int_{-\infty}^{\infty}\dd\omega\;\operatorname{Tr}\bigl[R_x(\ii\omega)\,R_y(-\ii\omega)\bigr]},\\[4pt]
 P(K)=2K\int_0^\infty h\,J_0\bigl(\sqrt2\,hK\bigr)\,C(h)\,\dd h,\\[4pt]
 E(k,t)=\Bigl(\frac{\nu}{\widetilde\nu}\Bigr)^{1/2}
 \frac{\widetilde\nu^{\,2}}{8L}\,\frac{P(kL)}{(kL)^2},
 \qquad L=\sqrt{\widetilde\nu\,\tau},\qquad \tau=t+t_0,
 \end{gathered}}
 \label{eq:intro-main}
\end{equation}
where $\nu$ is the physical viscosity, $\widetilde\nu$ the turbulent viscosity, $t_0>0$ the time offset of the similarity solution,
$R_L(\zeta)=(LH_L-\zeta)^{-1}$ the resolvents of the pendulums of the two arcs, which take the fractions $x$ and $y=1-x$ of the loop,
and $\mathcal T$ the trace with the vorticities inserted at the two ends of the hairpin (Sec.~\ref{sec:cond-resolvents}). The
denominator is the partition function of the hairpin.

What does this spectrum look like? It is positive at every computed wave number, and the enstrophy is finite. It falls as
$k^{-7/2}$ at large $k$. At small $k$ the answer depends on a question that we could not settle: is the plane of the polygon a
quenched or an annealed variable of the conditional correlation? In the first case, the order of Definition~\ref{post:cond} and of
\eqref{eq:intro-main}, the spectrum behaves as $k^{-1}\log(1/kL)$; in the second, as $k^{-2}/\log(1/kL)$. We compute both spectra
(Secs.~\ref{sec:3d-final} and~\ref{sec:3d-orders}, Fig.~\ref{fig:ev-spectrum}). The two solutions can be told apart. The decay
index of the energy above a fixed wave number varies with time: it is $1.1$ to $1.2$ in the first and about $1.45$ in the second,
and the energy and its dissipation depend on the size of the flow logarithmically in the first and almost linearly in the second
(Sec.~\ref{sec:3d-size}).

The most unexpected feature is common to the two solutions. At $kL=\sqrt2$ there is a threshold, produced by the hairpins with two
equal arcs, where the spectrum stays continuous while its slope diverges as a double logarithm. Since time enters only through
$L(t)$, the decay at a fixed wave number has a genuine finite-time singularity. In coordinate space the threshold becomes a wave:
the vorticity correlation oscillates with the period $\sqrt2\,\pi L\simeq4.44\,L$ (Sec.~\ref{sec:3d-osc} and
Fig.~\ref{fig:ev-corr}).

One thing changed while this paper was being written. The convergence of the whole ensemble to \eqref{eq:intro-main} is supported
here by exact sums and Monte Carlo simulations (Remark~\ref{rem:3d-matching}). It is now a theorem: the reduction of the Euler
ensemble to the quantum pendulum is proved, and formally verified in the proof assistant Lean~4, in the forthcoming paper by
Y.~Jiang, A.~Migdal et al.~\cite{JiangMigdalLean2026} (Sec.~\ref{sec:lean}).

\paragraph{How the picture was derived, and what is new here.} The loop equation was derived in 1993~\cite{M93}. Its exact solution,
the Euler ensemble, was found thirty years later, together with its relation to number theory~\cite{migdal2023exact}. The ensemble was
then written as a quantum trace over fermions and its spectrum evaluated by a saddle
point~\cite{migdal2024quantum,ReviewPaperAM,NTDT2}. Some steps of that work were heuristic. With the help of AI assistants, the
present paper goes much deeper into the mathematical structure of this solution and replaces several of them by exact ones. The operator identity makes the quantum mechanics in loop space a property of every
flow, and the Quantum Statistical Principle a general rule rather than an ansatz (Secs.~\ref{sec:qsp-identity} and~\ref{sec:qsp}).
The vorticity correlation acquires contact terms that make it invariant under translations and reparametrizations of the loop
(Sec.~\ref{sec:contact}), and it is normalized as a conditional average, since the loop functional of the hairpin is not one
(Supplement~\ref{sec:hairpin-not-one}). The saddle point is replaced by the exact reduction to the quantum pendulums, which computes the sum
over the spins directly (Sec.~\ref{sec:3d-fermion}). As a result the spectrum of the earlier versions is superseded, and we withdraw
the earlier claim that the zeros of the Riemann zeta function control the decay: in the local limit the arithmetic leaves only a flat
density of rational angles, and the singularity of the spectrum is geometric, not arithmetic. The corrections are listed in
Supplement~\ref{sec:S-intro-details}. Finally, the equal weight of the Euler states, a bare postulate in the earlier work, acquires a
mechanism, the spontaneous stochasticity of the equation and the stochastic model of ergodicity of
Secs.~\ref{sec:stoch-mechanism} and~\ref{sec:stoch}. This model was formulated in discussions with Claude Fable~5.2,
after the author's initial idea of stochastic perturbations from thermal noise along the zero mode of the momentum-loop equation.

\paragraph{Status of the results.} It is important to distinguish what is proved, what is computed, and what is assumed
(Table~\ref{tab:status}). The reduction of the Euler ensemble to the quantum pendulum is now a theorem, formally verified in
Lean~4~\cite{JiangMigdalLean2026}: the exact finite-$N$ statements; the convergence of the even ensemble at zero winding to the pendulum formulas, for
every $h>0$ (Supplement~\ref{sec:S-finiteN-proof}); and, in the grand canonical limit, the suppression of the odd sectors, of the
sectors with nonzero winding and of the angles away from the critical layer (Sec.~\ref{sec:lean}). There is only one physical
hypothesis: the equal probability of all the states of the Euler ensemble. We do not derive it, but we provide a mechanism for this
ergodicity (Secs.~\ref{sec:stoch-mechanism} and~\ref{sec:stoch}): the thermal fluctuations of the vertices of the polygon trigger a
diffusion over the Euler states, by infinitesimal moves between close states. Given this noise-triggered singular diffusion, the
ergodicity is proved, as an H-theorem. The Gibbs measure of the pendulums is not a hypothesis: it is derived from the equal weights,
whose sum is a trace over fermions. Nor is the division of the correlation by the amplitude of the hairpin: it is the definition of
a conditional probability, Bayes' formula (Definition~\ref{post:cond}). The matching of the renormalized amplitude is the
renormalization condition, a part of the definition of the theory (Sec.~\ref{sec:3d-renorm}). The open questions are whether the
limit is independent of the regularization, whether the random moves of the diffusion can be derived from the singular evolution, whether the global orientation of the polygon is quenched or annealed in
the conditional correlation, which changes the infrared law and the finite-size effects but neither the ultraviolet law nor the
threshold (Secs.~\ref{sec:stoch-rotation}, \ref{sec:3d-orders} and~\ref{sec:3d-size}, where both solutions are given), and whether the theory describes physical turbulence. We do not compare it with simulations and experiment in this paper: Sec.~\ref{sec:comparison} lists what the data can test, and the
comparison of both solutions with direct numerical simulations and with the Max Planck wind tunnel will be given with Wei
Wang~\cite{MigdalWang2026QPTF}.

\begin{table}[htp]
\centering\small
\caption{The status of the results.}
\label{tab:status}
\begin{tabular}{@{}>{\raggedright\arraybackslash}p{0.16\textwidth}>{\raggedright\arraybackslash}p{0.80\textwidth}@{}}
\toprule
Exact at finite $N$ & The operator identity \eqref{eq:qsp-identity}; the Euler solutions; the contact terms and the invariance of the
correlation; the sector counts; the zero-separation values \eqref{eq:3d-zero-separation}; the Fock trace and the transfer on
$\mathbb Z_q$. All are proved in Lean~4~\cite{JiangMigdalLean2026} (Sec.~\ref{sec:lean}).\\
Proved limits & The convergence of the even ensemble at zero winding to the pendulum formulas, $D^{\rm bulk}_N/N^{3/2}\to\widehat Z$,
$W_N/\sqrt N\to\widehat G$ and $NW_N/D^{\rm bulk}_N\to C$, for every $h>0$, and the grand canonical limit, formally verified in
Lean~4 in Ref.~\cite{JiangMigdalLean2026}. The rates of convergence, uniform on compact sets of $h>0$ with explicit errors, are
proved on paper but not formalized (Supplement~\ref{sec:S-finiteN-proof}).\\
Controlled asymptotics & The pendulum limit, with errors $O(1/N)$ at fixed $\xi$, $x$, $h$; the density of the rational angles near
$\pi$. In this limit, exactly: $C(0)=\frac18$ and the small-$h$ expansions, the growth $\log h$ of $h\widehat Z$, the forms of the
infrared, ultraviolet and threshold laws, $C_{\rm UV}$ as a quadrature.\\
Numerical evidence & The convergence of the whole ensemble to the pendulum source (exact sums to $N=120$) and of its critical layer
(Monte Carlo to $N=10^5$); the convergence of the ratio of the numerator to the bulk of the bare hairpin to its pendulum value in the critical layer and its
agreement with it within $1.6$ standard deviations at five values of $h$ over the whole even ensemble at zero winding (Monte Carlo to
$N=32000$, and $64000$ at $h=\frac14$); the contribution of the sector with $r\ne0$ of the grand ensemble to the conditional
correlation, which falls as $1/N$ (exact sums to $N=64$, Monte Carlo of the critical layer to $N=1024$); the quadratures of $\widehat G$, $\widehat Z$, $P$, and $P>0$ at every computed $K$, in both orders of the orientation average (Sec.~\ref{sec:3d-orders}); the fitted tail constants $F$,
$D_1$, hence $B_c$, $M$ ($E'$, $A_c$, $c_0$ and $D$ agree with their closed forms~\cite{JiangMigdalLean2026}, Supplement~\ref{sec:S-tail-closed}).\\
Definitions & The regularization on polygons with $N$ sides and the renormalization at fixed $\widetilde\nu=\nu_NN^2$, with the
matching of $N^{-1}$ with $(\nu/\widetilde\nu)^{1/2}$ as its condition (Remark~\ref{rem:intro-status}, Sec.~\ref{sec:3d-renorm}); the
vorticity correlation as the expectation conditioned on the hairpin, Bayes' formula (Definition~\ref{post:cond}). Its Gibbs measure
is derived, not postulated. Open: whether the orientation of the polygon is quenched or annealed in the conditioning
(Sec.~\ref{sec:3d-orders}).\\
Stochastic model & Proved (Secs.~\ref{sec:stoch-mechanism} and~\ref{sec:stoch}): the time derivative term is singular at every even state of zero winding;
no solution moves along the fixed points; in the equation with a discrete logarithmic time the residual of a step between two Euler states is its impulse, and the exchange of neighbouring spins is the step of the least impulse, $O(N^{-1/2})$ on two sides;
the discrete step, whatever its length, is valid for the distribution of the states, with errors of the order of $k/N$ for the correlations of $k$ spins, and the relaxation time is finite if the step decreases as $N^{-3}$; the relaxation rate is bounded by the intensity of the impulses; under symmetric spin exchanges and reflected shifts
$\pm\Delta\beta$ of $\beta$ the entropy grows in every step and the joint law becomes uniform
(Propositions~\ref{prop:stoch-singular}--\ref{prop:stoch-beta}). Assumed: that the instability acts as these moves.
Not formalized in Lean.\\
Physical hypothesis & One: the equal probability of all the Euler states. It is not derived. Its mechanism is the diffusion over
the Euler states triggered by the thermal noise, and given this diffusion it is proved, as an H-theorem (the stochastic model
above).\\
\bottomrule
\end{tabular}
\end{table}

The paper is organized as follows. Sections~\ref{sec:mle}--\ref{sec:contact} give the loop equation and its operator form, the
Quantum Statistical Principle and the vorticity correlation; Secs.~\ref{sec:three-dimensional}--\ref{sec:cond-pendulum} the Euler
ensemble, the mechanism of spontaneous stochasticity and the stochastic model of the equal weights, the conditional correlation, the reduction to the pendulums and their Gibbs state; Secs.~\ref{sec:3d-even}--\ref{sec:3d-limits}
the spectrum, the wave in coordinate space and the scaling limits; Sec.~\ref{sec:comparison} the predictions that simulations and experiment can test, and Sec.~\ref{sec:conclusion}
concludes. The Supplementary Material, whose sections, equations, figures and Remarks carry the prefix S, contains the other exact
solutions, the proofs, derivations and computations.

\section{The momentum-loop equation with advection}
\label{sec:mle}

The Navier--Stokes equation is nonlinear in the velocity, but for the loop functional it is a linear equation. In this section we
make this translation step by step. Its outcome is the momentum-loop equation \eqref{eq:rescaled-full-mle}, a finite system of
ordinary differential equations for the vertices of a polygon, and the rest of the paper is the study of its solutions.

Let us start from the Navier--Stokes equation and keep the advective term
throughout. For an
incompressible velocity field in $d$ dimensions we write
\begin{equation}
 \partial_t v_\beta=-v_\alpha\omega_{\alpha\beta}
 -\partial_\beta\left(p+\frac{v^2}{2}\right)
 +\nu\partial_\alpha\omega_{\alpha\beta},\qquad
 \omega_{\alpha\beta}=\partial_\alpha v_\beta-\partial_\beta v_\alpha ,
 \label{eq:eulerian-vorticity-form}
\end{equation}
where $p$ is the pressure per unit mass and repeated indices are summed.
The circulation characteristic functional and its momentum representation are
\begin{equation}
 \Psi[C,t]=\left\langle\exp\left(\frac{\ii}{\nu}
       \oint_C v_\alpha\,\dd C_\alpha\right)\right\rangle
 =\left\langle\Phi_P[C]\right\rangle_P,
 \qquad
 \Phi_P[C]=\exp\left(\ii\oint P_\alpha\,\dd C_\alpha\right).
 \label{eq:loop-fourier-convention}
\end{equation}
The second expression is an amplitude representation; in general it need not
be an average over a positive measure on real momentum loops. Real statistical
representatives will be constructed below. The operator formulation behind
\eqref{eq:loop-fourier-convention} was developed in
Refs.~\cite{M93,migdal2023exact,migdal2024quantum,ReviewPaperAM}.

The pressure gradient integrates to zero around a closed contour. For a fixed
spatial loop the exact equation is
\begin{equation}
 \partial_t\Psi[C,t]=\frac{\ii}{\nu}\oint\dd\theta\,C'_\beta(\theta)
 \left\langle\left[\nu\partial_\alpha\omega_{\alpha\beta}
             -v_\alpha\omega_{\alpha\beta}\right]_{C(\theta)}
             \exp\left(\frac{\ii}{\nu}\oint_Cv\cdot\dd C\right)
 \right\rangle .
 \label{eq:fixed-loop-equation}
\end{equation}
The velocity stays inside the average. Passing to a material contour moves
advection into the time derivative; it does not make advection vanish at a
fixed contour.

\subsection{The operator identity}
\label{sec:qsp-identity}

The loop functional of any flow is the ordered exponential of an operator, the covariant derivative. This is an exact identity, and
the momentum representation \eqref{eq:loop-fourier-convention} rests on it: it explains the correspondence between the covariant
derivative and the momentum loop that the rest of this section uses. Let $v(x)$ be the velocity field at a given time. The covariant derivative of the velocity,
\begin{equation}
 \hat D_\alpha(x_0)=\frac{\partial}{\partial x_{0\alpha}}+\frac{\ii}{\nu}\,v_\alpha(x_0),\qquad
 \bigl[\hat D_\alpha(x_0),\hat D_\beta(x_0)\bigr]=\frac{\ii}{\nu}\,\omega_{\alpha\beta}(x_0),
 \label{eq:qsp-covariant}
\end{equation}
is an operator in the Hilbert space $L^2(\mathbb R^d)$ of functions of a point $x_0$. Let the loop be attached at this point,
$C(\theta)=x_0+\int_{\theta_0}^{\theta}\dd C$, so that $x_0=C(\theta_0)$ and the shape of the loop is given by its tangent. The
loop-ordered exponential of $\hat D$ is the limit of the ordered products over the polygons inscribed in the loop, with the earlier
points to the left, Eq.~\eqref{eq:qsp-pexp}.

\begin{proposition}[Operator identity~\cite{ReviewPaperAM}]
\label{prop:qsp-identity}
The loop-ordered exponential of the covariant derivative is the circulation phase of the loop attached at $x_0$, times the unit
operator:
\begin{equation}
 \boxed{\mathcal P\exp\Bigl(\oint\dd C_\alpha\,\hat D_\alpha(x_0)\Bigr)=\exp\Bigl(\frac{\ii}{\nu}\oint_Cv_\alpha\dd C_\alpha\Bigr)\,\mathbb I .}
 \label{eq:qsp-identity}
\end{equation}
\end{proposition}

The proof follows the translations generated by $\hat D$ along the edges of the closed inscribed polygon, which return to $x_0$ and
leave only the phase (Supplement~\ref{sec:S-mle-proofs}).

The right side of \eqref{eq:qsp-identity} is a function of $x_0$, a multiplication operator. For a statistical ensemble of velocity
fields that is homogeneous in space its average does not depend on $x_0$, and the loop functional \eqref{eq:loop-fourier-convention}
appears as a number times the unit operator,
\begin{equation}
 \Bigl\langle\mathcal P\exp\Bigl(\oint\dd C_\alpha\,\hat D_\alpha(x_0)\Bigr)\Bigr\rangle=\Psi[C]\,\mathbb I .
 \label{eq:qsp-identity-average}
\end{equation}
The base point is therefore arbitrary: it is the point of the loop at which the operators are taken. The momentum loop of \eqref{eq:loop-fourier-convention} represents this operator: it replaces the covariant derivative inserted at a
point of the loop by $\ii P$ at that point.

\subsection{Operator correspondence and the calculus of momentum-loop discontinuities}
\label{sec:mle-calculus}

The momentum representation \eqref{eq:loop-fourier-convention} is a correspondence of operators, and the momentum-loop equation, the
velocity and the vorticity correlation of Sec.~\ref{sec:contact} are applications of one calculus. Let $e(\theta)=C'(\theta)$ be the tangent of the
loop, with the loop reconstructed from its base point, $C(\theta)=C(0)+\int_0^\theta e(s)\dd s$, and let $D_\alpha(\theta)=\delta/\delta e_\alpha(\theta)$. On the
two phases it acts as
\begin{equation}
 D_\alpha(\theta)\Phi_P=\ii P_\alpha(\theta)\,\Phi_P,\qquad
 D_\alpha(\theta)\Phi_v=\frac{\ii}{\nu}\,Q_\alpha(\theta)\,\Phi_v ,
 \label{eq:calc-correspondence}
\end{equation}
where $\Phi_v=e^{\ii\Gamma/\nu}$, $\Gamma=\oint v\cdot\dd C$, and the tangent response
$Q_\alpha(\theta)=v_\alpha(C(0))+\int_\theta^{2\pi}\omega_{\alpha\beta}(C)\,\dd C_\beta$ is continuous along the loop. In the operator form of Sec.~\ref{sec:qsp-identity},
$D_\alpha(\theta)$ inserts $\hat D_\alpha$ at the point $\theta$ of the ordered product. The momentum loop represents this insertion by $\ii P_\alpha(\theta)$:
this is the correspondence $\hat D_\alpha\leftrightarrow\ii P_\alpha$. Insertions at different points are ordered along the loop, the earlier points
to the left. The values $P(\theta)$ at different points are numbers and commute; the order is carried by $\theta$. What remains to be fixed is
the product of operators at one point.

\paragraph{Midpoints and jumps.} We regularize the parametric circle of the loop by $N$ cyclic momentum vertices $P_k$; the loop is piecewise
constant, and at every vertex it jumps. At an edge we introduce the midpoint and the jump,
\begin{equation}
 \bar P_k=\frac{P_{k+1}+P_k}{2},\qquad
 \Delta P_k=P_{k+1}-P_k,\qquad
 H_k=\bar P_k\cdot\Delta P_k,
 \label{eq:midpoint-jump}
\end{equation}
and the antisymmetric tensor (no factor $\frac12$)
\begin{equation}
 B_{\alpha\beta,k}
   =\bar P_{\alpha,k}\Delta P_{\beta,k}
      -\bar P_{\beta,k}\Delta P_{\alpha,k}.
 \label{eq:jump-bivector}
\end{equation}
All products of complex vectors are bilinear,
$z\cdot w=\sum_\alpha z_\alpha w_\alpha$, without complex conjugation.

\paragraph{Products at a jump.} At a jump every quantity has two one-sided values, $A_-$ on the earlier side and $A_+$ on the later one. The
commutator of two operators at the jump and the product rule are
\begin{equation}
 [A,B]=A_-B_+-B_-A_+=\bar A\,\Delta B-\Delta A\,\bar B ,
 \label{eq:jump-commutator}
\end{equation}
\begin{equation}
 \Delta(AB)=\bar A\,\Delta B+\bar B\,\Delta A,
 \qquad
 \Delta(P^2)=2\bar P\cdot\Delta P .
 \label{eq:bv-product-identity}
\end{equation}
Both are identities for the one-sided values. The first puts the earlier factor to the left, as the ordered product does; the second is the
product rule of functions of bounded variation with midpoint values at the jumps~\cite{Ambrosio2000,Volpert1967}.

\paragraph{Vorticity as a commutator.} The covariant derivatives obey $[\hat D_\alpha,\hat D_\beta]=\ii\omega_{\alpha\beta}/\nu$. With $\hat D\leftrightarrow\ii P$ and
\eqref{eq:jump-commutator}, $[\ii P_\alpha,\ii P_\beta]=-B_{\alpha\beta}$, so the vorticity at a jump is represented by
\begin{equation}
 \widehat\omega_{\alpha\beta,k}=\ii\nu B_{\alpha\beta,k}.
 \label{eq:vorticity-symbol}
\end{equation}
In the loop calculus the commutator of two tangent derivatives across a point is the area derivative,
$\mathcal L_{\alpha\beta}(\theta)=D_\alpha(\theta-0)D_\beta(\theta+0)-D_\beta(\theta-0)D_\alpha(\theta+0)$. By \eqref{eq:calc-correspondence} it gives $-B(\theta)\Phi_P$ on the
momentum phase and $(\ii/\nu)\,\omega(C(\theta))\Phi_v$ on the circulation phase, since $Q$ does not jump. Hence, on every loop and for every ensemble,
\begin{equation}
 \bigl\langle\omega_{\alpha\beta}(C(\theta))\,\Phi_v[C]\bigr\rangle=\bigl\langle\widehat\omega_{\alpha\beta}(\theta)\,\Phi_P[C]\bigr\rangle .
 \label{eq:calc-area}
\end{equation}

\paragraph{The gradient of an insertion.} The gradient of a function inserted at a point is another commutator,
$\partial_\lambda F=[\hat D_\lambda,F]$. At the jump that carries the insertion, $[\ii P_\lambda,\widehat F]=\ii P_{\lambda,-}\widehat F-\widehat F\,\ii P_{\lambda,+}=-\ii\Delta P_\lambda\widehat F$,
provided $\widehat F$ has no one-sided values of its own, that is, it is constant across its own jump. This is the central prescription, and the
loop calculus derives it (Supplement~\ref{sec:S-mle-proofs}):
\begin{equation}
 \boxed{\bigl\langle\partial_\lambda F(C(\theta))\,\Phi_v[C]\bigr\rangle=-\ii\,\bigl\langle\Delta P_\lambda(\theta)\,\widehat F(\theta)\,\Phi_P[C]\bigr\rangle .}
 \label{eq:calc-gradient}
\end{equation}
The momentum transferred by the gradient of an insertion is the jump at its mark.
The rule concerns one insertion at a jump; a product of two insertions at the same jump gives each factor its own slice, a
vertex inserted on an edge of zero length (Supplement~\ref{sec:S-mle-proofs}).

\paragraph{The velocity.} The same rule applied to $[\hat D_\alpha,v_\beta]-[\hat D_\beta,v_\alpha]=\omega_{\alpha\beta}$ and $[\hat D_\alpha,v_\alpha]=0$, for a velocity
insertion $V$ that does not jump where $P$ jumps, gives $\Delta P_\alpha V_\beta-\Delta P_\beta V_\alpha=\nu(\Delta P_\alpha\bar P_\beta-\Delta P_\beta\bar P_\alpha)$ and
$\Delta P\cdot V=0$. For $(\Delta P)^2\ne0$,
\begin{equation}
 V_\alpha=\nu\left(\bar P_\alpha
      -\frac{H}{(\Delta P)^2}\Delta P_\alpha\right).
 \label{eq:velocity-symbol}
\end{equation}
On an equal-radius configuration $H=0$, so $V=\nu\bar P$.

\paragraph{The momentum-loop equation.} Now we can write the equation itself. The two insertions of the loop equation \eqref{eq:fixed-loop-equation} are applications of
the calculus (Supplement~\ref{sec:S-mle-proofs}). In three dimensions
the advective term and the vorticity symbol are
\begin{equation}
 \left.\partial_t\bar P\right|_{\rm adv}
    =\frac{1}{\nu}V\times\widehat{\boldsymbol\omega},
 \qquad
 \widehat{\boldsymbol\omega}
    =\ii\nu\,\bar P\times\Delta P ,
 \label{eq:advection-fourier-translation}
\end{equation}
and in every dimension the complete midpoint momentum-loop equation, first derived in
Ref.~\cite{migdal2023exact}, is
\begin{equation}
 \boxed{\displaystyle
 \partial_t\bar P_k
   =\nu\left[H_k\Delta P_k-(\Delta P_k)^2\bar P_k\right]
    +\ii\nu\left[\frac{H_k^2}{(\Delta P_k)^2}-\bar P_k^2\right]
                 \Delta P_k .}
 \label{eq:full-mle}
\end{equation}
The first bracket is diffusion; the second is advection. This tensor equation
makes sense in every dimension $d>1$; no three-dimensional cross product is
needed. The identification of this finite equation with the complete physical loop operator, and the limitations of the calculus of
jumps for a continuum limit, are discussed in Supplement~\ref{sec:S-mle} and~\ref{sec:S-bv}.

\subsection{Rescaling}

We look for self-similar decaying solutions. Let us introduce
\begin{equation}
 \tau=t+t_0>0,\quad s=\log\frac{\tau}{\tau_*},\quad
 F_k(s)=\sqrt{2\nu\tau}\,P_k(t),\quad
 m_k=\frac{F_{k+1}+F_k}{2},\quad \delta_k=F_{k+1}-F_k .
 \label{eq:rescaled-vertices}
\end{equation}
Equation~\eqref{eq:full-mle} becomes autonomous,
\begin{equation}
 2\partial_s m_k=(1-\delta_k^2)m_k+(m_k\cdot\delta_k)\delta_k
   +\ii\left[\frac{(m_k\cdot\delta_k)^2}{\delta_k^2}-m_k^2\right]\delta_k .
 \label{eq:rescaled-full-mle}
\end{equation}
For even $N$ the midpoint equations carry an alternating compatibility condition (Supplement~\ref{sec:S-mle-proofs}).

On the equal-radius sector $m_k\cdot\delta_k=0$ the advective term reduces to
$-\ii m_k^2\delta_k$. In
three dimensions it can vanish, because a nonzero complex midpoint can be
bilinearly null, $m_k^2=0$. In two dimensions a nondegenerate circle has
$m_k^2\ne0$, and advection produces a complex angular rotation (Supplement~\ref{sec:two-dimensional}).
\section{The Quantum Statistical Principle}
\label{sec:qsp}

What is a statistical state of the fluid in loop space? Our answer is the Quantum Statistical Principle: it is a quantum state, a
density matrix in a Hilbert space, and the loop functional is its ordered trace along the loop. Why do we need such a rule? The
momentum representation \eqref{eq:loop-fourier-convention} writes the loop functional as an average of the phase
factor $\Phi_P[C]$ over momentum loops. For a conventional statistical distribution of the
velocity, such as the Gibbs distribution of thermal fluctuations, it is an ordinary average over random real momentum loops.
But already for the simplest deterministic flow, the uniform rotation, the Gaussian momentum loop that reproduces the loop
functional has an imaginary correlation function, and the amplitude is a quantum-mechanical phase factor rather than a
probability (Remark~\ref{rem:intro-positivity}); in the Euler ensemble the momentum loop is complex. The rule that says
what a statistical state is in the space of momentum loops should not be designed for one particular solution, and the loop
calculus provides it. By the operator identity of Sec.~\ref{sec:qsp-identity}, the loop functional of every flow is
the ordered exponential of the covariant derivative: a trace, with a definite Hilbert space and definite operators. This is the
general form of the statistics of the flow, and it can be recast in various representations (Sec.~\ref{sec:qsp-general}). The
correspondence between classical statistics and quantum mechanics, which mathematicians have established in great generality, makes
it a principle that holds in every representation (Sec.~\ref{sec:qsp-principle}), and the examples of Supplement~\ref{sec:S-qsp-examples} show conventional
statistical states in this form.

\subsection{The general form}
\label{sec:qsp-general}

\paragraph{The canonical representation.} With $\hat P_\alpha=-\ii\hat D_\alpha(x_0)$ and any density matrix $\hat\rho$ on $L^2(\mathbb R^d)$,
$\operatorname{Tr}\hat\rho=1$, the identity \eqref{eq:qsp-identity} reads
\begin{equation}
 \Psi[C]=\operatorname{Tr}\Bigl[\hat\rho\,\Bigl\langle\mathcal P\exp\Bigl(\ii\oint\hat P_\alpha\dd C_\alpha\Bigr)\Bigr\rangle\Bigr].
 \label{eq:qsp-canonical}
\end{equation}
This is the trace \eqref{eq:qsp} of the Quantum Statistical Principle, in a definite representation: the Hilbert space is that of
the functions of the base point, the momentum-loop operator is the same at every point of the loop, and the loop parameter enters
only through the order of the factors. The operators do not commute: $\ii\nu[\hat P_\alpha,\hat P_\beta]=\omega_{\alpha\beta}(x_0)$ is the
vorticity, as the jump commutator \eqref{eq:jump-commutator} and the vorticity symbol \eqref{eq:vorticity-symbol} of the momentum
loop say. The statistics of the velocity enters through the average, which is a density matrix of commuting operators, while, for a
homogeneous velocity ensemble, the state $\hat\rho$ on $L^2(\mathbb R^d)$ is arbitrary, since the averaged ordered exponential is then a
number times the unit operator, Eq.~\eqref{eq:qsp-identity-average}.

The form of the principle is therefore not a hypothesis: the loop functional of every flow has it. What the principle adds is
the freedom of representation. A statistical state of the fluid may be given by momentum-loop operators $\hat P_\alpha(\theta,t)$ that
depend on the point of the loop and act in another Hilbert space, with another density matrix, in place of the operator
$-\ii\hat D(x_0)$, constant along the loop. For the states of a known velocity distribution the ordered traces of such a
representation are those of \eqref{eq:qsp-canonical}. The momentum loop $P(\theta,t)$ of Sec.~\ref{sec:mle}, which represents the
covariant derivative by $\ii P$, is a representation of this kind, and so are the examples of Supplement~\ref{sec:S-qsp-examples}, the free field of the
Gibbs distribution and the oscillator of the uniform rotation, which is the canonical representation itself,
the Landau problem, and the fermionic vertex operators of the Euler ensemble
(Sec.~\ref{sec:qsp-euler}). The Euler ensemble is constructed in loop space, and whether its traces are those of a velocity
distribution is the question of positivity of Remark~\ref{rem:intro-positivity}.

\subsection{The principle}
\label{sec:qsp-principle}

\paragraph{Classical statistics and quantum mechanics.} A probability distribution is determined by its characteristic
function, a continuous positive-definite function~\cite{Bochner1933,Minlos1959}, and by the Gelfand--Naimark--Segal
construction~\cite{GelfandNaimark1943,Segal1947} such a function is the expectation value of a unitary representation in a state
of a Hilbert space: classical statistics is the commutative case of one structure with the quantum characteristic
functions~\cite{Kastler1965,LoupiasMiracleSole1966}. The transfer matrix of a chain and the Feynman--Kac formula turn a
classical statistical sum along a line into the trace of an ordered product of operators, with the line in the role of
time~\cite{Kac1949,SchultzMattisLieb1964} (Supplement~\ref{sec:S-qsp-correspondence}).

\paragraph{The principle.} The loop functional is the characteristic functional of the velocity field in the sense of
Hopf~\cite{Hopf1952}, evaluated on the current of the loop, and the loop supplies the time: its parameter $\theta$ orders the factors
$e^{\ii P_k\cdot\Delta C_k}$ of $\Phi_P$, and the closed loop makes the trace. We elevate the formal correspondence to a principle of
fluid statistics.

\emph{Quantum Statistical Principle. The statistical distribution of the fluid is a quantum state of an equivalent quantum
mechanics in loop space: a Hilbert space $\mathcal H$, momentum-loop operators $\hat P_\alpha(\theta)$ acting in it, with the loop
parameter in the role of time, and a density matrix $\hat\rho\ge0$, $\operatorname{Tr}\hat\rho=1$. The loop functional is the ordered trace}
\begin{equation}
 \boxed{\Psi[C]=\operatorname{Tr}\Bigl[\hat\rho\;\mathcal T\exp\Bigl(\ii\oint\hat P_\alpha(\theta)\,\dd C_\alpha(\theta)\Bigr)\Bigr].}
 \label{eq:qsp}
\end{equation}
The ordering $\mathcal T$ puts the earlier points of the loop to the left, as do the one-sided values in the jump commutator
\eqref{eq:jump-commutator}; on a polygon the ordered exponential is the product
$e^{\ii\hat P_0\cdot\Delta C_0}e^{\ii\hat P_1\cdot\Delta C_1}\cdots e^{\ii\hat P_{N-1}\cdot\Delta C_{N-1}}$, with $\Delta C_k=C_{k+1}-C_k$. When the
operators commute and are Hermitian, $\hat\rho$ is a probability distribution of real momentum loops, and \eqref{eq:qsp} is the
average in \eqref{eq:loop-fourier-convention}: this is classical statistics. In general the operators need not commute, and they
need not be Hermitian.

\paragraph{A representation, not a hypothesis.} The principle is a language for statistical states rather than a hypothesis
about them. It assumes nothing about the flow: a thermal distribution and a deterministic initial field, such as the uniform rotation,
are written in it on equal terms, and so are their combinations (Supplement~\ref{sec:qsp-gibbs}--\ref{sec:qsp-rotating}); a probability
distribution of momentum loops is the special case of commuting operators. Its advantage for fluid mechanics is the dynamics. The
loop functional obeys the loop equation \eqref{eq:fixed-loop-equation}, whose right side is the expectation value of the vorticity and
velocity insertions; the loop calculus writes these insertions as derivatives of $\Psi$ with respect to the loop, so the equation is
linear in the statistical state~\cite{M93,MigdalDuality}. The nonlinearity of the Navier--Stokes equation stays in the trajectories of
the momentum loop, Eq.~\eqref{eq:full-mle}; the trace \eqref{eq:qsp} evolves linearly, as the distribution of a nonlinear dynamical
system does under the Liouville equation, and every initial state written in this form, thermal or deterministic, is carried by the
same linear evolution in loop space. A hypothesis enters only with the choice of the state, and for decaying turbulence there is one: the equal weight of the Euler
states (Sec.~\ref{sec:3d-ensemble}). The Gibbs state that this weight induces on a hairpin near the critical angle is derived from
it, and the observables of the hairpin are its conditional expectations (Secs.~\ref{sec:qsp-euler} and~\ref{sec:cond-pendulum}).

\paragraph{Examples.} Supplement~\ref{sec:S-qsp-examples} shows that conventional statistical states of the fluid have the form
\eqref{eq:qsp}: the Gibbs distribution is the vacuum of a free bosonic field, the commutative case, and the uniform rotation, a
deterministic flow, needs noncommuting operators, $[\hat P_\alpha,\hat P_\beta]=-\ii\omega_{\alpha\beta}/\nu$, the Landau problem. The
principle is not an ad hoc rule designed for this solution.

\subsection{The Euler ensemble}
\label{sec:qsp-euler}

The Euler ensemble falls in line. It is a classical sum over the spin sequences of the polygon, and, like the Ising sum, it
becomes a trace over fermions, one fermion per step of the polygon (Sec.~\ref{sec:3d-fermion}). The density
matrix is the normalized projector onto the admissible states, $\hat\rho=\hat\Pi_N/\operatorname{Tr}\hat\Pi_N$ in
\eqref{eq:fermion-qsp-trace}: every admissible state has the same weight, as in the microcanonical ensemble
(Sec.~\ref{sec:3d-ensemble}). The conditioning on a loop has
no analogue in the examples: for the hairpin that measures the vorticity correlation, the bulk of the equal-weight trace is a Gibbs
partition function of two pendulums, and its normalized weight, a positive path measure of the slow angle, is the
statistical state of the observables at the ends of the hairpin, by the definition of a conditional probability (Sec.~\ref{sec:cond-pendulum}; Supplement~\ref{sec:S-qsp-euler}).

The state is positive in all these cases (on the hairpin as a positive path measure, not as the ordered operator product,
Sec.~\ref{sec:cond-state}), but this is a property of the quantum state in loop space. Whether the
velocity statistics that it describes is positive is a separate question. For the Gibbs distribution, the uniform rotation and the
rotating fluid (Supplement~\ref{sec:qsp-gibbs}--\ref{sec:qsp-rotating}) the answer is yes by construction. For the Euler ensemble it is tested on the observables (Remark~\ref{rem:intro-positivity}): the
vorticity correlation at zero separation is nonnegative for every state of the even ensemble at zero winding
(Sec.~\ref{sec:3d-zero-separation}), and so is the limiting spectrum at every computed wave number (Sec.~\ref{sec:3d-final}).

\paragraph{Spontaneous stochasticity.} In this representation a deterministic flow is the particular case in which the operators
are tuned so that all the exponentials but one cancel, with $|\Psi[C]|=1$ on every loop. Spontaneous
stochasticity~\cite{BernardGawedzkiKupiainen1998,FalkovichGawedzkiVergassola2001,ThalabardBecMailybaev2020,BandakMailybaevEyinkGoldenfeld2024}
then becomes a question of dynamics: as $\hat P(\theta,t)$ evolves, the tuning may be lost, and turbulence would correspond to a
universal decaying trajectory of the momentum-loop operator, the Euler ensemble studied here (Supplement~\ref{sec:qsp-stochasticity}).
\section{Vorticity correlations and contact terms in the second area derivative}
\label{sec:contact}

\begin{figure}[t]
 \centering
 \includegraphics[width=0.6\textwidth]{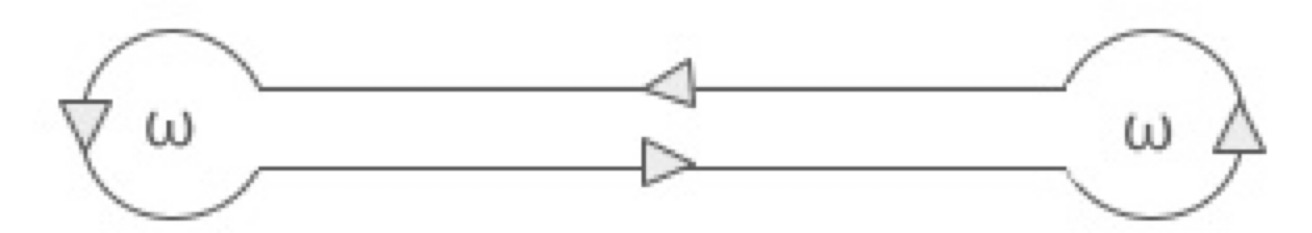}
 \caption{The hairpin loop of the vorticity correlation. The small loops at the two ends are the two area derivatives,
 which insert the vorticity $\omega$ at the two points; between them the loop is contracted to a thin wire that runs from one
 point to the other and back.}
 \label{fig:hairpin}
\end{figure}

The second area derivative of the loop functional is not the product of two vorticity insertions: it has contact terms. This is
the point of this section. Why does it matter? The energy spectrum is measured by the two-point correlation of vorticity, and in
the loop language this correlation is a second area derivative of the loop functional, taken at
two marked points of the loop, after which the loop is contracted to a thin hairpin that
runs between the two points and back~\cite{migdal2024quantum,ReviewPaperAM} (Fig.~\ref{fig:hairpin}). The
calculus of Sec.~\ref{sec:mle-calculus} turns each area derivative
into an insertion of the vorticity symbol, Eq.~\eqref{eq:calc-area}. But the first
area derivative inserts the vorticity at a point of the loop, and the second one varies the loop,
and with it the position of that point. In coordinate space this is the elementary product
rule. In momentum space it produces contact terms of the same order as the product, and
without them the correlation is neither invariant under translations of the momentum loop
nor sign-definite where it must be. We derive these terms, give their momentum image, and
normalize the correlation by the loop functional of the same hairpin, in the limit by its bulk part; Sec.~\ref{sec:3d-hairpin} and Supplement~\ref{sec:2d-positivity} use
them to test the solutions.

\subsection{The second area derivative of a marked loop}

We use the tangent derivative $D_\alpha(s)$ and the correspondence \eqref{eq:calc-correspondence} of Sec.~\ref{sec:mle-calculus}. The area
derivative $\mathcal L_{\alpha\beta}$, the commutator of two tangent derivatives across a point, integrated around the loop, inserts minus
the area bivector $A[P]$ of the momentum loop, which for a polygon is the sum of the jump bivectors \eqref{eq:jump-bivector}, and the
vorticity operator $-\ii\nu\mathcal L$ is the sum of the symbols \eqref{eq:vorticity-symbol}
(Supplement~\ref{sec:S-contact}, Eq.~\eqref{eq:ct-area-operator}). The two marks are treated as follows. The loop is arbitrary while both area derivatives
are taken, and only afterwards is it set to the hairpin with endpoints at the marks. The
circulation is therefore a functional of $C$, not an explicit function of the marks, and the
hairpin factor is never differentiated. What is differentiated is the vorticity inserted by
the first derivative at the point $C(\theta)$, because that point moves with the loop. The product rule of the loop
calculus then gives the following relation, derived in Supplement~\ref{sec:S-contact}. Let $F_\theta=\omega_{\gamma\delta}(C(\theta))$ be the vorticity inserted by the first derivative, and
$\Phi_v=e^{\ii\Gamma/\nu}$ the physical circulation phase, with the tangent response
$Q_\alpha(s)=\delta\Gamma/\delta e_\alpha(s)$ of \eqref{eq:calc-correspondence}. With
$\Omega_{\alpha\beta}=-\ii\nu\mathcal L_{\alpha\beta}$,
\begin{equation}
 \boxed{\Omega_{\alpha\beta}\bigl(F_\theta\Phi_v\bigr)=F_\theta\,\Omega_{\alpha\beta}\Phi_v+\mathcal E_{\alpha\beta;\gamma\delta}(\theta),\quad
 \mathcal E_{\alpha\beta;\gamma\delta}(\theta)=2\Bigl[\partial_\alpha\omega_{\gamma\delta}\bigl(\bar Q_\beta(\theta)-\bar Q_\beta(0)\bigr)
 -(\alpha\leftrightarrow\beta)\Bigr]\Phi_v .}
 \label{eq:ct-physical}
\end{equation}

This is the basic relation. In momentum space the two area derivatives compose to
$\mathcal L\mathcal L\Phi_P=A[P]A[P]\Phi_P$, because $A[P]$ does not depend on $C$. The composition is
the second area derivative of the loop functional, and it already contains the action of the
second derivative on the first insertion. The correlation of two physical vorticities is the
composition minus the contact term $\mathcal E$. Identifying the momentum product directly with the two-vorticity correlation, as in our earlier work~\cite{migdal2024quantum}, omits $\mathcal E$.

\subsection{The momentum image}

At a jump of the momentum loop the vorticity is the symbol \eqref{eq:vorticity-symbol}. As a
measure on the parameter circle it is
\begin{equation}
 \widehat\omega(\dd\theta)=\ii\nu\,\widetilde P(\theta)\times\dd P(\theta),\qquad
 \widetilde P=\bar P-\langle P\rangle_\mu,\qquad \langle P\rangle_\mu=\oint\bar P\,\dd\mu ,
 \label{eq:ct-measure}
\end{equation}
in three-dimensional notation (in $d$ dimensions $\times$ stands for the antisymmetric
product \eqref{eq:jump-bivector}). Here $\mu$ is a normalized measure on the loop. The reference momentum $\langle P\rangle_\mu$ is needed because on closed loops
the tangent derivative is defined only up to a constant operator (Supplement~\ref{sec:S-contact}). In the Euler
ensemble the jumps are equally spaced and of equal length, and $\mu$ is the counting measure of
the jumps; it coincides with $|\dd P|/\oint|\dd P|$ and with $\dd\theta/2\pi$.

The gradient of the inserted vorticity is represented by the jump that carries it, Eq.~\eqref{eq:calc-gradient}, and the tangent
response by $\nu P$. The average over the base point, which the cyclic symmetry of the ensemble requires, gives the antisymmetric
sawtooth kernel
\begin{equation}
 \kappa(s,\theta)=\tfrac12-\mu\bigl(\hbox{arc }s\to\theta\bigr),\qquad
 \dd_s\kappa(s,\theta)=\dd\mu(s)-\delta(s-\theta)\dd s .
 \label{eq:ct-kappa}
\end{equation}
The contact term of the second area derivative at $s$, acting on the vorticity inserted at the
jump $\theta$, is then the measure in $s$
\begin{equation}
 \boxed{\mathcal C(\dd s;\theta)=-\nu^2\Bigl\{\Delta P(\theta)\times\bigl[\kappa(s,\theta)\,\dd P(s)-\widetilde P(s)\,\dd_s\kappa(s,\theta)\bigr]\Bigr\}
 \cdot\Bigl\{\widetilde P(\theta)\times\Delta P(\theta)\Bigr\}.}
 \label{eq:ct-contact}
\end{equation}
The numerator of the two-point function measured by the hairpin with endpoints at $s$ and
$\theta$ and hairpin factor $\mathcal F(s,\theta)$ is
\begin{equation}
 \boxed{\mathcal W[\mathcal F]=\oint\!\!\oint\mathcal F(s,\theta)\,\widehat\omega(\dd s)\cdot\widehat\omega(\dd\theta)
 +\oint\!\!\oint\mathcal F(s,\theta)\,\mathcal C(\dd s;\theta),}
 \label{eq:ct-W}
\end{equation}
symmetrized in the two marks; the first term is the composition with centred symbols, the
second is minus the image of $\mathcal E$.

\paragraph{Invariance.} The closed-loop phase is invariant under $P(\theta)\mapsto P(\theta)+U$ with $U$ constant on the loop, and so is
every factor in \eqref{eq:ct-measure}--\eqref{eq:ct-W}: the correlation is invariant sample by sample, term by term. Only Stieltjes
measures, values at jumps and $\mu$-masses of arcs enter, so it is also unchanged under a reparametrization of the loop, provided
$\mu$ is defined by the loop itself; the integrals over the marks in \eqref{eq:ct-W} are the average over reparametrizations
(Supplement~\ref{sec:S-contact-main}).

\paragraph{Zero separation.} With $\mathcal F\equiv1$, integration by parts gives
$\oint[\kappa\,\dd P-\widetilde P\,\dd_s\kappa]=2\widetilde P(\theta)$, and
\begin{equation}
 \boxed{\mathcal W[1]=-\nu^2\Bigl(\oint\bar P\times\dd P\Bigr)^2+2\nu^2\sum_{\rm jumps}\bigl(\widetilde P\times\Delta P\bigr)^2 .}
 \label{eq:ct-zero}
\end{equation}
The first term is the square of the total vorticity of the loop; the contact terms add twice the
sum of the squares of the centred jump vorticities. Without them only the first term remains.
The identity holds for every momentum loop and is manifestly invariant.

\subsection{The polygon and the normalization}
\label{sec:ct-polygon}

For a polygonal momentum loop the discrete form of \eqref{eq:ct-contact}--\eqref{eq:ct-W} is the pair weight $\Phi_{AB}$ of
the endpoints at the vertices $A$ and $B$, Eq.~\eqref{eq:ct-polygon}. The sum over the unordered pairs, one half of \eqref{eq:ct-W},
averaged over the ensemble, is the numerator $\mathcal W_N[\mathcal F]$ of the correlation, Eq.~\eqref{eq:ct-normalized}; both are
written out in Supplement~\ref{sec:S-ct-polygon}.

The same sum without the two vorticities is the bare hairpin, the loop functional of the same hairpins with the same endpoints, and the
correlation is the conditional average, the ratio of the two:
\begin{equation}
 \begin{gathered}
 \mathcal D_N[\mathcal F]=\Bigl\langle\sum_{0\le A<B<N}\mathcal F_{AB}\Bigr\rangle=\frac N2\sum_{n=1}^{N-1}\bigl\langle\mathcal F_{0n}\bigr\rangle,\\
 \frac{\mathcal W_N[\mathcal F]}{\mathcal D_N[\mathcal F]}\;\longleftrightarrow\;
 \frac{\bigl\langle\omega(\boldsymbol r_1)\,\omega(\boldsymbol r_2)\,e^{\ii\Gamma/\nu}\bigr\rangle}{\bigl\langle e^{\ii\Gamma/\nu}\bigr\rangle},
 \end{gathered}
 \label{eq:ct-bare}
\end{equation}
with $\Gamma$ the circulation of the hairpin. The two vorticities are weighted by the phase of the hairpin that measures them and
normalized by the weight of the hairpin, as a conditional probability is by Bayes' formula.

The correlation is defined by this ratio, Definition~\ref{post:cond} of Sec.~\ref{sec:cond-def}. Our earlier
work~\cite{migdal2024quantum} and the previous versions of this paper set its denominator, the loop functional of the bare hairpin, to
one; for the Euler ensemble it is not one (Supplement~\ref{sec:hairpin-not-one}).

\paragraph{Positivity.} At zero separation $\mathcal W$ is the enstrophy of the flow, up to a positive factor, and the identity
\eqref{eq:ct-zero} makes the positivity test exact. Every momentum loop whose total vorticity is zero and whose centred jump
vorticities $\widetilde P\times\Delta P$ are real, as for the Euler states, has $\mathcal W[1]=2\nu^2\sum(\widetilde P\times\Delta P)^2\ge0$; a
loop with a large total vorticity has $\mathcal W[1]<0$, which no real flow can produce (Sec.~\ref{sec:3d-zero-separation},
Supplement~\ref{sec:S-contact-main}).
\section{The null-midpoint Euler polygon}
\label{sec:three-dimensional}

Here is the exact solution of the momentum-loop equation. The vertices of the polygon lie on a circle, at the corners of a regular
star polygon with unit sides; the polygon walks over them, one step forward or backward for each spin $\sigma_k=\pm1$, and it is
shifted as a whole along the normal to its plane by an imaginary constant. On every side the viscous and the advective terms of the
equation vanish separately.

The original three-dimensional solution~\cite{migdal2023exact} is complex, even
though its closed-loop phase has a real planar representative. When advection is
kept, this complex structure is essential. Let
\begin{equation}
 \beta=\frac{2\pi p}{q},\quad (p,q)=1,\quad 0<p<q,\quad q>2,
 \qquad
 R=\frac{1}{2\sin(\beta/2)},\qquad
 B=\frac12\cot(\beta/2),
 \label{eq:3d-arithmetic-parameters}
\end{equation}
and choose a signed angular walk
\begin{equation}
 \alpha_{k+1}-\alpha_k=\sigma_k\beta,\qquad
 \sigma_k\in\{-1,1\},\qquad
 \sum_{k=0}^{N-1}\sigma_k=qr,\quad r\in\mathbb Z .
 \label{eq:3d-spin-closure}
\end{equation}
With $\mathbf e_3$ normal to the real plane, put
\begin{equation}
 f_k=R(\cos\alpha_k,\sin\alpha_k,0),\qquad
 G_k=f_k+\ii B\,\mathbf e_3,\qquad
 P_k(t)=\frac{G_k}{\sqrt{2\nu\tau}} .
 \label{eq:3d-complex-euler}
\end{equation}
The third component is constant along the loop; in physical units it decays as
$P_z=\ii B/\sqrt{2\nu\tau}$, and in rescaled units it is the constant $\ii B$. A
fixed common rotation may be applied to the plane and its normal. The amplitude
$B$ of the normal is the same $B$ that will reappear in the vorticity
(Sec.~\ref{sec:3d-hairpin}).

\subsection{Null midpoint and the separate cancellation of advection}

Let $m_k=(G_{k+1}+G_k)/2$ and $\delta_k=G_{k+1}-G_k=f_{k+1}-f_k$. The real polygon
has unit chords and equal radii, and $B^2=R^2-\frac14$. Therefore
\begin{equation}
 \delta_k^2=1,\qquad m_k\cdot\delta_k=0,
 \qquad m_k^2=0,
 \qquad G_k^2=\frac14 .
 \label{eq:3d-null-identities}
\end{equation}
The midpoint is not zero; it is null in the bilinear sense, which says nothing
about its positive Hermitian norm.

Substitution into \eqref{eq:rescaled-full-mle} proves the solution edge by edge:
the viscous and the advective expressions vanish separately,
\begin{equation}
 (1-\delta_k^2)m_k+(m_k\cdot\delta_k)\delta_k=0,
 \qquad
 \left[\frac{(m_k\cdot\delta_k)^2}{\delta_k^2}-m_k^2\right]\delta_k=0 .
\end{equation}
In unscaled variables diffusion is not zero, $\partial_t\bar P_k=-\bar P_k/(2\tau)$;
advection is zero separately. The cancellation holds for every admissible spin
sequence, before any averaging and without assuming odd $N$. The same
finite-dimensional verification is given in Ref.~\cite{DeLellisInprep},
Eq.~(153) and Theorems~10.7--10.8.

\subsection{The statistical weight and the even ensemble at zero winding}
\label{sec:3d-ensemble}

The solutions need a statistical weight, and here the only physical hypothesis of the theory enters: every admissible state counts
once. In the Quantum Statistical Principle of Sec.~\ref{sec:qsp}, the density matrix
of the Euler ensemble gives equal weight to the distinct admissible states at finite cutoff, and the
normalized Haar measure to their global orientation:
\begin{equation}
 \langle\mathcal O\rangle_N
 =\frac{1}{Z_N}\sum_{a\in\mathcal E_N}
       \int \dd\mu(\Omega)\,\mathcal O(a,\Omega),
 \qquad Z_N=\#\mathcal E_N .
 \label{eq:finite-statistical-trace}
\end{equation}
The state space $\mathcal E_N$, including spin multiplicities, is specified
below in three dimensions and in Supplement~\ref{sec:two-dimensional} in two. Equal weight per state is not equal
weight per sector when the sectors have different dimensions. Nor does it define
a uniform probability measure on a countably infinite state space; only
normalized finite-cutoff limits are used.

Which states are admissible? In three dimensions they are the rational turning angles $\beta=2\pi p/q$ of
\eqref{eq:3d-arithmetic-parameters}, the windings $r$ and the spin sequences of total spin $M=qr$ in \eqref{eq:3d-spin-closure}. They fall into three sectors: odd $N$,
where the winding is odd and never zero; even $N$ with $r\ne0$; and even $N$ at zero winding. At zero winding $M=0$, and the
closure $\alpha_N-\alpha_0=\beta M=0$ holds for every turning angle, not only for rational $\beta/2\pi$. With the denominators
$3\le q<N$, as in Ref.~\cite{migdal2024quantum}, every angle is admissible with every balanced sequence, and the number of
states, with Euler's totient $\varphi$, is
\begin{equation}
 2^{-N}Z^{(0)}_N=2^{-N}\binom N{N/2}\,\Phi_N\simeq\frac{3\sqrt2}{\pi^{5/2}}\,N^{3/2},
 \qquad
 \Phi_N=\sum_{3\le q<N}\varphi(q)\simeq\frac{3N^2}{\pi^2} .
 \label{eq:grand-Z0}
\end{equation}
This sector dominates. In the grand canonical ensemble of Ref.~\cite{migdal2023exact}, $\sum_Ne^{-\mu N}2^{-N}Z_N$ with $\mu\to0$, the zero-winding states
carry the weight $N^{3/2}e^{-\mu N}$, while the odd states and the even states with $r\ne0$ carry $N^{1/2}e^{-\mu N}$: the zero-winding
sector outnumbers each of the other two by the factor $3N/\pi^2$, Eq.~\eqref{eq:grand-ratio}, and the grand ensemble becomes the even
ensemble at zero winding (Supplement~\ref{sec:3d-grand}). Basak and Zaharescu computed the exact partition functions of the odd and
the even ensembles and proved this dominance~\cite{Zah23}. So, we do not choose this ensemble: the counting does.

It is also the only sector whose vorticity correlation at zero separation is nonnegative for every state
(Sec.~\ref{sec:3d-zero-separation}). The odd ensemble and the two-dimensional solution are exact solutions of the
momentum-loop equation, but their correlations are negative at zero separation at finite $N$ (in the odd ensemble at every $N$ computed); that of the planar solution, conditioned on its
hairpin, is negative at every separation in the limit (Supplement~\ref{sec:2d-positivity}), and that of the odd ensemble is positive at zero
separation in the limit, but its spectrum is not (Supplement~\ref{sec:S-odd-cond}). They are presented, with the rigidity of the Euler
polygons, in the Supplementary Material (Secs.~\ref{sec:S-3d-odd}, \ref{sec:rigidity3d} and \ref{sec:two-dimensional}).

\paragraph{The limit.} We take the limit $N\to\infty$ inside this sector, at fixed turbulent viscosity
$\widetilde\nu=\nu_NN^2$, with $\nu_N$ the bare viscosity of the regularized equation, the viscosity of an auxiliary flow in which
the limit is computed (Sec.~\ref{sec:3d-renorm}). In Ref.~\cite{migdal2023exact} the limit was taken with
$\nu\to0$ and $\mu\to0$ at fixed $\nu/\mu^2$. Is this the same limit? The distribution of $N$ in the zero-winding
sector, $N^{3/2}e^{-\mu N}$, is not sharp: its relative width stays finite as $\mu\to0$. But the grand ensemble reduces to this
sector as $\mu\to0$ (Supplement~\ref{sec:3d-grand}; the last estimate is proved in Supplement~\ref{sec:S-finiteN-proof}), and at
fixed $\widetilde\nu$ the scaling functions of $kL$ and $r/L$ do not depend on $N$; only powers of $N$ stand in front of them, so the spread of $N$ in the grand ensemble leaves the renormalized correlation unchanged (Supplement~\ref{sec:S-3d-main}). The conditional correlation decreases as $N^{-1}$, the order of the two vorticity insertions
(Sec.~\ref{sec:cond-def}; Supplement~\ref{sec:S-3d-main}).

From now on the ensemble is the even ensemble at zero winding.

% ===== New section 2026-11-12 (revised): a stochastic model of ergodicity; between Sec. 5 and the vorticity correlation =====
\section{The physical mechanism of spontaneous stochasticity in the momentum-loop equation}
\label{sec:stoch-mechanism}

The equal weight of the Euler states (Sec.~\ref{sec:3d-ensemble}) was introduced as a postulate, the only physical hypothesis of
this theory and a counterpart of the ergodic hypothesis of Boltzmann, by which a trajectory of a Hamiltonian system covers its surface of constant energy uniformly in time. Can
it be derived from the momentum-loop equation (MLE) \eqref{eq:rescaled-full-mle}? Not in the literal sense, and it is important to
say why. The Euler states are fixed points of the MLE. A trajectory that starts on one of them stays there, so every distribution of
the turning angle $\beta$ and of the spin sequences is invariant, Eq.~\eqref{eq:rig3d-invariant-measure} of
Supplement~\ref{sec:rigidity3d}; and a trajectory that starts near them does not return: it leaves them and grows
(Sec.~\ref{sec:stoch-singular}). Whoever integrates the system \eqref{eq:rescaled-full-mle} numerically will find an instability,
not ergodicity.

We argue that this instability is the mechanism of ergodicity and not an obstacle to it. In the even ensemble at zero winding, the
sector that this paper solves, the time derivative term of the MLE is not invertible, and at every Euler state of this sector the
equation leaves a part of the velocity undetermined: an arbitrarily small displacement produces a velocity of order one in a
direction that the displacement chooses. A deterministic evolution with this property amplifies any noise, however weak. This is the
situation of spontaneous stochasticity (Sec.~\ref{sec:stoch-picture}), and we interpret the singularity of the MLE as its source: in the limit $N\to\infty$, which
is the limit of vanishing viscosity $\nu_N=\widetilde\nu/N^2$, a vanishing noise results in a random motion of the polygon over the
manifold of the Euler states.

The MLE cannot supply this motion itself. Its time derivative term is the velocity of the midpoints of the sides, and a family
of fixed points cannot move without moving them: a solution of the MLE that passes only through fixed points is at rest
(Proposition~\ref{prop:stoch-rigid}). In particular, the alternating velocity that the time derivative term does not see cannot
carry the polygon from one Euler state to another. Every motion between two Euler states therefore violates the equation
somewhere, and the natural question is how to realize the random moves with minimal distortion of the MLE while preserving exact
symmetry of the transitions so that an H-theorem drives the ensemble to uniformity.

\begin{figure}[t]
 \centering
 \includegraphics[width=\textwidth]{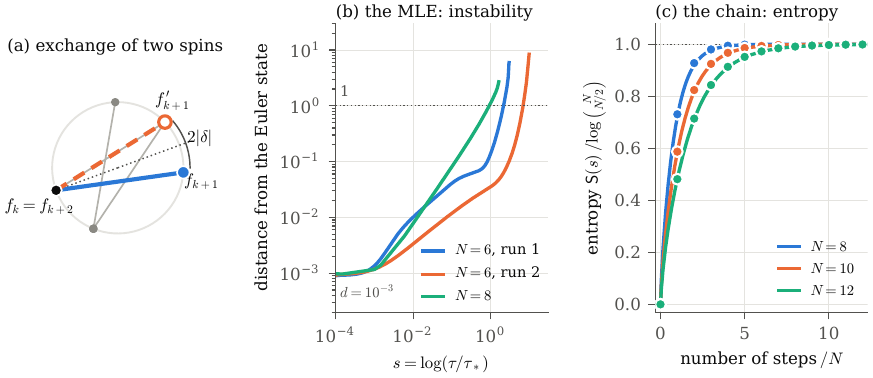}
 \caption{(a) The elementary move of the model, Eq.~\eqref{eq:stoch-exchange}: the exchange of two neighbouring opposite spins
 reflects one vertex of the Euler polygon in the radius through its two neighbours, which coincide, and leads in one step from one
 exact solution of the MLE to another (here $N=12$ and
 $\delta=\beta-\pi=-0.42$; in the critical layer $|\delta|=\xi/\sqrt N$). (b) The equation itself, integrated from a point at a
 distance $d$ from a zero-winding Euler state: the distance $\|F(s)-G\|$ from the state $G$ in three runs with $d=10^{-3}$, for $N=6$
 (\texttt{+++---}, $\beta=0.37\cdot2\pi$) and $N=8$ ($\beta=\pi-0.5$), on logarithmic scales; the polygon leaves the Euler state and
 does not return. (c) The model: the entropy of the distribution of the spin sequences, starting from a single sequence, computed on
 the complete state space: in the Markov chain (points, every $N$-th step) and in the equation with a
 continuous $s$ and the same transition probabilities (lines); see Propositions~\ref{prop:stoch-exchange}
 and~\ref{prop:stoch-discrete}.}
 \label{fig:stoch}
\end{figure}

This section describes the mechanism, and Sec.~\ref{sec:stoch} a model of the random motion that it produces.
Section~\ref{sec:stoch-picture} gives the physical picture, the amplification of the thermal noise at the singular fixed points of
the equation, with its predecessors, and shows that an exchange of two spins is an infinitesimal move in the space of the loops.
Section~\ref{sec:stoch-singular} describes the singularity, Sec.~\ref{sec:stoch-move} the exchange of two neighbouring spins, the
equation with a discrete time and the impulse of a step, and Sec.~\ref{sec:stoch-trigger} the thermal noise as the trigger that
selects the exchange of the least impulse. What is proved in this section are properties of the equation itself: the singularity
(Proposition~\ref{prop:stoch-singular}), the absence of solutions that move along the fixed points
(Proposition~\ref{prop:stoch-rigid}), the impulse of a step and the least impulse of the exchange of neighbours
(Propositions~\ref{prop:stoch-residual} and~\ref{prop:stoch-least}). The interpretation as spontaneous stochasticity and the rule
of the trigger are physical arguments, supported by simulations. Figure~\ref{fig:stoch} shows a move, the instability of the
equation and the growth of the entropy in the model.

\subsection{The physical picture: thermal noise, spontaneous stochasticity and a diffusion over the Euler states}
\label{sec:stoch-picture}

\paragraph{The thermal noise and its amplification.} A fluid is never free of noise. The molecular motion that makes it a fluid, and
gives it its viscosity, makes its velocity fluctuate, and these thermal fluctuations are a part of its hydrodynamics. That they
matter for turbulence is an old idea. Lorenz~\cite{Lorenz1969} found that in a flow with many scales of motion an error of the
smallest scales reaches the largest ones in a finite time, which a smaller initial error does not make longer, and
Ruelle~\cite{Ruelle1979} pointed out that the thermal fluctuations are such an error, amplified by the sensitive dependence of a
turbulent flow on its initial state. In its present form the idea is spontaneous
stochasticity~\cite{BernardGawedzkiKupiainen1998,FalkovichGawedzkiVergassola2001,EyinkBandak2020,ThalabardBecMailybaev2020}: in
the limit of vanishing viscosity the solutions of the limiting equations, those of a fluid particle in a rough velocity field or of
the ideal fluid itself, are not unique, so a noise that vanishes in this limit leaves behind a random solution, whose statistics do
not depend on the noise. Bandak et al.\ found that the thermal noise changes a turbulent flow already at the scales just below the
Kolmogorov length~\cite{BandakGoldenfeldMailybaevEyink2022}, and that in a shell model with the noise of fluctuating
hydrodynamics it reaches the largest scales within a few turnover times, in a time that does not depend on its
amplitude~\cite{BandakMailybaevEyinkGoldenfeld2024}. The mathematical prototype is an ordinary differential equation that loses the
uniqueness of its solutions at a singular point: a vanishing noise selects among the solutions with definite
probabilities~\cite{BaficoBaldi1982}, and a vanishing regularization of the singularity can select a continuation that does not
depend on the regularization~\cite{DrivasMailybaev2021}. The limit is then a fixed point of a renormalization group that acts on the transition
probabilities~\cite{EyinkBandak2020,MailybaevRaibekas2023,MailybaevMoriconi2026}, in the same framework as the period-doubling
transition to chaos~\cite{MailybaevMoriconi2026}, at which the sensitivity to an external noise is known to obey a universal
scaling~\cite{CrutchfieldNauenbergRudnick1981,ShraimanWayneMartin1981}.

\paragraph{Its realization in the momentum-loop equation.} Our mechanism is a particular realization of this idea, in the space of
the momentum loops. The loop equation is the Navier--Stokes equation for the loop functional, and the MLE is its momentum
representation, discretized on polygons (Sec.~\ref{sec:mle}): the dynamics is the same, and only its representation is different.
In this representation the amplification of the noise is not a cascade through the scales but a singularity of the equation at its
fixed points. The singular points are the Euler states. They are fixed points of the MLE, and in the even ensemble at zero
winding the equation is singular at every one of them: its time derivative term does not contain the alternating velocity of the
vertices, and the matrix that should determine this velocity vanishes (Proposition~\ref{prop:stoch-singular}). In the studies
quoted above the singularity belongs to a model, a rough velocity field, a shell model or a model equation, or to a singular
initial state; here it is not assumed, it is a proved property of the equation. Near such a state
the velocity is a ratio of two linear forms of the displacement, Eq.~\eqref{eq:stoch-u}. It does not vanish with the displacement,
and its direction is set by the direction of the displacement. This is the amplifier: a thermal displacement of the vertices,
however small, sets the polygon in motion with a velocity of order one, in a direction that the noise chooses, and the equation
neither chooses this direction itself nor returns the polygon to its fixed points (Sec.~\ref{sec:stoch-singular},
Fig.~\ref{fig:stoch}b). The fixed points are, moreover, degenerate. At a given orientation they form $\binom N{N/2}$ lines, one for
each balanced spin sequence and parametrized by the turning angle $\beta$, and all these lines pass through one point, the
degenerate zig-zag at $\beta=\pi$ (Sec.~\ref{sec:stoch-move}). The critical layer $|\beta-\pi|=\xi/\sqrt N$, which alone
contributes in the statistical limit, is the neighbourhood of this branch point. In it the lines of the neighbouring spin sequences
are at a distance that vanishes as $N^{-1/2}$, they are joined by segments of fixed points, with no barrier between them
(Remark~\ref{rem:stoch-bridge}), and the thermal noise chooses the neighbour to which the polygon passes, the one for which the
passage costs the least impulse (Sec.~\ref{sec:stoch-trigger}). These are the weak directions of the equation: in them it exerts no
force, and the noise is not damped.

\paragraph{The exchange is an infinitesimal move.} In the labels $(\beta,\boldsymbol\sigma)$ of the Euler states the exchange of two
spins looks like a finite move. In the space of the momentum loops, where the probability distribution evolves, it is
infinitesimal. An exchange moves one vertex out of $N$ by $2|\sin\frac\delta2|\simeq\xi/\sqrt N$, Eq.~\eqref{eq:stoch-exchange}.
For the maximum, the root mean square and the mean of the displacement of the vertices over the loop,
$\|F\|_\infty=\max_k|F_k|$ and $\|F\|_p^p=\frac1N\sum_k|F_k|^p$, this gives
\[
 \|\Delta F\|_\infty=2\Bigl|\sin\frac\delta2\Bigr|\simeq\frac\xi{\sqrt N},\qquad
 \|\Delta F\|_2=\frac2{\sqrt N}\Bigl|\sin\frac\delta2\Bigr|\simeq\frac\xi N,\qquad
 \|\Delta F\|_1=\frac2N\Bigl|\sin\frac\delta2\Bigr|\simeq\frac\xi{N^{3/2}} ,
\]
while the ensemble has a size of order one in each of these norms. By \eqref{eq:stoch-distance}, whose norm is
$\sqrt N\,\|\cdot\|_2$, two independent states of the uniform ensemble with the same $\beta$ and the same orientation are at the
mean square distance
$\|G-G'\|_2^2\simeq\frac12\int_0^1\bigl(1-e^{-\xi^2x(1-x)}\bigr)\dd x$, which is $(0.48)^2$ at $\xi=2$ and $\xi^2/12$ at small
$\xi$. In the root mean square an exchange is thus a step of about $4/N$ of the distance between two states, for $\xi$ of the
order of one. The observables change accordingly. A cyclically symmetric correlation function of $n$ points of the loop,
\[
 \Phi(F)=\frac1N\sum_{k=0}^{N-1}F_k^{a_0}F_{k+r_1}^{a_1}\cdots F_{k+r_{n-1}}^{a_{n-1}} ,
\]
with offsets $0,r_1,\dots,r_{n-1}$ that are distinct modulo $N$ and components $a_i$, obeys $|\Phi(F)-\Phi(F')|\le n\rho^{n-1}\|F-F'\|_1$ for any two
polygons with $|F_k|\le\rho$ and $|F_k'|\le\rho$, because a product of $n$ factors changes by at most $\rho^{n-1}$ times the sum of
the changes of its factors. On the Euler states $|G_k|=\rho=(R^2+B^2)^{1/2}\to\frac12$, so one exchange changes every such
function by
\[
 |\Delta\Phi|\le\frac{2n}N\,\rho^{\,n-1}\Bigl|\sin\frac\delta2\Bigr|\simeq\frac{n\,\xi}{2^{\,n-1}N^{3/2}} .
\]
The same bound holds for the change that the exchange of one step of the chain makes in the correlation function
$\langle F_{k_1}^{a_1}\cdots F_{k_n}^{a_n}\rangle$ of any $n$ different vertices in the distribution $W$, because a step moves a
given vertex with a probability of $1/N$ at most. The phase of the loop functional of a loop $C$, approximated by a polygon with sides not longer
than $|C|/N$, changes in an exchange by $\sqrt2\,|\sin\frac\delta2|\,|C|/L\simeq\xi|C|/(\sqrt{2N}\,L)$ at most
(Sec.~\ref{sec:stoch-move}). The spin sequences are thus discrete labels of loops that lie closer and closer to each other as $N$
grows, and the Markov chain of the exchanges is a walk over them by steps that vanish as $1/N$.

\paragraph{A diffusion over the fixed points.} Such a walk is a diffusion. The MLE has neither a drift nor a diffusion of its own
along its fixed points; the motion and its randomness are supplied by the noise, which the singularity amplifies and the weak
directions do not damp. In the critical layer the Euler states are parametrized by the slow angle $\theta(x)$, $x=k/N$,
Eq.~\eqref{eq:stoch-zigzag}, an exchange changes it at one point by $2\delta$, and for large $N$ the chain of Sec.~\ref{sec:stoch}
becomes the linear
stochastic equation \eqref{eq:stoch-EW} for $\theta(x,s)$: a diffusion in a space of functions, whose relaxation time is finite in
the statistical limit if the step of the chain decreases as $N^{-3}$, Eq.~\eqref{eq:stoch-step}. The symmetry of the steps gives
an H-theorem, and the distribution tends to the uniform one over the spin sequences (Propositions~\ref{prop:stoch-H}
and~\ref{prop:stoch-exchange}), for which the slow angle is a Brownian bridge (Sec.~\ref{sec:stoch-attractor}). This stationary
measure is the one that the rest of the paper reduces to the quantum pendulum: the pendulum is the attractor of a diffusion that the
thermal noise drives over the degenerate fixed points of the MLE. The rule of the exchange is a model, but the attractor does not
depend on it: every set of symmetric moves that connects the states of the Euler ensemble leads to the same uniform distribution,
and hence to the same pendulum (Propositions~\ref{prop:stoch-H} and~\ref{prop:stoch-beta}). A local rule that is not symmetric leads
to the Ising weights of Proposition~\ref{prop:stoch-ising}, which rescale the Hamiltonian of the pendulum, by a factor that tends to
one for the rule of the least impulse (Sec.~\ref{sec:stoch-trigger}).

\paragraph{The closest analogues.} Two limits of dynamical systems with a small noise are close to this picture. In a system whose
fixed points are saddles joined by heteroclinic orbits a small noise decides, at each saddle, along which orbit the system leaves
it, and in the limit of a vanishing noise the motion is a random jump process over the network of the saddles, a random walk in the
simplest case~\cite{StoneHolmes1990,Bakhtin2011}. In a system with a strong drift towards a manifold of fixed points the noise
survives, in the limit of the strong drift, as a diffusion along this manifold~\cite{Katzenberger1991}. Our model has the features
of both: the noise decides at a fixed point, as in the first, and the result is a diffusion over the fixed points, as in the
second. It differs from them in three respects. At a saddle the velocity vanishes linearly with the distance, and the time of a
passage grows as the logarithm of the inverse noise. The fixed points of the MLE are singular and not hyperbolic: the velocity is
homogeneous of degree zero in the displacement, Eq.~\eqref{eq:stoch-u}, it does not vanish with it, and the time of a passage has
no such logarithm. This is what
distinguishes spontaneous stochasticity from ordinary chaos, in which the time of the amplification grows as the noise
decreases~\cite{BandakMailybaevEyinkGoldenfeld2024}. The number of the fixed points grows as $\binom N{N/2}$ while the distance
between the neighbouring ones vanishes, so the walk over a network becomes a diffusion in a space of functions. And the fixed
points do not attract: the return of the displaced polygon to the Euler ensemble, by one exchange in a step $\Delta s$, is the
assumption of the model and not a property of the equation (Sec.~\ref{sec:stoch-attractor}). Among the studies of the thermal noise
in turbulence the closest one is Ref.~\cite{BandakMailybaevEyinkGoldenfeld2024}. There the noise is amplified by the cascade of the
error from the small scales to the large ones in the physical space; here it is amplified by the singular time derivative term of
the momentum-loop equation, at its fixed points.

\subsection{The singular time derivative at even \texorpdfstring{$N$}{N}}
\label{sec:stoch-singular}

The time derivative enters the MLE through the midpoints, $2\partial_sm_k=\partial_sF_{k+1}+\partial_sF_k$. For even $N$ this map from
the velocities of the vertices to the velocities of the midpoints is not invertible: it annihilates the alternating velocities
$\partial_sF_k=(-1)^ku$ with an arbitrary $u\in\mathbb C^3$ (Supplement~\ref{sec:S-mle-proofs}). In terms of the staggered polygon
$g_k=(-1)^kF_k$ the midpoints are differences, $m_k=\frac12(-1)^k(g_k-g_{k+1})$: the time derivative term of the MLE contains only
the differences of the staggered polygon and does not determine the velocity of its translation. Let $\rho_k(F)$ be the right side of \eqref{eq:rescaled-full-mle}. The equation is
equivalent to a constraint and an incomplete evolution,
\begin{equation}
 \Xi(F)\equiv\sum_{k=0}^{N-1}(-1)^k\rho_k(F)=0,\qquad
 \partial_sF_k=v_k+(-1)^ku,\qquad v_{k+1}+v_k=\rho_k ,
 \label{eq:stoch-dae}
\end{equation}
where $v$ is any particular solution. The MLE does not contain the alternating velocity $u$ at all. It is determined only
indirectly, by the requirement that the constraint is preserved in time, $\partial_s\Xi=0$:
\begin{equation}
 \mathsf K(F)\,u=-\sum_k\frac{\partial\Xi}{\partial F_k}\,v_k,\qquad
 \mathsf K(F)=\sum_k(-1)^k\frac{\partial\Xi}{\partial F_k} .
 \label{eq:stoch-K}
\end{equation}
This is a differential-algebraic system, regular where the $3\times3$ matrix $\mathsf K$ is invertible. The alternating shift
$F_k\to F_k+(-1)^kW$ leaves the midpoints in place and changes the chords by $-2(-1)^kW$; at a point with $\delta_k^2=1$,
$m_k\cdot\delta_k=0$ and $m_k^2=0$, as on the Euler states \eqref{eq:3d-null-identities}, the advective bracket of
\eqref{eq:rescaled-full-mle} is of second order, and the viscous terms give
\begin{equation}
 \mathsf K=2\sum_{k=0}^{N-1}\bigl(2\,m_k\otimes\delta_k-\delta_k\otimes m_k\bigr).
 \label{eq:stoch-K-euler}
\end{equation}

\begin{proposition}[The zero-winding states are singular points]
\label{prop:stoch-singular}
At every Euler state of even $N$ and zero winding, with any $\beta$ and any balanced spin sequence, $\mathsf K=0$.
\end{proposition}
\begin{proof}
The vertex angles are $\alpha_k=\phi+\beta S_k$, where the running spin $S_k=\sum_{l<k}\sigma_l$ is a closed walk on the integers,
$S_N=S_0$. Such a walk crosses every bond $(l,l+1)$ as many times upwards as downwards. A step that crosses this bond has the midpoint
$m=B\bigl[e\bigl(\phi+(l+\tfrac12)\beta\bigr)+\ii\,\mathbf e_3\bigr]$, where $e(\alpha)=(\cos\alpha,\sin\alpha,0)$, whichever its direction,
and the chord $\delta=\pm e'\bigl(\phi+(l+\tfrac12)\beta\bigr)$, with the sign of its direction. Both terms of \eqref{eq:stoch-K-euler}
are odd in the chord at a fixed midpoint, and they cancel in pairs.
\end{proof}

On an Euler state $\rho_k=0$, so $v=0$, and \eqref{eq:stoch-K} reads $0\cdot u=0$: the equation is silent about the alternating
velocity. This is a genuine ambiguity of the evolution and not an accident of the parametrization, as the following facts show.

\emph{The linearized system.} In the system linearized at a zero-winding state the three components of $u(s)$ are arbitrary functions
of $s$: the linearized MLE has solutions for every $u(s)$ (verified at seven states with $N=4,\dots,12$; ancillary note
\texttt{ergodicityNotes}, Sec.~7).

\emph{The velocity near a state.} At a point $F=G+d\,X$ of the constraint manifold, at a small distance $d$ from a zero-winding state
$G$, both sides of \eqref{eq:stoch-K} are of the order of $d$ and linear in the displacement, $\mathsf K=d\,\mathsf K_1[X]+O(d^2)$
and $-\sum_k(\partial\Xi/\partial F_k)v_k=d\,r_1[X]+O(d^2)$, so that
\begin{equation}
 u\to\mathsf K_1[X]^{-1}r_1[X]\qquad(d\to0),
 \label{eq:stoch-u}
\end{equation}
a ratio of two linear forms, homogeneous of degree zero in the displacement. It does not vanish with $d$, and it depends on the
direction $X$. We computed $u$ at four zero-winding states with $N=6$, $8$, $12$ and $16$, for sixteen random directions each:
$|u|$ ranges from $0.15$ to $2.1$, with the median between $0.30$ and $0.36$, and these numbers are the same for $d=10^{-2}$, $10^{-3}$,
$10^{-4}$ and $10^{-5}$; at a fixed direction $u$ changes by less than $0.3$ per cent between the last two. The velocity field of the MLE
has no limit at the zero-winding Euler states.

\emph{What a direct integration finds.} We integrated the system \eqref{eq:stoch-dae}--\eqref{eq:stoch-K}
from points of the constraint manifold at a distance $d$ from zero-winding Euler states, by an implicit method (the start is stiff, because
the rates of the alternating mode grow as $1/d$). In every run the polygon leaves the Euler state, and in none does it return
(Fig.~\ref{fig:stoch}b). For $N=6$ (the state \texttt{+++---}, $d=10^{-3}$, three runs) the displacement is first amplified in the alternating
direction, ten to fifty-fold within $\Delta s=0.1$, and then grows exponentially: the distance from the state is $0.8$ at $s=2$ and
$6$ at $s=3$ in one run, $2$ and $5$ at $s=8$ and $8$ and $10$ at $s\simeq9.7$ in the two others, and the norm of the polygon grows with
it; with $d=10^{-4}$ the displacement is amplified $150$-fold by $s=1$ and is still growing. For $N=8$ ($\beta=\pi-0.5$, two runs) the
polygon moves away in the alternating direction at once, with a velocity of order one: the distance is $0.09$ at $s=0.1$ and $1.1$ and
$0.6$ at $s=1$. For odd $N$, where the MLE is a regular system, every Euler state is linearly unstable, with a growth
exponent of at least $(N+2)/(2N-2)>\frac12$, and in 141 of 144 runs the perturbed trajectories ended on a collinear configuration
$F_k=\varphi_ke$, which grows as $e^{s/2}$ and carries no vorticity (ancillary note \texttt{ergodicityNotes}, Secs.~3 and~4); the
unrestricted linearization has growing modes in every sector (Remark~\ref{rem:rig3d-linear}). The deterministic evolution near the
Euler ensemble is an instability, not a relaxation.

This does not contradict the rigidity of the Euler polygons. Proposition~\ref{prop:rig3d-even} concerns the trajectories that
stay in the fixed complex lift \eqref{eq:rig3d-fixed-lift}; the lift removes the alternating mode, and inside it every state is at rest.
Rigidity is a property of the restricted system. The unrestricted equation is singular precisely on the states that the restriction
keeps at rest, and a perturbation that leaves the lift, however small, meets the undetermined velocity.

\paragraph{No solution moves along the fixed points.} The alternating velocity is the only one that the time derivative term does
not see, and it is natural to ask whether it can carry the polygon from one Euler state to another: is there a velocity $u(s)$
such that $\partial_sF_k=(-1)^ku(s)$ leads from the vertices of one Euler state to those of another, for instance of the state
with two spins exchanged? More generally, the fixed points form families $G(\lambda)$ with parameters $\lambda$, such as the angle
$\beta$, a complex rotation or the position on the bridge of Sec.~\ref{sec:stoch-move}, and one may look for a motion
$\lambda(s)$ of the parameters on which the time derivative term
$\partial_sG_k+\partial_sG_{k+1}=2\,\partial_s\lambda\,\partial_\lambda m_k$ vanishes: this would be an exact solution of the MLE
that moves along its fixed points. The answer to both questions is negative, and it is exact. A motion on which the time derivative
term vanishes keeps every midpoint in place, so it is an alternating shift of the vertices, $F_k=G_k+(-1)^kW(s)$, and on these
polygons the right side of the MLE is known in closed form.

\begin{proposition}[No solution moves along the fixed points]
\label{prop:stoch-rigid}
Let the null identities \eqref{eq:3d-null-identities} hold on every side of a polygon $G$, and let $W\in\mathbb C^3$. On the polygon
$G_k+(-1)^kW$, which has the midpoints $m_k$ of $G$ and the chords $\delta_k'=\delta_k-2(-1)^kW$, the right side of
\eqref{eq:rescaled-full-mle} is
\begin{equation}
 \boxed{\rho_k=4\Bigl[(-1)^k\,\delta_k\cdot W-W^2\Bigr]m_k-2\,(-1)^k\,(m_k\cdot W)\,\delta_k'
 +4\ii\,\frac{(m_k\cdot W)^2}{\delta_k'^{\,2}}\,\delta_k' .}
 \label{eq:stoch-shift}
\end{equation}
Let $G$ be an Euler state of even $N$ and zero winding with $\beta\ne\pi$, whose running spin $S_k$ takes at least four values. Then:
(a) among the polygons $G_k+(-1)^kW$ with $\delta_k'^{\,2}\ne0$ for all $k$, on which the equation is defined, the only fixed point of the MLE is that with $W=0$;
(b) a path of fixed points through $G$ on which the time derivative term of the MLE vanishes is constant.
\end{proposition}
\begin{proof}
The shift does not change the midpoints. With $\delta_k^2=1$, $m_k\cdot\delta_k=0$ and $m_k^2=0$ the new chords have
$\delta_k'^{\,2}=1-4[(-1)^k\delta_k\cdot W-W^2]$ and $m_k\cdot\delta_k'=-2(-1)^km_k\cdot W$, and the three terms of the right side of
\eqref{eq:rescaled-full-mle} become the three terms of \eqref{eq:stoch-shift}. On an Euler state with $\beta\ne\pi$ the midpoint
$m_k\ne0$ is a null vector, and the chord $\delta_k'$ is not, where the equation is defined; the two are linearly independent, so
$\rho_k=0$ requires $(-1)^k\delta_k\cdot W=W^2$. A step of the walk $S_k$ across the bond $(l,l+1)$ has
$(-1)^k\delta_k=(-1)^le'\bigl(\phi+(l+\tfrac12)\beta\bigr)$, whichever its direction: a unit vector in the plane of the polygon, which
turns by the angle $\beta-\pi$ from one bond to the next. A walk that takes four values crosses three bonds, and for $\beta\ne\pi$
their three unit vectors are different. The function $a\mapsto a\cdot W-W^2$ of a point $a$ of the plane is affine, and it vanishes
at three points of a circle, which are not on one line; so it vanishes identically, the projection of $W$ onto the plane is zero,
$W^2=0$, and hence $W=0$. This is (a). On a path of fixed points the time derivative term of the MLE is
$\partial_sF_k+\partial_sF_{k+1}=2\,\partial_sm_k$. If it vanishes, the velocities are alternating, $\partial_sF_k=(-1)^ku(s)$, the
path is $G_k+(-1)^kW(s)$ with $W=\int u\,\dd s$, and $W=0$ by (a).
\end{proof}

This is the rigidity of the Euler states in its sharpest form: the midpoints of an Euler state determine it among the fixed
points, so the fixed points do not form a family along which the time derivative term vanishes, whatever its parameters. The
alternating shift is the zero mode of the time derivative term, and the directions tangent to the manifold of fixed points are the
zero modes of the right side; no direction is both. To the first order in $W$ the right side \eqref{eq:stoch-shift} is
$4(-1)^k(\delta_k\cdot W)\,m_k-2(-1)^k(m_k\cdot W)\,\delta_k$, and it vanishes on every side only if $W\cdot G_k=0$ for all $k$,
while the vertices of the Euler state span $\mathbb C^3$. The exceptions are degenerate: the two alternating sequences
\texttt{+-+-}$\dots$, whose polygon has two vertices, so that the shift $W=t\,m_0$ along its null midpoint is a complex rotation of
the same polygon; the sequences with three values of $S_k$, which have one more, isolated, fixed point with the same midpoints; and
$\beta=\pi$. We verified \eqref{eq:stoch-shift} numerically to $10^{-16}$, and we searched for the zeros of the right side as a
function of $W$ from $400$ random complex starting points at each of four states with $N=6$, $8$, $12$ and $32$: the only zero
is $W=0$.

Any motion from one Euler state to another thus violates the MLE, and one can only choose where. A motion along the alternating
mode keeps the time derivative term equal to zero and violates the right side, in the first order of the shift, on all the $N$
sides; it cannot reach another Euler state at all, since it does not move the midpoints, while an exchange of two spins moves two
of them (Sec.~\ref{sec:stoch-move}). A step from one fixed point to another keeps the right side equal to zero at both of its ends
and violates the time derivative term, by twice the displacement of the midpoints, whatever the way from one end to the other. The
model of Sec.~\ref{sec:stoch} takes the second way, in its simplest form: the time $s$ is discrete, and a step of the polygon is a step of
the time.

\paragraph{The interpretation.} A deterministic evolution in which an infinitesimal displacement chooses a velocity of order one is
the situation in which spontaneous stochasticity
arises~\cite{BernardGawedzkiKupiainen1998,FalkovichGawedzkiVergassola2001,ThalabardBecMailybaev2020,BandakMailybaevEyinkGoldenfeld2024}:
the randomness of a vanishing noise survives at finite times (Sec.~\ref{sec:stoch-picture}). At even $N$ and zero winding the MLE therefore does not define the
evolution near the Euler ensemble by itself, and it has to be completed by a rule for the motion that it leaves undetermined.
Proposition~\ref{prop:stoch-rigid} limits what such a rule can be: restricted to its fixed points the equation has no motion at
all, so a rule that leads from one Euler state to another must violate it on the way. We choose the rule that violates it
least. The exchange of two neighbouring opposite spins is the step of the least impulse between two Euler states with different
spin sequences (Sec.~\ref{sec:stoch-move}, Propositions~\ref{prop:stoch-residual} and~\ref{prop:stoch-least}), and the thermal
noise selects the side at which it is made, every side with the same probability in the statistical limit
(Sec.~\ref{sec:stoch-trigger}). Section~\ref{sec:stoch} turns this rule into a Markov chain in the discrete time $s$ and proves
its H-theorem, its convergence to the uniform ensemble and the validity of the discrete step.
What the linear loop equation requires of the ensemble is less than a solution of the MLE for every polygon: it requires that the
stationary ensemble consist of exact solutions with constant weights, which holds in the attractor
(Sec.~\ref{sec:stoch-attractor}).

\subsection{The Euler states near \texorpdfstring{$\beta=\pi$}{beta=pi} and the exchange of two neighbouring spins}
\label{sec:stoch-move}

A state of the ensemble is a turning angle $\beta$, a balanced spin sequence $\boldsymbol\sigma=(\sigma_0,\dots,\sigma_{N-1})$,
$\sum_k\sigma_k=0$, and an orientation. Let $\delta=\beta-\pi$, as in Sec.~\ref{sec:3d-fermion} (the detuning $\delta$ carries no
index and should not be confused with the chords $\delta_k$). Since the running spin $S_k$ has the parity of $k$, the polygon
\eqref{eq:3d-complex-euler} is, for every $\beta$,
\begin{equation}
 f_k=R\,(-1)^ke(\theta_k),\qquad \theta_k=\phi+\delta S_k,\qquad
 m_k=B\Bigl[(-1)^k\sigma_k\,e'(\bar\theta_k)+\ii\,\mathbf e_3\Bigr],\qquad \bar\theta_k=\frac{\theta_k+\theta_{k+1}}2,
 \label{eq:stoch-zigzag}
\end{equation}
with $e(\theta)=(\cos\theta,\sin\theta,0)$, $e'(\theta)=(-\sin\theta,\cos\theta,0)$, $R=1/(2\cos\frac\delta2)$ and $B=-\frac12\tan\frac\delta2$;
near $\beta=\pi$ this is the zig-zag \eqref{eq:3d-zigzag}, and $e'(\theta)$ is the unit vector across it.

\paragraph{All the spin sequences meet at $\beta=\pi$.} Two Euler states with the same $\beta$, $\phi$ and plane and with the running
spins $S_k$ and $S_k'$ are at the distance
\begin{equation}
 \bigl\|G-G'\bigr\|^2=4R^2\sum_k\sin^2\frac{\delta\,(S_k-S_k')}2 ,
 \label{eq:stoch-distance}
\end{equation}
which vanishes at $\delta=0$. At a fixed orientation the Euler states of zero winding thus form $\binom N{N/2}$ lines, one for each
balanced sequence and parametrized by $\beta$, and all these lines pass through one point: the degenerate zig-zag
$f_k=\frac12(-1)^ke(\phi)$, a segment of unit length traversed back and forth, with vanishing midpoints. This point is excluded from
the ensemble and from the rigidity statements, which require $\beta\ne\pi$, but the critical layer $\delta=\pm\xi/\sqrt N$ of
\eqref{eq:3d-xi}, which alone contributes in the limit, is its neighbourhood: there the lines of different spin sequences are close to each
other, and the nearest of them, which differ by one exchange, are at the distance $\xi/\sqrt N$. A noise of a fixed strength, however
small, cannot keep them apart as $N\to\infty$; the thermal noise, which decreases as $N^{-1/2}$ in these variables, stays in a fixed
proportion to this distance (Sec.~\ref{sec:stoch-trigger}).

\paragraph{The exchange.} Take two neighbouring steps with opposite spins, $\sigma_k=-\sigma_{k+1}=\sigma$, and exchange the spins. The
running spin changes only at the vertex between the two steps, $S_{k+1}\to S_{k+1}-2\sigma$, so only this vertex moves, and with it
the two adjacent midpoints:
\begin{equation}
 \boxed{\Delta F_{k+1}=2\sigma(-1)^k\sin\frac\delta2\;e'(\theta_k),\qquad
 \Delta m_k=\Delta m_{k+1}=\tfrac12\,\Delta F_{k+1},\qquad \Delta F_j=0\quad(j\ne k+1).}
 \label{eq:stoch-exchange}
\end{equation}
The vertex is reflected in the radius through its two neighbours, which coincide, $f_k=f_{k+2}$ (Fig.~\ref{fig:stoch}a). In terms of
the slow angle the sequence $\theta_k,\theta_k+\sigma\delta,\theta_k$ becomes $\theta_k,\theta_k-\sigma\delta,\theta_k$: a corner of
its graph is flipped. In the critical layer the move is small: $|\Delta F_{k+1}|=2|\sin\frac\delta2|$, so one vertex out of $N$ moves by
$\xi/\sqrt N$, while the vertices themselves are of order one, $|f_k|=R\to\frac12$. The loop functional changes little as well: with
$P_k=F_k/\sqrt{2\nu_N\tau}=NF_k/(\sqrt2L)$ and the sides $\Delta C_k=O(|C|/N)$ of the polygon that approximates a loop of length $|C|$,
one exchange changes the phase $\sum_kP_k\cdot\Delta C_k$ of $\Phi_P[C]$ by an amount of the order of $\xi|C|/(\sqrt N\,L)$. In the
continuum limit the random moves become a continuous motion of the momentum loop.

The new polygon is again an Euler state, with the same $\beta$, the same plane and zero winding, so the right side of
\eqref{eq:rescaled-full-mle} vanishes before and after the move (Sec.~\ref{sec:three-dimensional}).

\paragraph{The discrete equation and the impulse of a step.} Let the time $s$ of the rescaled MLE \eqref{eq:rescaled-full-mle}
advance by steps $\Delta s$, and let $\Delta_sX\equiv X(s+\Delta s)-X(s)$. With the right side taken as the mean of its values at the two ends
of the step, the equation becomes
\begin{equation}
 \boxed{2\,\Delta_sm_k=\frac{\Delta s}2\Bigl[\rho_k\bigl(F(s)\bigr)+\rho_k\bigl(F(s+\Delta s)\bigr)\Bigr]+\mathcal J_k(s),}
 \label{eq:stoch-residual-def}
\end{equation}
where $\rho_k(F)$ is the right side of \eqref{eq:rescaled-full-mle}, and $\mathcal J_k(s)$ is the residual of the discrete equation
on the side $k$ in this step, which we call the impulse of the step; $|\cdot|$ denotes the Hermitian norm. Only the time $s$ of the
rescaled equation is discretized. A step of $s$ multiplies the physical time $\tau$ by $e^{\Delta s}$, so the steps are not uniform in
the physical time, and no discretization of the loop equation in that time is implied: the loop equation and the MLE keep their
time derivatives, and \eqref{eq:stoch-residual-def} is a model of the evolution inside the Euler ensemble, which the MLE leaves
undetermined.

\begin{proposition}[The impulse of a step]
\label{prop:stoch-residual}
(a) For a step from a fixed point of the MLE with the midpoints $m_k$ to a fixed point with the midpoints $m_k'$ the first term on
the right side of \eqref{eq:stoch-residual-def} vanishes, whatever the step $\Delta s$, and
\begin{equation}
 \mathcal J_k=2\,(m_k'-m_k) .
 \label{eq:stoch-impulse}
\end{equation}
(b) For the exchange \eqref{eq:stoch-exchange} of two neighbouring opposite spins of an Euler state,
\begin{equation}
 \boxed{\begin{gathered}
 \mathcal J_k=\mathcal J_{k+1}=\Delta F_{k+1},\qquad\mathcal J_j=0\quad(j\ne k,k+1),\\
 |\mathcal J_k|=2\Bigl|\sin\frac\delta2\Bigr|\simeq\frac\xi{\sqrt N},\qquad
 \frac{|\mathcal J_k|}{|G_k|}\simeq\frac{2\,\xi}{\sqrt N},\qquad\frac{|\mathcal J_k|}{|2m_k|}=\sqrt2\,\cos\frac\delta2 .
 \end{gathered}}
 \label{eq:stoch-residual}
\end{equation}
(c) The same impulse belongs to every continuous motion between the two states. On a path of polygons $F(s')$, $0\le s'\le1$, with
the residual $r_k=\partial_{s'}F_k+\partial_{s'}F_{k+1}-\rho_k(F)$ of the MLE itself,
$\int_0^1r_k\,\dd s'=2(m_k'-m_k)-\int_0^1\rho_k\,\dd s'$, which is \eqref{eq:stoch-impulse} on every path of fixed points, whatever
the path and the law of the motion along it.
\end{proposition}
\begin{proof}
(a) Both $\rho_k(F(s))$ and $\rho_k(F(s+\Delta s))$ vanish. (b) Only the vertex $k+1$ moves, so
$2\Delta_sm_k=2\Delta_sm_{k+1}=\Delta F_{k+1}$, with $|\Delta F_{k+1}|=2|\sin\frac\delta2|$ by \eqref{eq:stoch-exchange};
$|G_k|^2=R^2+B^2=(1+\sin^2\frac\delta2)/(4\cos^2\frac\delta2)$ and $|2m_k|=2\sqrt2\,|B|=\sqrt2\,|\tan\frac\delta2|$.
(c) $\partial_{s'}F_k+\partial_{s'}F_{k+1}=2\,\partial_{s'}m_k$.
\end{proof}

The discrete equation thus needs no path between the two states, and its residual is known exactly. It is confined to the two
sides of the reflected vertex, and there, compared with the vertices, it is $2\xi/\sqrt N$: for $\xi=2$ it is $0.491$, $0.249$,
$0.125$, $0.0625$ and $0.03125$ at $N=64$, $256$, $1024$, $4096$ and $16384$. It vanishes in the statistical limit. Compared with
the midpoints of the two sides it does not: there it is about $\sqrt2$ at every $N$, since the exchange turns a midpoint of the critical
layer into another null direction. But this is the residual on the two sides that the step touches. In the chain of
Sec.~\ref{sec:stoch-H} the bond is chosen at random among $N$, so a given side is touched with a probability of $2/N$ at most, and
the mean square of the impulse over the sides is about $2/N$ of that of the midpoints, Eq.~\eqref{eq:stoch-langevin}. Part (c)
says that nothing is gained by a continuous motion in place of the step: the impulse is the price of the step and not of the way in
which it is made (Remark~\ref{rem:stoch-chord}). The violation of the equation can be reduced in one way only, by a smaller step,
and here the answer is exact.

\begin{proposition}[The step of the least impulse]
\label{prop:stoch-least}
Let $0<|\delta|<\pi$, and let $G$ and $G'$ be two different Euler states of zero winding with the same $\beta$, in the same
plane, which have at least one vertex in common, $G'_{j_0}=G_{j_0}$. Then on at least two sides their midpoints differ by
$|m_j'-m_j|=|\sin\frac\delta2|$, so that the impulses $\mathcal J_j=2(m_j'-m_j)$ of the step from $G$ to $G'$ obey
\begin{equation}
 \boxed{\sum_{j=0}^{N-1}|\mathcal J_j|^2\ge8\sin^2\frac\delta2\simeq\frac{2\xi^2}N .}
 \label{eq:stoch-least}
\end{equation}
For $|\delta|\ne\frac\pi3$ the equality holds if and only if $G'$ is obtained from $G$ by an exchange of two neighbouring opposite
spins in which only the vertex between them moves, Eq.~\eqref{eq:stoch-exchange}.
\end{proposition}
\begin{proof}
The two polygons have the form \eqref{eq:stoch-zigzag} with the slow angles $\theta_j$ and $\theta_j'$ and the same $R>0$. Let
$D_j=\theta_j'-\theta_j$ and $u_j=e(\theta_j')-e(\theta_j)=2\sin\frac{D_j}2\,e'(\theta_j+\frac{D_j}2)$. Then $\Delta F_j=R(-1)^ju_j$,
the vertex $j$ is in place if and only if $D_j\in2\pi\mathbb Z$, the difference $D_{j+1}-D_j=\delta(\sigma_j'-\sigma_j)$ is $0$ or
$\pm2\delta$, and $\mathcal J_j=\Delta F_j+\Delta F_{j+1}=R(-1)^j(u_j-u_{j+1})$.

(i) The inequality. The vertex $j_0$ is in place and the polygons differ, so there is a maximal run of consecutive vertices that
are moved, $k+1,\dots,l$, with the vertices $k$ and $l+1$ in place. Then $D_k\in2\pi\mathbb Z$ and $D_{k+1}\notin2\pi\mathbb Z$,
so $D_{k+1}=\pm2\delta$ modulo $2\pi$, and $|\mathcal J_k|=R\,|u_{k+1}|=2R|\sin\delta|=2|\sin\frac\delta2|$. The same holds on
the side $l$, which is another side. This gives \eqref{eq:stoch-least}.

(ii) The run of an equality. The equality requires $\mathcal J_j=0$, that is $u_j=u_{j+1}$, on every other side. Outside the run
this means that no other vertex is moved, since $u_{l+1}=0$, and inside it that $u_j$ is constant. Suppose that the run contains
the vertices $k+1$ and $k+2$, and let $D_{k+1}=2\eta\delta$ modulo $2\pi$, $\eta=\pm1$, so that
$u_{k+1}=2\eta\sin\delta\;e'(\theta_{k+1}+\eta\delta)$. If $\sigma_{k+1}'=\sigma_{k+1}$, then $D_{k+2}=D_{k+1}$, and $u_{k+2}$ is
$u_{k+1}$ turned by the angle $\sigma_{k+1}\delta$, which is not a multiple of $2\pi$. If $\sigma_{k+1}'=-\sigma_{k+1}$, then
$\sigma_{k+1}=-\eta$, because the vertex $k+2$ is moved, so $D_{k+2}=4\eta\delta$ and
$u_{k+2}=2\eta\sin2\delta\;e'(\theta_{k+1}+\eta\delta)=2\cos\delta\;u_{k+1}$. In both cases $u_{k+2}\ne u_{k+1}$ for
$|\delta|\ne\frac\pi3$. So the run is the single vertex $k+1$.

(iii) The exchange. Now $D_j\in2\pi\mathbb Z$ for all $j\ne k+1$, and $2\delta$ is not a multiple of $2\pi$. Hence
$\sigma_j'=\sigma_j$ for $j\ne k,k+1$, while $\sigma_k'=-\sigma_k$ and $\sigma_{k+1}'=-\sigma_{k+1}$. Since the total spin of
$\boldsymbol\sigma'$ is zero, $\sigma_{k+1}=-\sigma_k$: the step is the exchange \eqref{eq:stoch-exchange} of these two spins.
Conversely, this exchange attains the bound by Proposition~\ref{prop:stoch-residual}(b).
\end{proof}

The exchange of two neighbouring spins is thus singled out among all the steps inside the ensemble by the MLE itself: it is the step
that violates the equation on the least number of sides, two, and by the least amount on each. The common vertex excludes the
rotations of the whole polygon about its normal, which are symmetries of the ensemble and not changes of its state. We verified the
proposition by an enumeration of all the pairs of states with a common vertex for $N=6$, $8$ and $10$: the least value
$8\sin^2\frac\delta2$ is attained by the exchanges of neighbours and by no other pair, and the next value, that of the exchange of
two opposite spins at the distance $2$, is larger by the factor $1+2\sin^2\frac\delta2$, Eq.~\eqref{eq:stoch-arc} below. The same
holds outside the critical layer, at $\delta=0.3$, $-0.9$, $1.5$ and $2.5$ for $N=6$ and $8$. At $|\delta|=\frac\pi3$ the bound
is attained by other steps as well, such as the exchange of two neighbouring pairs, \texttt{++--}$\,\to\,$\texttt{--++}, in which
three vertices move with $u_j$ constant, as the proof shows.

\begin{remark}[The bridge of fixed points: no barrier between the two states]
\label{rem:stoch-bridge}
The two states of an exchange are not separated by any barrier of the right side of the MLE. Let $G$ be an Euler state of zero
winding with $\sigma_k=-\sigma_{k+1}$, and $\boldsymbol\sigma'$ the sequence after the exchange of these two spins. Keep all the
vertices in place except $F_{k+1}$. Then the null identities \eqref{eq:3d-null-identities} hold on the edges $k$ and $k+1$ exactly
on two complex lines,
\begin{equation}
 F_{k+1}=-G_k+2\tau\,m_k(\boldsymbol\sigma)\qquad\hbox{and}\qquad F_{k+1}=-G_k+2\tau\,m_k(\boldsymbol\sigma'),\qquad\tau\in\mathbb C,
 \label{eq:stoch-bridge}
\end{equation}
which meet at $F_{k+1}=-G_k$, where the two midpoints vanish; the point $\tau=1$ of the first line is the state $\boldsymbol\sigma$,
and that of the second line is the state $\boldsymbol\sigma'$. Indeed, at a pair of opposite spins the outer vertices coincide,
$G_{k+2}=G_k$. With $F_{k+1}=-G_k+w$ the two edges have $m_k=m_{k+1}=\frac12w$ and $\delta_k=-\delta_{k+1}=w-2G_k$, and since
$G_k^2=\frac14$ the identities $\delta^2=1$, $m\cdot\delta=0$, $m^2=0$ reduce to $G_k\cdot w=0$ and $w^2=0$. The bilinear form is
nondegenerate on the plane $G_k\cdot w=0$, because $G_k^2\ne0$, so its null vectors form two lines, and the midpoints
$m_k(\boldsymbol\sigma)$ and $m_k(\boldsymbol\sigma')$ span them: they are null, orthogonal to $G_k=m_k-\frac12\delta_k$ and not
proportional, since by \eqref{eq:stoch-zigzag}
$m_k(\boldsymbol\sigma)\cdot m_k(\boldsymbol\sigma')=-2B^2\cos^2\frac\delta2=-\frac12\sin^2\frac\delta2\ne0$. At every point of the
two lines the right side of \eqref{eq:rescaled-full-mle} vanishes on every edge, and so do the constraint $\Xi$ and the matrix
$\mathsf K$ of \eqref{eq:stoch-K-euler}, in which the edges $k$ and $k+1$, with equal midpoints and opposite chords, cancel, and the
other edges cancel as in Proposition~\ref{prop:stoch-singular} (verified numerically to $10^{-16}$). The two segments
$0\le\tau\le1$ join the two Euler states through exact fixed points; their Hermitian length is $4\sqrt2\,|B|\simeq\sqrt2\,\xi/\sqrt N$,
and they leave the fixed complex lift \eqref{eq:rig3d-fixed-lift}: the normal component of the moving vertex, $\ii B(2\tau-1)$,
passes through $-\ii B$ where the segments meet. The model does not use this path. It shows that a step of the chain does not jump
over a region where the equation exerts a force.
\end{remark}

\begin{remark}[Continuous motions between the two states: the mirror chord]
\label{rem:stoch-chord}
If a step is made as a continuous motion in a time $\Delta s$, the residual $r_k$ of the MLE has the time integral of
Proposition~\ref{prop:stoch-residual}(c), and the paths differ only in how they spend it. On the bridge \eqref{eq:stoch-bridge},
travelled at a constant speed, $|r_k|=4\sqrt2\,|B|/\Delta s$ on the two sides: $\sqrt2/\cos\frac\delta2$ times more than the
displacement requires, because of the corner of the bridge. The direct planar chord between the two positions of the vertex,
inside the fixed complex lift,
\begin{equation}
 F_{k+1}(u)=\frac{1-u}2\,G_{k+1}(\boldsymbol\sigma)+\frac{1+u}2\,G_{k+1}(\boldsymbol\sigma'),\qquad u\in[-1,1],
 \label{eq:stoch-chord}
\end{equation}
is not a path of fixed points, but its right side is of the second order in $\delta$. With
$\eta(u)\equiv(1-u^2)\sin^2\frac\delta2$, the midpoint $m_k(u)=m_{k+1}(u)$ and the chord
$\delta_k(u)=-\delta_{k+1}(u)=F_{k+1}(u)-G_k$, without approximation,
\begin{equation}
 \begin{gathered}
 \rho_{k}=\eta\,m_k-\frac\eta2\left[1-\frac{\ii}{2(1-\eta)}\right]\delta_k,\qquad
 \rho_{k+1}=\eta\,m_k-\frac\eta2\left[1+\frac{\ii}{2(1-\eta)}\right]\delta_k,\\
 |\rho_k|^2=|\rho_{k+1}|^2=\eta^2\left(2B^2+\frac14+\frac1{16(1-\eta)}\right),
 \end{gathered}
 \label{eq:stoch-chord-rhs}
\end{equation}
since $w(u)=F_{k+1}(u)+G_k$ obeys $G_k\cdot w=0$ and $w^2=-\eta$, so that $\delta_k^2=1-\eta$, $m_k\cdot\delta_k=-\frac12\eta$ and
$m_k^2=-\frac14\eta$. Hence $|\rho_k|\simeq\sqrt5\,(1-u^2)\,\xi^2/(16N)$, and at a constant speed
$|r_k|=2|\sin\frac\delta2|/\Delta s+O(\xi^2/N)$, with one half of the integral $\int\sum_j|r_j|^2\dd s$ of the bridge. In the
critical layer this is the least possible, to the leading order. For every smooth path between the two states whose vertices stay
within a distance $d=O(|\delta|)$ of them, to the leading order in $d$ and $|B|$,
\begin{equation}
 \begin{gathered}
 \int_0^{\Delta s}|r_j|\,\dd s\ge2\Bigl|\sin\frac\delta2\Bigr|-6\,d\,(|B|+d)\,\Delta s\quad(j=k,k+1),\\
 \int_0^{\Delta s}\sum_j|r_j|^2\,\dd s\ge\frac{8\sin^2\frac\delta2}{\Delta s}-O\Bigl(\frac{\xi^3}{N^{3/2}}\Bigr).
 \end{gathered}
 \label{eq:stoch-bound}
\end{equation}
Indeed, with $F_j=G_j+X_j$ and $|X_j|\le d$ one has $|1-\delta_j^2|\le4d$, $|m_j\cdot e'(\theta_k)|\le|B|+d$,
$|m_j\cdot\delta_j|\le d$ and $|\delta_j\cdot e'(\theta_k)|\le2|B|+2d$, so that the component of $\rho_j$ along the displacement
$e'(\theta_k)$ of the vertex is $6d(|B|+d)$ at most, while the integral of $2\,\partial_sm_j\cdot e'(\theta_k)$ is
$\pm2\sin\frac\delta2$; the second bound follows by the Cauchy--Schwarz inequality. We verified \eqref{eq:stoch-chord-rhs} to
$10^{-16}$ and minimized $\int\sum_j|r_j|^2\dd s$ numerically over the paths of all the vertices in a tube around the chord, in
which every real coordinate of every vertex deviates from the chord path by at most $|B|$ (by at most $3|B|$), for $\Delta s=1$ and
for $\delta=-0.25$ and $\delta=-0.0625$, the values of $\xi=2$ at $N=64$ and $N=1024$. Among the paths without an alternating shift
$F_j\to F_j+(-1)^jW(s)$ of the polygon the bound $8\sin^2\frac\delta2/\Delta s$ is not passed, and the chord attains it: the least
value is $1.0004$ and $1.00002$ of the bound, against $1.0007$ and $1.00004$ for the chord. The alternating shift, which the time
derivative term does not see and which acts through the right side on all the sides at once, reduces the action to the fraction
$0.92$ ($0.77$) of the bound at $\delta=-0.25$ and $0.98$ ($0.95$) at $\delta=-0.0625$: a gain proportional to $|\delta|$, which
vanishes in the statistical limit. (Without the tube the minimization has no meaning at even $N$: the residual does not contain the
alternating velocity, and the minimizing sequences make large alternating shifts between the steps of the discretized time.) None
of this is needed for the chain, in which a step of the polygon is a step of the time; it shows that the impulse
\eqref{eq:stoch-impulse} cannot be avoided by a continuous motion in the critical layer either.
\end{remark}

\begin{remark}[An exact solution with one moving vertex]
\label{rem:stoch-onevertex}
The MLE does move a single vertex between two coinciding neighbours, but away from the Euler states and not between them. Let $G$ be an Euler
state with $\sigma_k=-\sigma_{k+1}$, so that $G_{k+2}=G_k$, let $n$ be a null vector with $n\cdot G_k=0$, and keep all the vertices
in place except $F_{k+1}$. Then $F_{k+1}=\alpha(s)\,G_k+\nu(s)\,n$ is an exact solution of \eqref{eq:rescaled-full-mle} if
\begin{equation}
 \partial_s\alpha=\tfrac12(1+\alpha),\qquad\partial_s\nu=\tfrac14(1+\alpha)\,\nu ,
 \label{eq:stoch-onevertex}
\end{equation}
that is, $1+\alpha=\epsilon_0e^{s/2}$ and $\nu=\nu_0\exp\bigl[\tfrac12(1+\alpha-\epsilon_0)\bigr]$: on the two sides that contain the
vertex the right sides are equal, and their advective brackets vanish. The line $\alpha=-1$ is the bridge \eqref{eq:stoch-bridge} of Remark~\ref{rem:stoch-bridge}, and every point of it is at rest.
A vertex displaced from it by $\epsilon_0G_k$ leaves it as $e^{s/2}$, at the growth rate $\frac12$ of the collinear configurations
quoted above (verified by numerical integration of the complete equation to $10^{-16}$).
\end{remark}

The exchange is the smallest move inside the ensemble: the flip of a single spin changes the total spin, and the polygon no longer
closes. More generally, if $S_k=S_l$, the reversal of all the spins between the vertices $k$ and $l$ reflects this arc of the polygon
in the radius through $f_k=f_l$ and leaves the other vertices in place; the exchange is the shortest such move, $l=k+2$.
The exchange of two distant opposite spins, $\sigma_k=-\sigma_l$, is a move of another kind, and a small one: it turns the arc of
the polygon between the two steps by the angle $2\delta$ about the normal, and moves every vertex of the arc by
$2R|\sin\delta|\simeq\xi/\sqrt N$. The displacements of two neighbouring vertices of the arc nearly cancel in their midpoint, so the
impulse of this step is $2|\sin\frac\delta2|$ on its two end sides and only $4\sin^2\frac\delta2$ on each of the $\ell-1$ sides
inside the arc, $\ell=l-k$:
\begin{equation}
 \sum_j|\mathcal J_j|^2=8\sin^2\frac\delta2\,\Bigl[1+2(\ell-1)\sin^2\frac\delta2\Bigr] .
 \label{eq:stoch-arc}
\end{equation}
It has a path of exact fixed points as well, on which the arc is turned as a rigid body by a
complex rotation while its two end vertices move along null lines (verified numerically for $N$ up to $2048$ to $10^{-15}$), and a
path inside the fixed complex lift, the real rotation of the arc about the normal, on which the right side of the MLE vanishes on
all the sides inside the arc and is $O(\delta^2)$ on its two end sides. Unlike the exchange of neighbours it
displaces a macroscopic fraction of the vertices at once; Remark~\ref{rem:stoch-scaling} shows what it gains.

\subsection{The thermal noise as a trigger: the exchange of the least impulse}
\label{sec:stoch-trigger}

The random choice of the exchange does not have to be postulated as a new element of the theory (Sec.~\ref{sec:stoch-picture}). The
velocity of a real fluid always has thermal fluctuations: the molecular motion that makes it a
fluid, and gives it its viscosity, also makes its velocity fluctuate. In the momentum-loop representation the thermal fluctuations
are a white noise of the momentum loop, and on a polygon they are independent isotropic Gaussian displacements of its vertices
(Supplement~\ref{sec:qsp-gibbs}, Eq.~\eqref{eq:qsp-white}); they add to the momentum loop of the flow when they are independent of
it, as in the rotating fluid in equilibrium (Supplement~\ref{sec:qsp-rotating}). For a regular evolution such a noise would be a
correction that vanishes with its amplitude. At the Euler states of the MLE it is not, because an arbitrarily small displacement
chooses there a velocity of order one (Sec.~\ref{sec:stoch-singular}). The noise then acts as a trigger: it decides what happens,
and in the statistical limit its amplitude does not enter the result.

\paragraph{The rule.} A simple rule makes this explicit. The step of the least impulse is an exchange of two neighbouring spins
(Proposition~\ref{prop:stoch-least}), and its impulse is proportional to the midpoint of the side at which it is made,
$|\mathcal J_k|=\sqrt2\,\cos\frac\delta2\,|2m_k|$ by \eqref{eq:stoch-residual}. On an Euler state all the midpoints have the same
length, so the least impulse is the same on all the $N$ sides, and the equation does not choose among them. The noise does. Let the
vertices be displaced by independent isotropic real Gaussian vectors, $F_k=G_k+\epsilon\,\eta_k$. The midpoints become
\[
 2m_k\;\to\;2m_k+\epsilon\,(\eta_k+\eta_{k+1}) ,
\]
they are no longer equal, and one of them is the smallest; by the same proportionality, an exchange at this side would cost the
least impulse. The rule is that the spins at the two ends of this side, $\sigma_k$ and $\sigma_{k+1}$, are exchanged, whatever they are, and the
polygon becomes the Euler state of the new sequence; if the two spins are equal, nothing changes. Here the spin $\sigma_j$ is
attached to the vertex $j$ as the direction of the step that leaves it, $S_{j+1}=S_j+\sigma_j$, so the exchange is that of the bond
$k$ of the model of Sec.~\ref{sec:stoch} (Definition~\ref{def:stoch-model}), and for opposite spins it reflects the vertex $k+1$. The side is chosen by its midpoint
and not by its spins: a choice among the pairs of opposite spins only would be made with a probability that depends on the number of
such pairs, and the weights of the states would be proportional to this number instead of being equal.

In the variables of the slow angle this side is the shortest link. By \eqref{eq:stoch-zigzag} the points
$(-1)^kf_k=R\,e(\theta_k)$ step around the circle of radius $R$ by the angles $\pm\delta$, and the link between the consecutive
points $k$ and $k+1$ is, up to the sign, the real part of the midpoint $2m_k$ of the side $k$:
$R\,[e(\theta_{k+1})-e(\theta_k)]=-2(-1)^k\,{\rm Re}\,m_k$. On an Euler state all the links have the length $|\tan\frac\delta2|$,
and an exchange at a link costs the impulse $2|\sin\frac\delta2|$, which is $2\cos\frac\delta2$ times its length. The noise
displaces the points and leaves the imaginary part $\ii B\,\mathbf e_3$ of the midpoints in place, so the side with the smallest
midpoint is the shortest link between two neighbouring points, and the rule exchanges the spins at its two ends. A link that
vanishes is halfway through an exchange: the vertex $k+1$ is then at $-f_k$, on the mirror in which the exchange of two opposite
spins reflects it; and where the midpoint vanishes altogether, at $F_{k+1}=-G_k$, the two states of the exchange meet through
fixed points of the equation (Remark~\ref{rem:stoch-bridge}). The shortest link is thus the one nearest to its exchange.

\paragraph{The amplitude of the thermal noise.} How large is $\epsilon$ for the thermal noise? The vertices enter the loop functional
as $P_k=NF_k/(\sqrt2L)$ (Sec.~\ref{sec:stoch-move}), so the displacements $\epsilon\eta_k$ add to the phase $\sum_kP_k\cdot\Delta C_k$
of a loop $C$ a Gaussian variable with the variance $N^2\epsilon^2\sum_k|\Delta C_k|^2/(2L^2)$. For a polygon with $N$ equal
sides $\sum_k|\Delta C_k|^2=|C|^2/N$, and the noise multiplies the loop functional by
\[
 \Bigl\langle\exp\Bigl(\frac{\ii N\epsilon}{\sqrt2L}\sum_k\eta_k\cdot\Delta C_k\Bigr)\Bigr\rangle=\exp\Bigl(-\frac{N\epsilon^2|C|^2}{4L^2}\Bigr) .
\]
For the thermal noise this factor is the perimeter law $e^{-m|C|}$ of Supplement~\ref{sec:qsp-gibbs}, hence
\[
 \epsilon=2L\sqrt{\frac m{N|C|}},\qquad
 \frac\epsilon{|\tan\frac\delta2|}\simeq\frac{4L}\xi\sqrt{\frac m{|C|}},\qquad
 e^{-m|C|}=\exp\Bigl(-\frac{\xi^2|C|^2}{16L^2}\,\frac{\epsilon^2}{\tan^2\frac\delta2}\Bigr) .
\]
The length $|C|$ of the loop enters $\epsilon$ through the parametrization. The white noise of Supplement~\ref{sec:qsp-gibbs},
Eq.~\eqref{eq:qsp-white}, is defined in the arc length of the loop: a vertex of the polygon stands for a piece of the loop of the
length $|C|/N$, and each component of its thermal momentum has the variance $2mN/|C|$. In an arbitrary parametrization the same
distribution is a mixture of such noises with different amplitudes, which does not contain the loop
(Eq.~\eqref{eq:qsp-gibbs-W} there, with the proper time). Neither the loop nor the amplitude reaches the model: as shown below,
in the statistical limit the links are selected with the probability $1/N$ at every value of the ratio
$\epsilon/|\tan\frac\delta2|$.
In the variables of the Euler polygon the thermal displacement of a vertex decreases as $N^{-1/2}$, as the links do, and its
ratio to the links does not depend on $N$. A noise with a fixed $\epsilon$ would not be a small one: its factor in the loop
functional vanishes as $N$ grows. A noise that only triggers the exchanges, and does not change the loop functional of a loop of
the size $L$ by itself, has $m|C|\ll1$, and it is therefore weak compared with the links, $\epsilon\ll|\tan\frac\delta2|$. For a
loop of the length $L$ and $\xi=2$ the thermal factor is $0.996$, $0.98$, $0.94$, $0.78$ and $0.37$ at
$\epsilon/|\tan\frac\delta2|=\frac18$, $\frac14$, $\frac12$, $1$ and $2$. The weak noise is thus the case of a vanishing noise, and
we consider it first.

\paragraph{The weak noise.} For $\epsilon\ll|\tan\frac\delta2|$ the noise cannot cancel a link. The lengths of the links change in
the first order, by $\epsilon X_k$ with $X_k=n_k\cdot(\eta_k+\eta_{k+1})$, where $n_k$ is the unit vector of
${\rm Re}\,m_k$, and the shortest link is that of the least $X_k$. The $X_k$ have the variance $2$, and consecutive ones are
correlated through their common vertex: $\langle X_kX_{k+1}\rangle=1$ if $\sigma_k=-\sigma_{k+1}$, where the walk of the slow angle
turns back, and $-\cos\delta$ if $\sigma_k=\sigma_{k+1}$, where it goes on. A link is the shortest less often if a neighbour is
positively correlated with it, so the choice depends on the neighbouring spins. A choice that depends on the neighbouring spins
is the case of Proposition~\ref{prop:stoch-ising}; here $A/B-1=0.109$, $0.060$, $0.034$ and $0.023$ at $N=64$, $256$,
$1024$ and $4096$. The Gaussian tail gives an estimate. The link of the pattern \texttt{+-+-} has two positively correlated
neighbours and that of \texttt{++--} has one, so $A/B-1$ is about the probability that such a neighbour is shorter than the
shortest link: if a standard normal variable exceeds $z_N$ with the probability $1/N$, it is the probability that it exceeds
$z_N/\sqrt3$, which is $0.107$, $0.062$, $0.037$ and $0.022$ at these $N$ and decreases as $N^{-1/3}$ up to a power of $\log N$.
The coupling $J\simeq\frac14(A/B-1)$ of the Ising weights \eqref{eq:stoch-ising} thus vanishes in the statistical limit, and the
Hamiltonian $H_b^{(J)}$ of the pendulum with these weights (Sec.~\ref{sec:stoch-H}) tends to that of the uniform ensemble. For
$N=8$, $10$ and $12$ we found the stationary distribution of the chain with this rule in this limit on the complete state space: its
weights differ by a factor of up to $2.3$, and they follow \eqref{eq:stoch-ising} with $J\simeq0.04$ within $7\%$. The rule is not
symmetric under the reflection of the loop: the selected link exchanges its step with the next one, and it is selected less often
if the previous step is opposite to its own, since it then has two positively correlated neighbours. The probabilities $r_1$
and $r_2$ of the move \texttt{++-+}$\,\to\,$\texttt{+-++} and of its inverse therefore differ ($1.00/N$ and $0.89/N$ at $N=64$);
by Proposition~\ref{prop:stoch-ising} this does not change the stationary distribution, which depends on $A/B$ only.

The decrease of the coupling uses the Gaussian distribution of the thermal noise, in which two correlated variables are rarely
extreme together. For an isotropic noise with exponential tails $A/B-1$ decreases only slowly, $0.19$, $0.14$, $0.11$ and $0.10$ at
the same $N$; for a noise with power-law tails it does not decrease at all ($0.25$, $0.26$, $0.33$ and $0.37$ for the Student
distribution with three degrees of freedom), since one large displacement then decides the choice; for a bounded noise it
decreases as it does for the Gaussian one ($0.080$, $0.042$ and $0.020$ at $N=64$, $256$ and $1024$ for the uniform distribution
in a disc).

\paragraph{A noise comparable with the links.} If the noise is able to cancel a link, that is, if $\epsilon$ is not small compared
with $|\tan\frac\delta2|$, the shortest link of the displaced polygon is one that the noise has nearly cancelled, and the
probability of this is the same for every $k$, because the links of the Euler state have the same length and the noise is isotropic
and the same at all the vertices. Each link is then selected with the probability $1/N$, up to the effect of the correlation of the
consecutive links, which decreases as $N$ grows. As $\epsilon/|\tan\frac\delta2|\to\infty$ the selected link becomes that of the
smallest $|\eta_k+\eta_{k+1}|$, whatever the state, and the probability $1/N$ becomes exact, for every noise that is statistically
the same at all the vertices. In a simulation with $\xi=2$ the links with opposite spins at their ends were selected less often
than the links with equal spins by $3.4\%$, $1.0\%$ and $0.2\%$ at $\epsilon=\frac12|\tan\frac\delta2|$ for $N=64$, $256$ and
$1024$, by $1.0\%$, $0.2\%$ and $0.0\%$ at $\epsilon=|\tan\frac\delta2|$, and by $0.2\%$ or less at
$\epsilon\ge2|\tan\frac\delta2|$, with a statistical error of $0.14\%$ or less; at $\epsilon\ge2|\tan\frac\delta2|$ the frequencies
of the individual links are compatible with $1/N$ as well ($\chi^2$ per degree of freedom from $0.94$ to $1.19$). Between the two
cases the dependence on the spins decreases monotonically as the noise grows: $A/B-1=0.100$, $0.082$, $0.034$ and $0.013$ at
$\epsilon/|\tan\frac\delta2|=\frac18$, $\frac14$, $\frac12$ and $1$ for $N=64$, against $0.109$ in the limit of the weak noise,
and $0.030$, $0.008$ and $0.002$ at the first three of these ratios for $N=1024$, against $0.034$. At a fixed ratio it decreases
with $N$ faster than in that limit.

\paragraph{The result.} In the statistical limit each link is thus selected with the probability $1/N$, which is the rule (i) of
the model of Sec.~\ref{sec:stoch} (Definition~\ref{def:stoch-model}), whatever the amplitude of the noise: for a weak noise because
the coupling that it induces vanishes, and for a noise comparable with the links because it does not distinguish them. The
randomness of the model is that of the thermal noise, of which the isotropy, the homogeneity along the loop and, for a weak noise,
the Gaussian distribution are used. The temperature of the fluid enters the theory only here, and its value drops out; it is not
the temperature of the pendulum, which is a property of the loop and of its fast spins and not of the fluid
(Sec.~\ref{sec:cond-thermal}). What is still assumed is the rule, that the displaced polygon passes to its next Euler state by the
exchange of the least impulse, and the time $\Delta s$ between two such passages; to derive them from the singular evolution is the
open problem of Sec.~\ref{sec:stoch-attractor}.

\section{Models of ergodic diffusion over the Euler ensemble}
\label{sec:stoch}

The mechanism of Sec.~\ref{sec:stoch-mechanism} calls for a random motion of the polygon over the Euler states, which the MLE
cannot supply itself. The question is how to realize the random moves with minimal distortion of the equation, while preserving the
exact symmetry of the transitions, so that an H-theorem drives the ensemble to uniformity. We answer it by a discretization of the
rescaled MLE in its own time. Only the logarithmic time $s=\log(\tau/\tau_*)$ of the
rescaled equation \eqref{eq:rescaled-full-mle} is made discrete, with a step $\Delta s$; the loop equation in the physical time and the
derivation of the MLE from it, with their time derivatives, are not touched. The derivative $2\partial_sm_k$ is replaced by the
difference $2[m_k(s+\Delta s)-m_k(s)]$ over a step, and the right side by the mean of its values at the two ends of the step. In one step the polygon
passes from an Euler state to an Euler state: by an exchange of two neighbouring opposite spins, which reflects one vertex, and by
a shift of the turning angle by $\pm\Delta\beta$ with mirror reflection at the ends of the admissible interval. At both ends of
every step the right side of the MLE vanishes, so the right side of the discrete equation is zero, and no path between the two
states has to be drawn. The left side is not zero: it is twice the displacement of the midpoints, which we call the impulse of the
step (Proposition~\ref{prop:stoch-residual}). For an exchange it differs from zero on the two sides of the reflected vertex only,
where it is $2|\sin\frac\delta2|\simeq\xi/\sqrt N$, with the scaling variable $\xi=\sqrt N|\beta-\pi|$ of the critical layer
\eqref{eq:3d-xi}; among all the steps between two Euler states with different spin sequences the exchange of two neighbours has the
least impulse (Proposition~\ref{prop:stoch-least}). The impulses are the noise of the model, a stochastic force added to the equation. A single polygon does not obey the
MLE with them, and it is not required to: every state that it visits is an exact solution, and the loop functional and the
vorticity correlations are averages over the probability distribution of these states. It is this distribution, and not the
position of a vertex, that the discretization has to reproduce.

The model is thus a Markov chain: the next state is a function of the present state and of a random variable, the number of the
bond and the sign of the shift. Its transition probabilities are symmetric, a step and its inverse are equally probable, and this
is the only property of them that the result needs. The probability distribution $W(\boldsymbol\sigma,\beta,s)$ of the state obeys a
linear difference equation, which we solve. For this chain ergodicity is a theorem
(Propositions~\ref{prop:stoch-H}, \ref{prop:stoch-exchange} and~\ref{prop:stoch-beta}). Starting from any distribution, even from a
single polygon, the entropy grows in every step, and the distribution converges to equal weights of the balanced spin sequences and
to a uniform density of $\beta$; along almost every trajectory the time averages converge to the averages over this ensemble, which
is ergodicity in the sense of Boltzmann. This uniform ensemble of random spins is what the reduction to the quantum pendulum uses
(Secs.~\ref{sec:3d-fermion}--\ref{sec:3d-even}). The exact solution also verifies the discretization. The distribution $W(s)$ of the chain is compared with the solution
$W^{\rm c}(s)$ of the master equation with a continuous $s$ and the same transition probabilities per unit of $s$. In a mode that a
step changes with the probability $\mu<1$ the two solutions differ by $\mu/[2e(1-\mu)]$ at most, at all times. The difference
quotient $\Delta_sW^{\rm c}/\Delta s=[W^{\rm c}(s+\Delta s)-W^{\rm c}(s)]/\Delta s$ of the continuous solution differs from its
derivative $\partial_sW^{\rm c}$ by the fraction $\frac12\mu$ at most; and the difference quotient $\Delta_sW/\Delta s$ of the chain,
which is equal to $\partial_sW^{\rm c}$ at $s=0$, differs from it later by $\mu/[2e(1-\mu)]$ of its initial value at most. For the
correlation functions of $k$ spins $\mu\le4k/N$, so all these errors are of the order of $k/N$ and vanish in the statistical limit
(Proposition~\ref{prop:stoch-discrete}). The length of
the step drops out of them: they are set by the part of the polygon that a step changes. The step is therefore a free
parameter, the unit of the clock of the model. The approach to the uniform ensemble takes $N^3/(4\pi^2)$ steps, that is, the time
$T=\Delta s\,N^3/(4\pi^2)$. With $\Delta s=c/N^3$ the relaxation time $T=c/(4\pi^2)$ is finite in the statistical limit, the error
of the distribution is $2k/N$ as before, and the slow angle of the polygon obeys a Langevin equation that does not contain $N$
(Remark~\ref{rem:stoch-step}). The price is paid by the single polygon and not by the averages: a finite relaxation time requires
impulses that displace the midpoints by their own size in this time, whatever the symmetric steps
(Proposition~\ref{prop:stoch-intensity}, Remark~\ref{rem:stoch-scaling}).

Let us stress what is claimed and what is not. The singularity of the MLE is proved (Proposition~\ref{prop:stoch-singular}), and
so are the absence of solutions that move along the fixed points, the impulse of a step and the least impulse of the exchange of
neighbours (Propositions~\ref{prop:stoch-rigid}, \ref{prop:stoch-residual} and~\ref{prop:stoch-least}); these are the results
of Sec.~\ref{sec:stoch-mechanism}. The
random step is a model: its choice is not unique, it is not derived from the singular evolution, and this section is not a
derivation of the equal weights from the Navier--Stokes equation. What the model shows is that the statistical state assumed in this
paper is the attractor of a minimal-distortion symmetric stochastic dynamics inside the Euler ensemble, and that the attractor is
insensitive to the details of this dynamics: it does not depend on the probabilities of the moves or on their step sizes, only on the
symmetry of the transitions, which we classify precisely (Propositions~\ref{prop:stoch-ising} and~\ref{prop:stoch-beta}).
Section~\ref{sec:stoch-H} describes the Markov chain, its H-theorem, its solution and the validity of the discrete step,
Sec.~\ref{sec:stoch-beta} the reflected shifts of the turning angle, Sec.~\ref{sec:stoch-attractor} the attractor, and
Sec.~\ref{sec:stoch-rotation} the randomization of the global orientation of the polygon;
Fig.~\ref{fig:stoch}c shows the growth of the entropy in the model.

\subsection{The model: a Markov chain in the discrete time \texorpdfstring{$s$}{s}, and its H-theorem}
\label{sec:stoch-H}

\begin{definition}[The stochastic model]
\label{def:stoch-model}
The state is a balanced spin sequence $\boldsymbol\sigma$ on a ring of even length $N$ and a turning angle $\beta$ in the admissible
interval $I_N=[\beta_{\min},2\pi-\beta_{\min}]\subset(0,2\pi)$ ($\beta_{\min}=2\pi/N$); the polygon is the Euler state
\eqref{eq:3d-complex-euler} with these parameters. The time $s=\log(\tau/\tau_*)$ advances by steps $\Delta s$, and in every step the
state changes by two independent random moves.
(i) A bond $k$ of the ring, the pair of the neighbouring steps $k$ and $k+1$, the pair $(N-1,0)$ included, is chosen at random, with
the probability $1/N$, and the spins $\sigma_k$ and $\sigma_{k+1}$ are exchanged. If they are opposite, the vertex between the two
steps is reflected, Eq.~\eqref{eq:stoch-exchange}; if they are equal, the polygon does not change.
(ii) With a probability $q\le1$ the turning angle is shifted, by $+\Delta\beta$ or by $-\Delta\beta$ ($0<\Delta\beta<|I_N|$) with
equal probabilities, with mirror reflection at both ends of $I_N$, Eq.~\eqref{eq:stoch-Tpm}; otherwise it keeps its value.
The step $\Delta s$ is a parameter of the model: the number of the steps in a unit of $s$ is $1/\Delta s$.
\end{definition}

The next state is thus a function of the present state and of a random variable $\omega_s=(k_s,\epsilon_s)$: a bond $k_s$
distributed uniformly over the ring and a sign $\epsilon_s=0,+1,-1$ with the probabilities $1-q$, $\frac q2$, $\frac q2$,
independent in different steps. The model is a Markov chain, and the polygon obeys the discrete equation
\eqref{eq:stoch-residual-def}, in which the right side of the MLE vanishes, with the random impulses of Proposition~\ref{prop:stoch-residual}. For the
exchanges
\begin{equation}
 2\,\Delta_sm_j=\mathcal J_j(s),\qquad
 \Bigl\langle\sum_j|\mathcal J_j|^2\Bigr\rangle\simeq\frac{\xi^2}N\simeq\frac2N\,\sum_j|2m_j|^2 ,
 \label{eq:stoch-langevin}
\end{equation}
because one bond out of $N$ is chosen, about one half of the bonds join opposite spins, an exchange has
$\sum_j|\mathcal J_j|^2=8\sin^2\frac\delta2\simeq2\xi^2/N$, and $\sum_j|2m_j|^2=8NB^2\simeq\xi^2/2$ in every state. The impulse
is of the order of the midpoints on the two sides that a step touches, and its mean square over the sides is $2/N$ of that of the
midpoints, in every step and for every $\Delta s$.

The random choice of the bond is not an independent postulate. The thermal noise of the fluid makes it: it selects the side at
which an exchange costs the least impulse, and in the statistical limit every side is selected with the probability $1/N$
(Sec.~\ref{sec:stoch-trigger}). The sign of the shift of the angle can be taken from the same noise (Sec.~\ref{sec:stoch-beta}).

Let $W_a(s)$ be the probability of the state $a$ at the time $s$, and $P_{ab}$ the probability of the transition from $a$ to $b$
in one step, $\sum_bP_{ab}=1$. The distribution obeys the linear equation
\begin{equation}
 W_b(s+\Delta s)=\sum_aW_a(s)\,P_{ab},\qquad\hbox{that is,}\qquad \Delta_sW_b=\sum_a\bigl(W_aP_{ab}-W_bP_{ba}\bigr).
 \label{eq:stoch-master}
\end{equation}
The transition probabilities are called \emph{doubly stochastic} if also $\sum_aP_{ab}=1$ for every $b$, \emph{irreducible} if every
state can be reached from every other one by a chain of moves, and \emph{aperiodic} if the numbers of steps in which a state can
return to itself have no common divisor. We first state the H-theorem in the generality in which it holds, for a finite set of $n$
states; the balanced spin sequences are such a set, with $n=\binom N{N/2}$, and the angle is treated in Sec.~\ref{sec:stoch-beta}.

\begin{proposition}[H-theorem for doubly stochastic moves]
\label{prop:stoch-H}
Let the transition probabilities be doubly stochastic. Then:
(a) The entropy $\mathsf S=-\sum_aW_a\log W_a$ does not decrease in a step,
\begin{equation}
 \boxed{\mathsf S(s+\Delta s)-\mathsf S(s)=\sum_{a,b}W_a(s)\,P_{ab}\,\log\frac{W_a(s)}{W_b(s+\Delta s)}\;\ge\;0 ,}
 \label{eq:stoch-H}
\end{equation}
with equality only if, for every $b$, all the $W_a(s)$ with $P_{ab}>0$ are equal.
(b) If the chain is irreducible and aperiodic, $W_a(s)\to1/n$ as $s\to\infty$ from every initial distribution.
(c) For almost every trajectory of an irreducible chain the fraction of the time spent in each state tends to $1/n$: the time
average of every function of the state tends to its average with equal weights.
(d) The same holds for the equation with a continuous time, $\Delta s\,\partial_sW_b=\sum_a(W_aP_{ab}-W_bP_{ba})$, with
$\Delta s\,\dd\mathsf S/\dd s=\sum_{a,b}W_aP_{ab}\log(W_a/W_b)\ge0$.
\end{proposition}
\begin{proof}
(a) Since the rows of $P$ sum to one, $\mathsf S(s)=-\sum_{a,b}W_aP_{ab}\log W_a$, and
$\mathsf S(s+\Delta s)=-\sum_{a,b}W_aP_{ab}\log W_b(s+\Delta s)$ by \eqref{eq:stoch-master}; this gives the sum in \eqref{eq:stoch-H}. Since the
columns sum to one as well, $W_b(s+\Delta s)$ is a mean of the $W_a(s)$ with the weights $P_{ab}$, and by the convexity of $x\log x$,
$\sum_aP_{ab}W_a\log W_a\ge W_b(s+\Delta s)\log W_b(s+\Delta s)$ for every $b$~\cite{CoverThomas2006}, with equality only if the $W_a$ that enter
are equal. (b), (c) The uniform distribution is stationary because the columns sum to one. An irreducible chain on a finite set has
one stationary distribution; every initial distribution converges to it if the chain is aperiodic, and the time averages converge
to the averages over it with probability one (the ergodic theorem for Markov chains). (d) Here
$\dd\mathsf S/\dd s=-\sum_b\partial_sW_b\log W_b$, since $\sum_b\partial_sW_b=0$, which is the sum quoted after the two indices are renamed in one of the terms,
and by the inequality $\log x\ge1-1/x$ it is not less than $\sum_{a,b}P_{ab}(W_a-W_b)=0$.
\end{proof}

\begin{proposition}[The random exchanges]
\label{prop:stoch-exchange}
For even $N\ge4$ the exchanges of Definition~\ref{def:stoch-model} have the transition matrix $P=\frac1N\sum_kT_k$, where $T_k$ exchanges the two
spins of the bond $k$. It is symmetric, irreducible and aperiodic, so Proposition~\ref{prop:stoch-H} applies: the entropy of the
distribution of the spin sequences grows in every step, until the weights are equal. The approach to the equal weights is
exponential,
\begin{equation}
 \sum_a\Bigl(W_a(s)-\frac1n\Bigr)^2\le\bigl(1-\Delta_N\bigr)^{2s/\Delta s}\sum_a\Bigl(W_a(0)-\frac1n\Bigr)^2,\qquad
 \boxed{\Delta_N=\frac2N\Bigl(1-\cos\frac{2\pi}N\Bigr)\simeq\frac{4\pi^2}{N^3} ,}
 \label{eq:stoch-gap}
\end{equation}
where $1-\Delta_N$ is the eigenvalue of $P$ next to one.
\end{proposition}
\begin{proof}
The exchange $T_k$ is its own inverse, so $P$ is symmetric. Exchanges of unequal neighbours bring every balanced sequence to
\texttt{+}$\cdots$\texttt{+-}$\cdots$\texttt{-}, as in a bubble sort, so all the sequences are connected, and for $N\ge4$ a sequence with two
equal neighbours keeps its place with a positive probability, so the chain is aperiodic. Further, $P=1+L/N$, where $L$ is the
generator of the symmetric exclusion process on a ring of $N$ sites with $N/2$ particles and the unit rate per
bond~\cite{Liggett1985}. Its gap is a theorem of Caputo, Liggett and Richthammer~\cite{CaputoLiggettRichthammer2010}: the gap of the
exclusion process on any graph equals the gap of a single particle walking on it, here $2(1-\cos\frac{2\pi}N)$. The
eigenvalues of $P$ are not less than $-\frac12$: in terms of the operators $\mathbf s_k$ of the spin one half,
$T_k=\frac12+2\,\mathbf s_k\cdot\mathbf s_{k+1}$, so $P$ is the Hamiltonian of the Heisenberg chain up to a constant, and
$\mathbf s_k\cdot(\mathbf s_{k-1}+\mathbf s_{k+1})\ge-1$. (The least eigenvalue is $-0.43$, $-0.41$, $-0.40$ and $-0.40$ for $N=6$,
$8$, $10$ and $12$, and tends to $1-2\log2=-0.386$, by the energy of the antiferromagnetic ground state.) Hence $1-\Delta_N$, which
is not less than $\frac12$, is the largest modulus of an eigenvalue other than one.
\end{proof}

The rule (i) of Definition~\ref{def:stoch-model} chooses a bond and not an exchange, and this matters. If one of the available
pairs of opposite neighbours were chosen at random instead, a move would be made in every step, but the transition probabilities
would not be doubly stochastic: the stationary weight of a sequence would be proportional to the number of its pairs of opposite
neighbours, which varies between $2$ and $N$, and the chain would have the period two, since every exchange changes the parity of
$\frac12\sum_kk\,\sigma_k$. The symmetry of the step requires the idle steps. We checked both propositions on the complete state
space for $N\le12$: $P$ is symmetric and doubly stochastic, its second eigenvalue equals $1-\Delta_N$ to all digits ($0.833333$,
$0.926777$, $0.961803$ and $0.977671$ for $N=6$, $8$, $10$ and $12$), and the entropy grows in every step. Figure~\ref{fig:stoch}c
shows the entropy for a start from a single sequence. For $N=12$ and the sequence \texttt{++++++------} it rises monotonically from
$0$ to $\log\binom{12}6=6.8287$; its deficit is $3.54$ after $N$ steps, $0.59$ after $4N$ and $7.8\times10^{-4}$ after $16N$ steps,
and it decays as $(1-\Delta_N)^{2n}$ with the number $n$ of the steps.

\paragraph{Which moves give equal weights.} The equal weights do not require equal probabilities of the moves; they require doubly
stochastic ones, and it is worth seeing how much freedom this leaves. The MLE contains only scalar products, so it is covariant
under $O(3)$, and a reflection in a plane that contains the normal $\mathbf e_3$ reverses all the spins. We therefore consider
probabilities that are invariant under the reversal of all the spins and local: when a bond is chosen, the probability of the
exchange may depend on the exchanged pair and on its two neighbouring spins.

\begin{proposition}[Local exchanges]
\label{prop:stoch-ising}
Let the pair $(\sigma,-\sigma)$ between the neighbouring spins $a$ and $b$ be exchanged, when its bond is chosen, with a probability $R(a,\sigma,b)>0$, where
$R(a,\sigma,b)=R(-a,-\sigma,-b)$. Let $A$ be this probability for the move \texttt{++--}$\,\to\,$\texttt{+-+-} and $B$ for the inverse
move. Then the only stationary distribution on the balanced sequences is the Gibbs distribution of an Ising chain,
\begin{equation}
 w(\boldsymbol\sigma)\propto\exp\Bigl(-J\sum_k\sigma_k\sigma_{k+1}\Bigr),\qquad e^{4J}=\frac AB ,
 \label{eq:stoch-ising}
\end{equation}
whatever the two other probabilities, those of \texttt{++-+}$\,\to\,$\texttt{+-++} and of its inverse. The weights are equal if and only if $A=B$.
\end{proposition}
\begin{proof}
With the symmetry there are four probabilities: $A$, $B$, and $r_1$ and $r_2$, those of \texttt{++-+}$\,\to\,$\texttt{+-++} and of its
inverse. The moves $A$ and $B$ change the number of the pairs of unequal neighbours by $\pm2$ and satisfy the detailed balance with
\eqref{eq:stoch-ising}. The moves $r_1$ and $r_2$ do not change this number, so they connect states of equal weight, and they are
doubly stochastic: a sequence has as many moves $r_1$ and $r_2$ out as in. Indeed, the patterns \texttt{++-+} and \texttt{--+-} are the runs of
length one preceded by a longer run, the patterns \texttt{+-++} and \texttt{-+--} are the runs of length one followed by a longer run,
and on a ring the sequence of the run lengths passes from the longer runs to the runs of length one as many times as back. The chain
is irreducible, so the stationary distribution is unique.
\end{proof}

We verified \eqref{eq:stoch-ising} on the complete state space for $N=8$, $10$ and $12$, to $10^{-16}$, including the cases
$r_1\ne r_2$. The proposition delimits the universality of the equal weights. They hold for arbitrary probabilities of the exchange
in an environment of equal neighbours, reversible or not ($r_1\ne r_2$ is a drift of the isolated spins along the loop), and for
arbitrary probabilities that depend only on the exchanged pair: then $A=B$ by the reflection, and even without the reflection a
ring has as many pairs $(+-)$ as $(-+)$, so the transition probabilities are doubly stochastic even for unequal probabilities of
the two exchanges. The equal weights fail only if the rule of the moves distinguishes \texttt{++--} from \texttt{+-+-}, that
is, if the move of a vertex depends on the positions of the vertices two steps away from it. The MLE couples a vertex only to its two
neighbours, through the two edges that it shares, and the two positions of the moving vertex are mirror images with respect to these
neighbours; this is the reason to expect $A=B$. If $A\ne B$, the model would lead to a random spin system with a nearest-neighbour
interaction, and the pendulum would change in a simple way. The correlation of two spins at the distance $r$ is then
$(-\tanh J)^r$, so the variance of the sum of the spins, which gives the diffusion of the slow angle, is multiplied by $e^{-2J}$,
and the variance of their alternating sum, which gives the fast noise, by $e^{2J}$. In the bulk of an arc the Hamiltonian
\eqref{eq:fermion-pair} becomes
\[
 H_b^{(J)}=-\frac12\,e^{-2J}\,\partial_\theta^2+e^{2J}\,\frac{b^2}8\,\sin^2\theta ,\qquad e^{4J}=\frac AB ,
\]
which is the Hamiltonian at $J=0$ with $\xi^2$ replaced by $e^{-2J}\xi^2$ and $b$ by $e^{2J}b$: the slow angle is a Brownian bridge
of the total time $e^{-2J}\xi^2$, and the separation $h$ of the hairpin is rescaled by $e^{2J}$. We obtained
this form from the second cumulant of the steps and checked it by summing the product of the transfers
\eqref{eq:fermion-alternating} over the correlated spins exactly: for $J=\pm0.3$ the difference from
$U_b\,e^{-n\delta^2H_b^{(J)}}U_b^{-1}$ falls as $1/n$, to $10^{-4}$ at $n=3200$ steps, while with the potential of $J=0$ it stays at
$0.12$ and $0.06$. We did not examine the operators at the ends of the arcs, so no statement about the correlation function at
$A\ne B$ is made here. The reversals of longer arcs can be added with any
probabilities that are equal for a move and its inverse; they make the approach faster, but they move many vertices at once, and we
do not use them. So do the exchanges of distant spins (Sec.~\ref{sec:stoch-move}), to which we return in
Remark~\ref{rem:stoch-scaling}.

\paragraph{The solution of the chain: the slow angle and the number of steps.} The chain is solved by its harmonics. The mean spin
obeys a closed equation, the heat equation on the ring with a discrete time,
$\Delta_s\langle\sigma_k\rangle=\frac1N\bigl(\langle\sigma_{k+1}\rangle+\langle\sigma_{k-1}\rangle-2\langle\sigma_k\rangle\bigr)$, so
the Fourier harmonic $j$ of the mean slow angle $\theta_k=\phi+\delta S_k$ is multiplied in every step by $1-\mu_j$,
\begin{equation}
 \mu_j=\frac2N\Bigl(1-\cos\frac{2\pi j}N\Bigr)\le\frac4N ,
 \label{eq:stoch-harmonics}
\end{equation}
and the gap $\Delta_N=\mu_1$ of \eqref{eq:stoch-gap} is the rate of the first harmonic, the slowest observable of the model. In
the same way the correlation functions of $k$ spins obey closed linear equations, those of $k$ particles of the exclusion process on
the ring (the duality of the exclusion process~\cite{Liggett1985}), in which a step moves a particle with a probability of $2k/N$ at
most. On the scale of the loop, $k=xN$, the exchanges are a diffusion of the slow angle: by the theory of the equilibrium
fluctuations of the exclusion process~\cite{KipnisLandim1999} its fluctuations obey, for large $N$, the linear stochastic equation
of Edwards and Wilkinson~\cite{EdwardsWilkinson1982},
\begin{equation}
 \partial_s\theta=\frac1{N^3\Delta s}\,\partial_x^2\theta+\frac{\sqrt2\,\xi}{N^{3/2}\sqrt{\Delta s}}\,\eta(x,s),\qquad
 \bigl\langle\eta(x,s)\eta(x',s')\bigr\rangle=\delta(x-x')\,\delta(s-s') ,
 \label{eq:stoch-EW}
\end{equation}
of which the chain is a discretization. The relaxation therefore takes $1/\Delta_N=N^3/(4\pi^2)$ steps, that is, the time
\begin{equation}
 \boxed{T=\frac{\Delta s}{\Delta_N}\simeq\frac{\Delta s\,N^3}{4\pi^2} .}
 \label{eq:stoch-T}
\end{equation}
From the worst
initial state the distance in total variation drops after $N^3\log N/(8\pi^2)$ steps~\cite{Lacoin2017}, and the entropy deficit
decays at a rate of the same order, $1/N^3$ per step~\cite{LeeYau1998} (both for a continuous time); locally, the spins come to
the equilibrium with the local mean spin in a number of steps of the order of $N$. This is the usual situation in statistical mechanics, where
the relaxation time of a conserved density grows as the square of the size of the system. Whether the time \eqref{eq:stoch-T} is
large or finite depends on the step, to which we return in Remark~\ref{rem:stoch-step}.

\paragraph{The discrete step and the continuous time.} Is the discrete step of $s$ a valid discretization? For the probability
distribution of the states the question has a definite meaning. The equation with a continuous $s$ and the same transition
probabilities per unit of $s$ is the master equation
\begin{equation}
 \partial_sW^{\rm c}_b=\frac1{\Delta s}\sum_a\bigl(W^{\rm c}_aP_{ab}-W^{\rm c}_bP_{ba}\bigr),
 \label{eq:stoch-master-c}
\end{equation}
and \eqref{eq:stoch-master} is its discretization with the step $\Delta s$, in which the derivative $\partial_sW$ is replaced by the
difference quotient $\Delta_sW/\Delta s$. Both equations are solved by the eigenvectors of the symmetric matrix $P$, so the two solutions, and the
difference quotient with the derivative, can be compared exactly.

\begin{proposition}[The validity of the discrete step]
\label{prop:stoch-discrete}
Let $P$ be symmetric, $W(s)=W(0)P^{n}$ the solution of the chain \eqref{eq:stoch-master} after $n=s/\Delta s$ steps, and
$W^{\rm c}(s)=W(0)\,e^{n(P-1)}$ the solution of \eqref{eq:stoch-master-c} with the same initial distribution.
(a) $W^{\rm c}(s)=\sum_{l\ge0}e^{-n}\frac{n^l}{l!}\,W(l\,\Delta s)$: the solution with a continuous time is the chain after a random number of
steps, which has the Poisson distribution with the mean $n=s/\Delta s$.
(b) In an eigenvector of $P$ with the eigenvalue $1-\mu$, $0\le\mu<1$, the chain decays as $(1-\mu)^n$ and the continuous solution
as $e^{-\mu n}$, and for all $s\ge0$
\begin{equation}
 \boxed{0\le e^{-\mu s/\Delta s}-(1-\mu)^{s/\Delta s}\le\frac{\mu}{2e\,(1-\mu)},\qquad
 1-\frac\mu2\le\frac{\Delta_s\,e^{-\mu s/\Delta s}/\Delta s}{\partial_s\,e^{-\mu s/\Delta s}}=\frac{1-e^{-\mu}}\mu\le1 .}
 \label{eq:stoch-discrete}
\end{equation}
Neither bound contains the step $\Delta s$.
(c) For the exchanges of Definition~\ref{def:stoch-model} the functions of $k$ spins span a space that $P$ maps into itself, and
on it $\mu\le4k/N$. Let $4k<N$, and let $F(s)$ and $F^{\rm c}(s)$ be the averages of a function $f$ of $k$ spins in the chain and
in the continuous equation, regarded as functions of the initial sequence. Then at every $s$ of the chain, and for every step $\Delta s$,
\[
 \begin{gathered}
 \bigl\|F^{\rm c}-F\bigr\|\le\frac{2k}{e\,(N-4k)}\,\bigl\|f-\langle f\rangle\bigr\|,\qquad
 \Bigl\|\frac{\Delta_sF^{\rm c}}{\Delta s}-\partial_sF^{\rm c}\Bigr\|\le\frac{2k}N\,\bigl\|\partial_sF^{\rm c}\bigr\|,\\
 \Bigl\|\frac{\Delta_sF}{\Delta s}-\partial_sF^{\rm c}\Bigr\|\le\frac{2k}{e\,(N-4k)}\,\bigl\|\partial_sF^{\rm c}(0)\bigr\| .
 \end{gathered}
\]
The second inequality compares the difference quotient and the derivative of one function, the continuous solution. The third
compares the difference quotient of the chain with the derivative of the continuous solution at the same $s$; the two are equal
at $s=0$, where $\Delta_sF/\Delta s=\partial_sF^{\rm c}$. The norm is the root
of the mean square over the sequences, and $\langle f\rangle$ is the average with equal weights.
\end{proposition}
\begin{proof}
(a) $e^{n(P-1)}=e^{-n}\sum_ln^lP^l/l!$. (b) First, $(1-\mu)^n\le e^{-\mu n}$. Next,
$e^{-\mu n}-(1-\mu)^n\le e^{-\mu n}\,n\,[-\log(1-\mu)-\mu]\le\mu n\,e^{-\mu n}\,\mu/[2(1-\mu)]\le\mu/[2e(1-\mu)]$, since
$-\log(1-\mu)-\mu\le\mu^2/[2(1-\mu)]$ and $x\,e^{-x}\le1/e$. The difference of $e^{-\mu s/\Delta s}$ over a step is
$(e^{-\mu}-1)\,e^{-\mu s/\Delta s}$, its derivative is $-(\mu/\Delta s)\,e^{-\mu s/\Delta s}$, and $1-\frac\mu2\le(1-e^{-\mu})/\mu\le1$. (c) By the duality of the exclusion
process~\cite{Liggett1985} the products of $k$ occupation numbers $\frac12(1+\sigma_i)$ evolve as $k$ particles of the exclusion
process on the ring. In its generator the diagonal elements are $-2k$ at least and the rows sum to zero, so its eigenvalues lie
between $-4k$ and $0$, and $P=1+L/N$. The eigenvectors of the symmetric matrix $P$ are orthogonal, so the bounds (b) of the modes
are bounds of the norms. For the third inequality, in a mode $\Delta_sF/\Delta s-\partial_sF^{\rm c}=(\mu/\Delta s)\,[e^{-\mu n}-(1-\mu)^n]$
times the amplitude, and $\mu/\Delta s$ times the amplitude is the modulus of the initial rate.
\end{proof}

\begin{table}[!htb]
\centering\small
\caption{The chain against the equation \eqref{eq:stoch-master-c} with a continuous $s$, on the exact
solutions from a single initial sequence. $A$: the largest difference of the two solutions over the sites and over $s$. $B$: the
largest relative difference of the difference quotient $\Delta_sX^{\rm c}/\Delta s$ and the derivative $\partial_sX^{\rm c}$ of the continuous solution. Neither depends on the
step $\Delta s$. $A$ and $B$ are the largest values over the sites. The bounds are those of
Proposition~\ref{prop:stoch-discrete}, which are proved for the root of the mean square; the loop functional is that of a circle of the radius $L$, $\xi=2$.}
\label{tab:stoch-discrete}
\begin{tabular}{@{}llccccc@{}}
\toprule
quantity & initial sequence & $N$ & $A$ & $N\!A$ & $B$ & $N\!B$ (bound)\\
\midrule
mean spins $\langle\sigma_k\rangle$ & \texttt{+}$\cdots$\texttt{+-}$\cdots$\texttt{-} & $64$ & $5.4\times10^{-3}$ & $0.35$ & $2.3\times10^{-2}$ & $1.47$ ($2$)\\
 & & $1024$ & $3.3\times10^{-4}$ & $0.34$ & $1.5\times10^{-3}$ & $1.50$ ($2$)\\
 & & $4096$ & $8.3\times10^{-5}$ & $0.34$ & $3.7\times10^{-4}$ & $1.50$ ($2$)\\
 & random & $64$ & $1.2\times10^{-2}$ & $0.78$ & $3.1\times10^{-2}$ & $1.96$ ($2$)\\
 & & $1024$ & $7.6\times10^{-4}$ & $0.78$ & $2.0\times10^{-3}$ & $2.00$ ($2$)\\
 & & $4096$ & $1.9\times10^{-4}$ & $0.78$ & $4.9\times10^{-4}$ & $2.00$ ($2$)\\
two spins & \texttt{+}$\cdots$\texttt{+-}$\cdots$\texttt{-} & $16$ & $2.3\times10^{-2}$ & $0.37$ & $1.1\times10^{-1}$ & $1.82$ ($4$)\\
 & & $32$ & $1.1\times10^{-2}$ & $0.35$ & $6.0\times10^{-2}$ & $1.91$ ($4$)\\
 & & $64$ & $5.3\times10^{-3}$ & $0.34$ & $3.1\times10^{-2}$ & $1.95$ ($4$)\\
loop functional & random & $64$ & $3.8\times10^{-2}$ & $2.4$ & & \\
 & & $256$ & $1.0\times10^{-2}$ & $2.6$ & & \\
 & & $1024$ & $2.6\times10^{-3}$ & $2.6$ & & \\
\bottomrule
\end{tabular}
\end{table}

The number $\mu$ measures the probability that a step changes the quantity in question, and the discrete step is valid for a quantity
that a step rarely changes, however short or long the step is in the time $s$: the rates of the decay per step, $-\log(1-\mu)$ in
the chain and $\mu$ in the continuous equation, differ by the fraction $\frac12\mu$. Part
(a) shows where the difference comes from: the continuous time is the chain after $n\pm\sqrt n$ steps.
For the functions of a few spins $\mu\le4k/N$, and the discretization error vanishes in the statistical limit. We verified this on
the exact solutions (Table~\ref{tab:stoch-discrete}). For the mean spins, from a single initial sequence, the largest difference
between the chain and the continuous equation over all the sites and all $s$ is $0.34/N$ for the sequence
\texttt{+}$\cdots$\texttt{+-}$\cdots$\texttt{-} and $0.78/N$ for a random sequence, and the largest relative difference of
$\Delta_s$ and $\partial_s$ in the continuous solution is $1.5/N$ and $2.0/N$, for $N$ from $64$ to $4096$; for the correlation functions of two spins the
two numbers are $0.34/N$ and $1.95/N$ at $N=64$. The difference quotient of the chain differs from the derivative of the
continuous solution at the same $s$ by $0.57/N$ to $0.59/N$ and by $0.62/N$ to $0.65/N$ of the initial rate for the two sequences,
in the root of the mean square over the sites, where the bound is $2/[e(N-4)]\simeq0.74/N$, and by $0.59/N$ for two spins at
$N=64$. The loop functional of a smooth loop is not a function of a few spins, but a step
changes its phase by $O(N^{-1/2})$ only: for a circle of the radius $L$ in the plane of the polygon and $\xi=2$, from a single
random polygon, it changes by $12/N$ to $15/N$ in the first step, and the chain and the continuous equation differ by $2.4/N$, $2.6/N$ and
$2.6/N$ at $N=64$, $256$ and $1024$ (a simulation of $4\times10^5$ and $2\times10^5$ runs, with the continuous time as the Poisson
average of the same runs). Thus in everything that enters the correlation functions the two solutions and their difference quotients agree with each other
and with the derivative $\partial_sW^{\rm c}$ up to $O(1/N)$, and so do the two solutions for the loop functional. This is not so for the
probabilities of the individual sequences in the first steps of a
start from a single sequence, where the modes with $\mu$ of order one are present. On the complete state space for $N=12$ the
relative difference of $\Delta_sW^{\rm c}/\Delta s$ and $\partial_sW^{\rm c}$, summed over the sequences, is $0.22$ at $s=0$, $0.08$ after $8$ steps and $0.011$,
the value $\frac12\Delta_N$ of the slowest mode, after about $100$ steps; the total variation between the two solutions is $0.04$ at
most for the start from \texttt{++++++------}, but $0.58$ for the start from the alternating sequence, in which every bond is
active and the first step is made with certainty in the chain and with the probability $1-1/e$ in the time $\Delta s$ of the continuous equation.
These fast modes do not enter a function of a few spins, and they decay in a few steps. The discretization of $s$
is therefore valid in the statistical limit for the quantities that the theory computes, for every step $\Delta s$.

How fast the chain can approach the equal weights, for a given intensity of the impulses, is settled by the following
proposition and remark.

\begin{proposition}[The relaxation rate and the intensity of the noise]
\label{prop:stoch-intensity}
Let a Markov chain on the balanced spin sequences at a given $\beta$ have symmetric and irreducible transition probabilities
$P_{ab}$. Let $m_k(a)$ be the midpoints of the Euler polygon of the sequence $a$, with the vertex $0$ at a fixed place, let
$\mathcal J_k(a\to b)=2[m_k(b)-m_k(a)]$ be the impulse of the step from $a$ to $b$, and let
\begin{equation}
 \mathcal D(a)=\sum_bP_{ab}\sum_k\bigl|\mathcal J_k(a\to b)\bigr|^2,\qquad
 \mathcal V=\sum_k\Bigl(\bigl\langle|2m_k|^2\bigr\rangle-\bigl|\langle2m_k\rangle\bigr|^2\Bigr),
 \label{eq:stoch-intensity}
\end{equation}
with the averages over the equal weights. Then the gap $\Delta$ of the chain, one minus the eigenvalue of $P$ next to one, obeys
\begin{equation}
 \boxed{\Delta\le\frac{\langle\mathcal D\rangle}{2\,\mathcal V}\simeq\frac{\langle\mathcal D\rangle}{\sum_k|2m_k|^2}\simeq\frac{2\,\langle\mathcal D\rangle}{\xi^2} ,}
 \label{eq:stoch-gap-bound}
\end{equation}
where the last two forms hold in the critical layer, up to a factor $1+O(1/N)$.
\end{proposition}
\begin{proof}
For symmetric transition probabilities the gap is the least value of the ratio of the Dirichlet form
$\frac1{2n}\sum_{a,b}P_{ab}[f(a)-f(b)]^2$ to the variance of $f$ over the real functions $f$ of the state that are not constant.
Apply this to the real and the imaginary parts of the $3N$ components of $2m_k(a)$ and add: the Dirichlet forms add up to
$\frac12\langle\mathcal D\rangle$, and the variances to $\mathcal V$. By \eqref{eq:stoch-zigzag}, $|2m_k|^2=8B^2$ in every state,
and $|\langle2m_k\rangle|^2=4B^2[1+O(1/N)]$: the normal part $2\ii B\,\mathbf e_3$ is the same in all the states, and the planar part,
whose sign is that of the spin, averages to a vector of the relative order $N^{-1/2}$. Hence $2\mathcal V=8NB^2[1-O(1/N)]$, and
$8NB^2=2N\tan^2\frac\delta2\simeq\xi^2/2$.
\end{proof}

The quantity $\mathcal D$ is the intensity of the noise of the model: the mean square of the impulses in one step, summed over the
sides, as in \eqref{eq:stoch-langevin}. The proposition says that the polygon cannot forget its state faster than the noise moves
its midpoints: whatever the symmetric steps, a relaxation in $1/\Delta$ steps requires impulses of the intensity
$\frac12\xi^2\Delta$ per step at least, and a relaxation in the time $T$ the intensity $\xi^2/(2T)$ per unit of $s$, which
displaces the midpoints by their own size in the time $T$.

\begin{remark}[The number of steps and the choice of the moves]
\label{rem:stoch-scaling}
For the chain of Definition~\ref{def:stoch-model}, $\langle\mathcal D\rangle\simeq\xi^2/N$ by \eqref{eq:stoch-langevin}, so the
bound \eqref{eq:stoch-gap-bound} is $2/N$, while the gap is $\Delta_N\simeq4\pi^2/N^3$:
\begin{equation}
 \boxed{\frac1{\Delta_N}\simeq\frac{N^3}{4\pi^2}\ \hbox{steps},\qquad
 \Delta_N\Bigm/\frac{\langle\mathcal D\rangle}{2\,\mathcal V}\simeq\frac{2\pi^2}{N^2} .}
 \label{eq:stoch-rate}
\end{equation}
The exchanges of neighbours fall short of the bound by the factor $2\pi^2/N^2$, because a local step changes a harmonic of the
slow angle with a wavelength of the order of the loop only by diffusion. The bound is attained, up to a factor of order one, by
steps at all distances. Let a pair of sites of the ring be chosen at random in every step, and their spins exchanged; if the spins
are opposite, the arc between them is turned about the normal by the angle $2\delta$ (Sec.~\ref{sec:stoch-move}). These steps are
symmetric as well, so the H-theorem and the equal weights hold as before. The mean spin is multiplied in every step by
$1-\frac2{N-1}$, so all the harmonics of the slow angle relax at the same rate, and the gap is
$\frac2{N-1}$~\cite{CaputoLiggettRichthammer2010}. By \eqref{eq:stoch-arc}, with one half of the pairs opposite and the mean
distance $N/3$ between them,
\begin{equation}
 \langle\mathcal D\rangle\simeq\frac{\xi^2}N\Bigl(1+\frac{\xi^2}6\Bigr),\qquad
 \boxed{\frac1\Delta=\frac{N-1}2\ \hbox{steps} ,}
 \label{eq:stoch-rate-all}
\end{equation}
so the gap is within the factor $1+\xi^2/6$ of the bound \eqref{eq:stoch-gap-bound}. Each step still has an impulse that vanishes
as $N^{-1/2}$ compared with the vertices, $2|\sin\frac\delta2|$ on the two end sides of the arc and $4\sin^2\frac\delta2$ on the
sides inside it, the discrete step is valid in the sense of Proposition~\ref{prop:stoch-discrete}, with $\mu=\frac2{N-1}$ for the mean
spins, and the chain relaxes in about $N/2$ steps in place of $N^3/(4\pi^2)$. In the language of the Langevin equation
\eqref{eq:stoch-EW} these steps are its Fourier acceleration~\cite{BatrouniEtAl1985}, a noise with the kernel
$(-\partial_x^2)^{-1}$, which has the same equilibrium.

Several exchanges can be made in one step, one after another or at once: the exchanges of bonds without a common spin commute,
and so do the turns of arcs about the normal. The chain is again symmetric, with the same H-theorem and the same attractor.
Proposition~\ref{prop:stoch-intensity} limits what this can gain: a relaxation in $n$ steps requires impulses whose mean square in a
step, summed over the sides, is $1/n$ of that of the midpoints at least, whatever the symmetric steps
(Remark~\ref{rem:stoch-step}). We verified \eqref{eq:stoch-gap-bound} on the complete state space for $N\le12$ and $\xi=2$: the ratio of the gap to
the bound is $0.38$, $0.24$, $0.16$ and $0.11$ for the exchanges of neighbours and $0.79$, $0.74$, $0.71$ and $0.69$ for the
exchanges of all pairs at $N=6$, $8$, $10$ and $12$, and for large $N$ it tends to $2\pi^2/N^2$ and to $1/(1+\xi^2/6)$.

The stationary state is the same in all these cases, since Proposition~\ref{prop:stoch-H} does not depend on the transition
probabilities. Which of them the singular evolution realizes is a part of the open problem stated in
Sec.~\ref{sec:stoch-attractor}. We keep the exchanges of neighbours in Definition~\ref{def:stoch-model}, as the steps of the least
impulse. The approach to the equal weights, as $(1-\Delta_N)^{s/\Delta s}\simeq
(\tau_*/\tau)^{1/T}$ with the time $T$ of \eqref{eq:stoch-T}, is a power of the physical time, whose exponent depends on the step.
\end{remark}

\begin{remark}[The step of the time and the relaxation time]
\label{rem:stoch-step}
The step $\Delta s$ does not enter the stationary state, the growth of the entropy or the errors of
Proposition~\ref{prop:stoch-discrete}. It is the unit of the clock of the chain: the relaxation time \eqref{eq:stoch-T} is
proportional to it. The loop functional and the vorticity correlations are averages over the distribution $W$ of states each of
which is an exact solution of the MLE, so this distribution is all that the discretization has to reproduce, and we are free to
choose the step so that the relaxation time is finite in the statistical limit:
\begin{equation}
 \boxed{\Delta s=\frac c{N^3}:\qquad T=\frac c{4\pi^2},\qquad
 \partial_s\theta=\frac1c\,\partial_x^2\theta+\xi\,\sqrt{\frac2c}\;\eta(x,s) ,}
 \label{eq:stoch-step}
\end{equation}
with a constant $c$. The error of the correlation functions of $k$ spins is $2k/N$ at most, as for every step, and it vanishes in
the limit. The Langevin equation \eqref{eq:stoch-EW} of the slow angle no longer contains $N$: the chain has a continuum limit, in
which the harmonic $j$ of the slow angle relaxes at the rate $4\pi^2j^2/c$, and the spins come to the local equilibrium in a time
of the order of $c/N^2$, which vanishes. Only the distance in total variation from the worst initial state still needs a growing time,
$c\log N/(8\pi^2)$~\cite{Lacoin2017}. We checked the limit on the exact solution: for $\Delta s=c/N^3$ and the initial sequence
\texttt{+}$\cdots$\texttt{+-}$\cdots$\texttt{-} the mean spins of the chain differ from the solution of the diffusion equation
$\partial_sm=\frac1c\partial_x^2m$ with the same initial profile by $2.9/N^2$ at $s=0.01\,c$ and by $1.4/N^2$ at $s=0.05\,c$, for
$N$ from $256$ to $4096$.

The step need not contain one exchange. With $M$ bonds chosen at random in a step, one after another, $W(s+\Delta s)=W(s)\,P^M$,
a mode with the eigenvalue $1-\mu$ of $P$ is multiplied in a step by $(1-\mu)^M$, the error of the functions of $k$ spins is
$2kM/N$ at most, and $T=\Delta s\,N^3/(4\pi^2M)$. The process \eqref{eq:stoch-step} is obtained with $\Delta s=Mc/N^3$, and its
error is $2kM/N=2k\,N^2\Delta s/c$: the step has to be small compared with $c/N^2$, and one exchange in a step is the shortest
one. For the exchanges at all distances of Remark~\ref{rem:stoch-scaling}, with one random pair of sites in a step, the error is
again $2k/N$ at most and $T=\Delta s\,(N-1)/2$, so
\begin{equation}
 \Delta s=\frac cN:\qquad T=\frac c2 .
 \label{eq:stoch-step-all}
\end{equation}

What the step does change is the equation \eqref{eq:stoch-residual-def} for a single polygon. Per unit of $s$ the intensity of the
impulses of the exchanges of neighbours, compared with the midpoints, is
\begin{equation}
 \frac1{\Delta s}\,\Bigl\langle\sum_k|\mathcal J_k|^2\Bigr\rangle\Bigm/\sum_k|2m_k|^2\simeq\frac2{N\,\Delta s} ,
 \label{eq:stoch-step-residual}
\end{equation}
which is $2N^2/c$ for the step \eqref{eq:stoch-step}, and the residual of the equation divided by the step,
$\mathcal J_k/\Delta s$, is not small on the sides that a step touches. This is the usual situation of a diffusion with a
noise. One bond out of $N$ is exchanged in a step, so the step \eqref{eq:stoch-step} updates a given site at the rate
$2/(N\Delta s)=2N^2/c=2/(c\,\Delta x^2)$, where $\Delta x=1/N$ is the spacing of the sites on the loop: the ordinary diffusive
scaling of the time with the square of the spacing. The growth of $\mathcal J_k/\Delta s$ is the lattice form of the fact that the
paths of the Langevin equation \eqref{eq:stoch-step}, with its noise white in $x$ and in $s$, are not differentiable in $s$, while
the harmonics of the slow angle and the averages over the paths obey smooth equations. By Proposition~\ref{prop:stoch-intensity} an intensity of
$1/T$ at least is the price of a relaxation in the time $T$, whatever the symmetric steps: the polygon cannot forget its state
without displacing its midpoints by their own size. The exchanges of neighbours exceed this least intensity by the factor
$N^2/(2\pi^2)$, and the exchanges at all distances with the step \eqref{eq:stoch-step-all} by the factor $1+\xi^2/6$ only. If one
required, in addition, that a single polygon obey the discrete MLE in the mean square, with a residual
$2M/(N\Delta s^2)$ of the midpoints not larger than some $\eta$, the relaxation time would be $N^2/(2\pi^2\eta)$ at least, for
the same tolerance $\eta$ of the error of the distribution, and this value would be attained at a step of order one. We do not require this: the positions of the vertices between the Euler
states are not observable, and the statistical averages contain only the distribution $W$. For the loop functional
$\Psi=\sum_aW_a\Psi_a$ the step sets the rate of the change during the relaxation, $\partial_s\Psi$, and not its total,
$\Psi(\infty)-\Psi(0)$, which no step changes. For the step \eqref{eq:stoch-step} this rate is of order one, after a first stage of
the duration of the order of $c/N^2$ in which a single polygon comes to the local equilibrium and its loop functional drops by a finite
amount (Sec.~\ref{sec:stoch-attractor}).

We checked the statements on several exchanges in a step for $M$ up to $N/16$. For the mean spins at $N=1024$, from a random
sequence, the largest relative difference of the difference quotient and the derivative of the continuous solution is $2.0$, $1.98$, $1.90$ and $1.63$ times
$M/N$ for $M=1$, $4$, $16$ and $64$. The mean square impulse of a step is $0.98$ and $0.92$ of $M\xi^2/N$ for $M/N=\frac1{64}$ and
$\frac1{16}$ (exchanges that meet at a vertex partly cancel). For the loop functional of Table~\ref{tab:stoch-discrete} the chain
and the continuous equation differ by $2.1\,M/N$ and $1.9\,M/N$ at $N=1024$ for $M=4$ and $16$.
\end{remark}

\subsection{The turning angle: symmetric reflected shifts}
\label{sec:stoch-beta}

The spin exchanges do not move $\beta$. The ensemble needs a flat density of $\beta$ across the critical layer
$\beta-\pi=O(N^{-1/2})$ (Sec.~\ref{sec:cond-state}), and at zero winding $\beta$ is a continuous parameter of exact fixed points: the
polygon closes for every $\beta$, and $\partial_\beta G$ is an exact zero mode of the linearized right side of the MLE
(Remark~\ref{rem:rig3d-family}). A noise acting on a line of fixed points makes the coordinate along the line wander, since
nothing restores it; this is the origin of part (ii) of Definition~\ref{def:stoch-model}.

No other source of randomness is needed
for it than the thermal noise that triggers the exchanges (Sec.~\ref{sec:stoch-trigger}). The component of the displacement
$\epsilon\,\eta$ of the vertices along the line, $\sum_k\eta_k\cdot{\rm Re}\,\partial_\beta G_k$, has a symmetric distribution, so
its sign, taken as the sign of the shift, is $+$ or $-$ with equal probabilities. In the statistical limit this sign does not
depend on the link that the same noise selects. The covariance of the component along the line with the change $X_k$ of the length of a link
(Sec.~\ref{sec:stoch-trigger}) is the derivative of this length, $|\tan\frac\delta2|$, with respect to $\beta$: it is
$1/(2\cos^2\frac\delta2)$ in absolute value, the same for all the links, because a displacement along the line changes all of
them by the same amount. For a weak Gaussian noise the component is therefore independent of the differences of the $X_k$, which
decide which link is the shortest, and its sign is independent of the selected link (its correlation with a single link is small
as well: the coefficient is of the order of $1/N$ in a typical state and $N^{-1/2}$ at most, in the two alternating sequences).
For a noise that dominates the links the sign is independent of the selected link too, because the selection is then an even
function of the noise and the component is an odd one. Between these two cases a dependence remains, of the order of $1/N$ in
a typical state. Given the component along the line, the vertices are displaced, to the first order, to the Euler state with a
shifted $\beta$, in which the links are equal again, and the rest of the noise does not depend on the component. The component
thus changes only the ratio of the noise to the links, by a fraction of the order of $1/N$, and with it the frequencies of the
links with opposite and with equal spins at their ends (Sec.~\ref{sec:stoch-trigger}). At $N=64$ the probabilities of the sign
$+$ at these two kinds of links differ by $3\times10^{-4}$ to $4\times10^{-4}$ for $\epsilon/|\tan\frac\delta2|$ from $\frac14$
to $1$, as this argument predicts.

The shift stands on the same discrete symmetric
footing as the exchanges of the spins: in a step of $s$ the angle is shifted by $+\Delta\beta$ or by $-\Delta\beta$
($0<\Delta\beta<|I_N|$), with mirror reflection at both ends $a$ and $b$ of the admissible interval $I_N=[a,b]\subset(0,2\pi)$
($a=\beta_{\min}=2\pi/N$, $b=2\pi-\beta_{\min}$, $|I_N|=b-a$),
\begin{equation}
 \boxed{T_+(\beta)=\begin{cases}\beta+\Delta\beta,&\beta\le b-\Delta\beta,\\ 2b-\Delta\beta-\beta,&\beta>b-\Delta\beta,\end{cases}\quad
 T_-(\beta)=\begin{cases}\beta-\Delta\beta,&\beta\ge a+\Delta\beta,\\ 2a+\Delta\beta-\beta,&\beta<a+\Delta\beta.\end{cases}}
 \label{eq:stoch-Tpm}
\end{equation}
Both ends of such a step are Euler states, so the right side of the discrete equation \eqref{eq:stoch-residual-def} vanishes, and
by Proposition~\ref{prop:stoch-residual}(a) the impulse is
$\mathcal J_k=2[m_k(T_\pm(\beta))-m_k(\beta)]\simeq\pm2\Delta\beta\,\partial_\beta m_k$ on every side, where by
\eqref{eq:stoch-zigzag}, at a fixed $\phi$,
\begin{equation}
 \bigl|\partial_\beta m_k\bigr|^2=\frac1{8\cos^4\frac\delta2}+B^2\bar S_k^2,\qquad\bar S_k=\frac{S_k+S_{k+1}}2 .
 \label{eq:stoch-dbeta}
\end{equation}
In the critical layer $|\partial_\beta m_k|$ is of order one, not less than $1/(2\sqrt2)\simeq0.354$ (between $0.35$ and $0.62$ in our
states with $\xi=2$ and $N$ up to $16384$). The impulse of a shift is therefore of the order of $\Delta\beta$ on every side.
Compared with the vertices it is small for every small step: for $\Delta\beta=N^{-1/2}$ it is $0.245$, $0.061$ and $0.019$ at
$N=64$, $1024$ and $16384$, and vanishes as the impulse \eqref{eq:stoch-residual} of an exchange. But a step of the order of
$N^{-1/2}$ is a finite fraction of the width $\xi/\sqrt N$ of the critical layer, and it changes the size of the zig-zag by a finite
factor: its impulse exceeds the midpoints themselves ($\max_k|\mathcal J_k|/|m_k|\simeq1.3$ for $\Delta\beta=N^{-1/2}$ and
$\xi=2$), and, since the shift acts on all the sides at once, the sum over the sides does not decrease with $N$. On the average
over the spin sequences
\begin{equation}
 \sum_k|\mathcal J_k|^2\simeq\Delta\beta^2\,N\Bigl(\frac12+\frac{\xi^2}{24}\Bigr),
 \label{eq:stoch-beta-impulse}
\end{equation}
which is about $0.6$ for $\Delta\beta=N^{-1/2}$, against $2\xi^2/N$ for an exchange of spins, Eq.~\eqref{eq:stoch-least}. Nothing
requires such a step, because Proposition~\ref{prop:stoch-beta} below holds for every step size. With $\Delta\beta\simeq2\xi/N$ the
impulse \eqref{eq:stoch-beta-impulse} of a shift of the angle is that of one exchange of spins, it is small compared with the
midpoints on every side ($\max_k|\mathcal J_k|/|m_k|=0.19$ and $0.05$ at $N=1024$ and $16384$), and the density of the angle becomes
flat across the critical layer in about $N/(4q)$ steps, long before the spins relax. This is the step of the least distortion that
we recommend; the uniform distribution is the same for all the steps.

\begin{proposition}[Symmetric reflected shifts of the turning angle]
\label{prop:stoch-beta}
Let $\beta\in I_N=[a,b]$ make in every step $\Delta s$, with a probability $0<q\le1$, one of the reflected shifts \eqref{eq:stoch-Tpm} with $0<\Delta\beta<|I_N|$,
$T_+$ and $T_-$ with equal probabilities, so that its density obeys
$\varrho(\beta,s+\Delta s)=(1-q)\,\varrho(\beta,s)+q\,(\mathcal T\varrho)(\beta,s)$ with
$(\mathcal T\varrho)(\beta)=\frac12\bigl[\varrho(T_+(\beta))+\varrho(T_-(\beta))\bigr]$. Then:
(a) The transition operator $\mathcal T$ is self-adjoint on $L^2(I_N)$ and doubly stochastic ($\mathcal T1=1$, $\int_{I_N}\mathcal T\varrho\,\dd\beta=\int_{I_N}\varrho\,\dd\beta$), and for a density with $\int_{I_N}|\varrho\log\varrho|\,\dd\beta<\infty$ the differential entropy $\mathsf S_\beta=-\int_{I_N}\varrho\log\varrho\,\dd\beta$ does not decrease in a step:
\begin{equation}
 \boxed{\mathsf S_\beta(s+\Delta s)-\mathsf S_\beta(s)\;\ge\;0 ,}
 \label{eq:stoch-H-beta}
\end{equation}
with equality only if $\varrho(T_+(\beta))=\varrho(T_-(\beta))$ almost everywhere, both being equal to $\varrho(\beta)$ if $q<1$.

(b) On the Neumann cosine basis $\psi_m(\beta)=\cos\bigl(\pi m(\beta-a)/|I_N|\bigr)$ ($m\ge0$),
$\mathcal T\psi_m=\cos(\pi m\Delta\beta/|I_N|)\,\psi_m$, so the harmonic $m$ of the density is multiplied in every step by
$1-\mu_m$, $\mu_m=q\,[1-\cos(\pi m\Delta\beta/|I_N|)]$. Whenever $\Delta\beta/|I_N|$ is irrational (or if $\Delta\beta$ is drawn
from a continuous density on $(0,\Delta\beta_{\max})$), equality in \eqref{eq:stoch-H-beta} holds only for the uniform density
$\varrho(\beta)=1/|I_N|$, every initial density $\varrho(\cdot,0)\in L^1(I_N)$ converges to $1/|I_N|$ as $s\to\infty$, and along every
trajectory starting from any $\beta_0\in I_N$ the fraction of time spent in each subinterval of $I_N$ converges to its normalized
length.

(c) The discrete step is valid for the smooth densities: Proposition~\ref{prop:stoch-discrete}(b) holds for every harmonic $m$ with
$\mu_m<1$, where $\mu_m\simeq\frac12q\,(\pi m\Delta\beta/|I_N|)^2$ for the low harmonics. These are all the harmonics if
$q<\frac12$, and if $q=\frac12$ and $\Delta\beta/|I_N|$ is irrational. For $q>\frac12$ there are harmonics with $\mu_m\ge1$; they
decay as well, since $|1-\mu_m|<1$, but the equation with a continuous $s$ does not describe them.
\end{proposition}
\begin{proof}
(a) On $[a,b-\Delta\beta]$ the map $T_+$ is a translation onto $[a+\Delta\beta,b]$ with inverse $T_-$ and Jacobian $1$; on the upper
collar $(b-\Delta\beta,b]$ it is an involution, $T_+(T_+(\beta))=\beta$, with $|\dd T_+/\dd\beta|=1$; and $T_-$ acts symmetrically on
$[a,a+\Delta\beta)$. Hence $\int_{I_N}f\,(\mathcal Tg)\,\dd\beta=\int_{I_N}(\mathcal Tf)\,g\,\dd\beta$ for all $f,g\in L^2(I_N)$, so
$\mathcal T$ is self-adjoint and preserves mass and the constant function. The density after a step is a mean of
$\varrho(\beta)$, $\varrho(T_+(\beta))$ and $\varrho(T_-(\beta))$ with the weights $1-q$, $\frac q2$, $\frac q2$; the function
$-x\log x$ is strictly concave, so $-\varrho\log\varrho$ at the time $s+\Delta s$ is not less than the same mean of its values, and the integral of
this mean is $\mathsf S_\beta(s)$, because $\mathcal T$ preserves the integral. Since $-x\log x\le1/e$, the entropy after the step
is finite as well.
(b) Extending $\psi_m$ evenly across $a$ and $b$ to a $2|I_N|$-periodic function $\tilde\psi_m$ gives
$\psi_m(T_\pm(\beta))=\tilde\psi_m(\beta\pm\Delta\beta)$, whence
$(\mathcal T\psi_m)(\beta)=\cos(\pi m\Delta\beta/|I_N|)\,\psi_m(\beta)$, and each mode $m\ge1$ is multiplied by $1-\mu_m$ in every step. When
$\Delta\beta/|I_N|\notin\mathbb Q$, $|\cos(\pi m\Delta\beta/|I_N|)|<1$ for every $m\ge1$, so $\varrho(\cdot,s)\to1/|I_N|$ in
$L^2(I_N)$ and by density in $L^1(I_N)$, while for any single trajectory the empirical averages of every $\psi_m$ ($m\ge1$)
converge to zero almost surely (their mean square over $M$ steps is $O(1/M)$, because on the circle of the length $2|I_N|$ the
walk is a translation by $\Delta\beta$ times a lazy random walk on the integers), which by Weyl's criterion gives the uniform
distribution on $I_N$. (c) is \eqref{eq:stoch-discrete} for the mode $m$.
\end{proof}

Note that mirror reflection \eqref{eq:stoch-Tpm} is essential: naive clamping at $a$ and $b$ is not invertible on the boundary collars
and piles weight onto the endpoints, whereas \eqref{eq:stoch-Tpm} is a measure-preserving bijection on each collar and keeps the
uniform distribution exact at every step size $\Delta\beta$ (symmetry and cosine eigenvalues verified numerically to $10^{-14}$).
For a rational step $\Delta\beta=|I_N|/M$ the walk does not fill the interval: it stays on a finite orbit, of $M$ points if it
starts at the centre of a cell of the lattice $a+j\Delta\beta$ and of $2M$ points otherwise (a start on this lattice itself is an
exception, with one half of the weight at each of the two ends of $I_N$). On the orbit the transition matrix is symmetric, so the
equal weights of its points are stationary. They are the attractor on the orbit of $M$ points, and for $q<1$ on that of $2M$
points, and they tend to the uniform density as $M\to\infty$.
The two moves of a step are independent, so the transition operator of the chain is the product of that of the spins and that of
the angle, its modes are the products of their modes, and the joint distribution $W(\boldsymbol\sigma,\beta,s)$ converges to the
uniform one.

\subsection{The attractor: the random spins of the pendulum}
\label{sec:stoch-attractor}

Let us collect the results. In the chain of Definition~\ref{def:stoch-model} the spins and the turning angle change in every
step of $s$ by independent symmetric moves, and Propositions~\ref{prop:stoch-H}, \ref{prop:stoch-exchange}
and~\ref{prop:stoch-beta} apply without approximation. In the limit $s\to\infty$ the probability
distribution of the state tends to the uniform one,
\begin{equation}
 \boxed{W(\boldsymbol\sigma,\beta,s)\,\dd\beta\;\to\;\binom N{N/2}^{-1}\,\frac{\dd\beta}{|I_N|} ,}
 \label{eq:stoch-attractor}
\end{equation}
where $|I_N|$ is the length of the interval of the admissible angles,
from every initial distribution of the spins and of the angle, a single polygon included: equal weights of the balanced spin sequences
and a flat density of the turning angle in the critical layer. The time averages along a trajectory tend to the averages over this
distribution. The plane of the polygon is not moved by the exchanges or by the shifts of the angle; its isotropic distribution in \eqref{eq:finite-statistical-trace} is the symmetry
of the equation, and symmetric random rotations on $O(3)$ produce it by the same H-theorem. How the orientation is randomized, and
how fast, is the subject of Sec.~\ref{sec:stoch-rotation}.

\paragraph{The random spin system of the pendulum.} This is the random spin system that the rest of the paper solves. The equal weights
of the sequences are the projector $\hat P_0$ of the fermionic trace \eqref{eq:fermion-qsp-trace}. The angles of the ensemble are the
rational ones, counted with Euler's totient \eqref{eq:grand-Z0}, and here $\beta$ takes the values that the shifts reach in $I_N$; the two agree where it matters, since
at zero winding every $\beta$ gives a fixed point (Remark~\ref{rem:rig3d-family}), and in the limit the density of the rational angles
near $\pi$ is uniform (Sec.~\ref{sec:ev-arith}). With the uniform spins the transfer operators of the zig-zag
\eqref{eq:fermion-alternating} multiply into the Euclidean evolution of the two pendulums \eqref{eq:3d-pendulum}; with the flat density
of $\xi$ and the return of the slow angle the measure is flat in $\xi^2$, Eq.~\eqref{eq:cond-state}; and the integrals over $\xi$ give
the energy denominators \eqref{eq:fermion-two-arc} of the source and of the partition function (Sec.~\ref{sec:3d-even}). Nothing else
about the state is used.

\paragraph{The thermal circle as a stationary state.} In the stationary state $S_{xN}/\sqrt N$ is a Brownian bridge, and the slow angle
is the Brownian bridge of the total time $\xi^2$ of Eq.~\eqref{eq:cond-survival}, with the weight $e^{-\mathcal S_0}$,
$\mathcal S_0[\theta]=\frac12\int_0^{\xi^2}\dot\theta^2\dd\tau$, $\tau=\xi^2x$. This is the kinetic term of the pendulum
\eqref{eq:3d-pendulum}, and \eqref{eq:stoch-EW} is its Langevin equation,
$\partial_s\theta=-\Gamma\,\delta\mathcal S_0/\delta\theta+\sqrt{2\Gamma}\,\eta$ with $\Gamma=\xi^2/(N^3\Delta s)$, which is $\xi^2/c$ for the step \eqref{eq:stoch-step}. In the language of
the stochastic quantization of Parisi and Wu~\cite{ParisiWu1981}, the logarithmic time $s$ of the decay is the fictitious time, and the
Euclidean path measure of the free rotor on the thermal circle is the equilibrium that the random exchanges reach. The potential of the
pendulum is not a part of the state: it is produced by the hairpin, as the average of its phase over the fast spins
(Sec.~\ref{sec:cond-state}). The pendulum is in this sense the attractor of the stochastic motion.

\paragraph{The loop functional.} The loop equation is linear, and each Euler state solves it, so a mixture of the Euler states with
constant weights is a solution. With weights that depend on the time, $\Psi=\sum_aW_a(s)\Psi_a$, the equation has the residual
$\sum_a\partial_sW_a\Psi_a$, the time derivative of the loop functional itself; in the chain it is
$\Delta_s\Psi=\sum_a\Delta_sW_a\Psi_a$ per step. For a smooth loop it is of the order of $1/N$ in a step ($12/N$ to $15/N$ in the first step
from a single polygon, for a circle of the radius $L$ and $\xi=2$, where $|\Psi|$ falls from $1$ to $0.47$ in $N/11$ steps and to
$0.23$ in $6N/11$ steps at $N=1024$, before the slow relaxation of the slow angle), it agrees with the derivative in the equation with a continuous
$s$ up to $O(1/N)$ (Table~\ref{tab:stoch-discrete}), and it decays as $(1-\Delta_N)^{s/\Delta s}\simeq(\tau_*/\tau)^{1/T}$, a power of
the physical time; summed over the steps it is $\Psi(\infty)-\Psi(0)$, whatever the step. In the stationary state the residual vanishes: the loop functional of the chain, the average of $\Phi_P[C]$ over
its realizations at a given time, is an exact solution of the loop equation, although no single realization, with its moves, is an
exact solution of the MLE (Propositions~\ref{prop:stoch-rigid} and~\ref{prop:stoch-residual}). A process that starts
from one polygon starts from a loop functional of unit modulus, a deterministic flow in the sense of
Supplement~\ref{sec:qsp-stochasticity}, and ends in the Euler ensemble, with $|\Psi|<1$. The model does not replace the loop equation
during the relaxation; it selects the weights of the solution that remains.

\paragraph{What is assumed and what is proved.} The singularity of the time derivative term (Proposition~\ref{prop:stoch-singular}),
the limit \eqref{eq:stoch-u}, the instability of the trajectories, the absence of solutions that move along the fixed points
(Proposition~\ref{prop:stoch-rigid}), the impulse of a step and the least impulse of the exchange of neighbours
(Propositions~\ref{prop:stoch-residual} and~\ref{prop:stoch-least}), the fixed-point bridge and the bound for the continuous paths
(Remarks~\ref{rem:stoch-bridge} and~\ref{rem:stoch-chord}) are properties of the MLE itself. The exchange formula
\eqref{eq:stoch-exchange}, the H-theorems and the convergence for both the spins and the reflected shifts of $\beta$
(Propositions~\ref{prop:stoch-H}, \ref{prop:stoch-exchange} and~\ref{prop:stoch-beta}), the classification of the local
probabilities of the exchange (Proposition~\ref{prop:stoch-ising}), the validity of the discrete step
(Proposition~\ref{prop:stoch-discrete}) and the bound of the relaxation rate by the intensity of the impulses
(Proposition~\ref{prop:stoch-intensity}) are exact statements about the model, directly suitable for formal verification in Lean~4.
Three things are assumed in the model. The first is that the undetermined motion is completed by symmetric local exchanges inside the Euler
ensemble whose probabilities do not distinguish \texttt{++--} from \texttt{+-+-}, one in a step $\Delta s$ of the logarithmic time $s$.
Their randomness is that of the thermal noise, which is present in every fluid and acts here as a trigger: it selects the side
with the smallest midpoint, where an exchange costs the least impulse (Sec.~\ref{sec:stoch-trigger}); what is assumed is this rule.
The step is a free parameter of the model, the unit of its clock: it enters neither the attractor nor the errors of the
probability distribution (Remark~\ref{rem:stoch-step}). Among the steps between different Euler states these exchanges have the least impulse; the right side of the MLE
vanishes at both ends of a step, and the impulse is confined to two sides, where it is $O(N^{-1/2})$ relative to the vertices, with a mean
square over the sides of $2/N$ of that of the midpoints; the final state does not depend on the probabilities of the moves. The
second is the independent symmetric reflected shift \eqref{eq:stoch-Tpm} of $\beta$ on $I_N$ with an incommensurate (or continuously
distributed) step $\Delta\beta$. The third concerns the two limits. With exchanges of
neighbours the number of steps of the relaxation grows as $N^3$, and as $N$ if the exchanges act at all distances along the loop
(Remark~\ref{rem:stoch-scaling}). Either the limit $s\to\infty$ is taken before $N\to\infty$, or the step decreases as $c/N^3$ (as
$c/N$), and then the relaxation time of the correlation functions is finite, with the same errors of the distribution; the
impulses per unit of $s$ are then not small, which concerns the single polygon and not the averages
(Remark~\ref{rem:stoch-step}). A derivation of the moves and of
their probabilities from the singular evolution \eqref{eq:stoch-dae}--\eqref{eq:stoch-K} is an open problem; so is the use of the model beyond
the equal-time ensemble, for the correlations at different times, which it also defines. The programs and the numerical checks of
Secs.~\ref{sec:stoch-mechanism} and~\ref{sec:stoch} are in the directory \texttt{stoch/} of the archive of programs
(Supplement~\ref{sec:S-code}); those of the chain, of the thermal trigger and of the global rotation are among the ancillary files.

\subsection{Quenched or annealed global rotation}
\label{sec:stoch-rotation}

The average \eqref{eq:finite-statistical-trace} over the ensemble contains, besides the equal weights of the states, the Haar measure
of their global orientation in $O(3)$. The chain of Definition~\ref{def:stoch-model} produces the equal weights. How the orientation
is randomized, and how fast, decides its place in an average that is a ratio, as the conditional correlation
\eqref{eq:cond-postulate} is. A variable that reaches its equilibrium together with the spins is annealed: it is averaged in the
numerator and in the normalization. A variable that stands still while the spins relax is quenched: the ratio is taken at its fixed
value and averaged afterwards. In a linear average, such as the loop functional, the two orders coincide. A rotation of the polygon
consists of a rotation in its plane and of a tilt of the plane, and the two behave differently.

\paragraph{The rotation in the plane.} The exchanges themselves produce it. An exchange of the spins $\sigma_k=-\sigma_{k+1}$ turns
the vertex $k+1$ by the angle $-2\delta\sigma_k$ (Sec.~\ref{sec:stoch-move}), so the mean slow angle
$\Theta=N^{-1}\sum_k\theta_k$ changes by $-2\delta\sigma_k/N$. On a ring the bonds \texttt{+-} and \texttt{-+} are equally
numerous, so $\Theta$ has no drift, and in the stationary state, where a bond joins opposite spins with the probability
$N/[2(N-1)]$,
\begin{equation}
 \bigl\langle(\Theta_n-\Theta_0)^2\bigr\rangle=\frac{2\delta^2n}{N(N-1)}\simeq\frac{2\xi^2n}{N^3}
 \label{eq:stoch-inplane}
\end{equation}
after $n$ steps, exactly. At the relaxation time $n_{\rm rel}=N^3/(4\pi^2)$ of the slowest harmonic of the spins,
Eq.~\eqref{eq:stoch-gap}, this is $\xi^2/(2\pi^2)$. The rotation in the plane is thus randomized by the exchanges, in the same
time as the spins and at a rate that does not depend on the noise. It is the zero mode of the slow angle of the pendulum, and it
is annealed: in the Gibbs state \eqref{eq:cond-C} it is inside the trace.

\paragraph{The plane.} Neither move of the model tilts the plane: an exchange reflects a vertex within it, and a shift of $\beta$
changes only $R$, $B$ and the slow angles. The chain is therefore ergodic at a fixed plane, the attractor
\eqref{eq:stoch-attractor} is reached in every plane separately, and the distribution of the planes remains that of the initial
ensemble, which is isotropic for an isotropic turbulence. In the model the plane is quenched.

\paragraph{How the thermal noise tilts the plane.} The noise that triggers the exchanges also contains a rigid rotation of the
polygon. A displacement $\epsilon\,\eta_k$ of the vertices (Sec.~\ref{sec:stoch-trigger}) is closest, in the mean square, to the
rotation $\omega\times f_k$ with
\begin{equation}
 \omega=\epsilon\,I^{-1}\sum_kf_k\times\eta_k,\qquad I=\sum_k\bigl(|f_k|^2-f_k\otimes f_k\bigr),\qquad
 \langle\omega_a\omega_b\rangle=\epsilon^2\bigl(I^{-1}\bigr)_{ab},
 \label{eq:stoch-omega}
\end{equation}
where $I$ is the tensor of inertia of the vertices. Its principal values are $NR^2$ for the normal and $NR^2w$ and $NR^2(1-w)$ for
two axes in the plane, where $w=\frac12\bigl(1-|N^{-1}\sum_ke^{2\ii\theta_k}|\bigr)$ measures the spread of the slow angles; in
the uniform ensemble $\langle w\rangle=0.08$, $0.24$ and $0.38$ at $\xi=1$, $2$ and $4$, and $\xi^2/12$ for small $\xi$. Every
displacement thus tilts the plane, by an angle of the order of $\epsilon/\sqrt N$. On a manifold of fixed points nothing restores
the polygon. If the rotations of the successive independent displacements add up, the polygon performs a rotational Brownian
motion, and its normal $\mathbf n$ forgets its direction at the mean rate $1/n_{\rm tilt}$,
\begin{equation}
 \langle\mathbf n_0\cdot\mathbf n_j\rangle\simeq e^{-j/n_{\rm tilt}},\qquad
 n_{\rm tilt}=\frac{2NR^2}{a\,\epsilon^2}\simeq\frac N{2a\,\epsilon^2},\qquad a=\Bigl\langle\frac1{w(1-w)}\Bigr\rangle,
 \label{eq:stoch-tilt}
\end{equation}
where $j$ counts the displacements; $a=18.7$, $6.5$ and $4.3$ at $\xi=1$, $2$ and $4$, and $a\simeq16.8/\xi^2$ for small $\xi$.
A simulation of this motion, with the chain of the exchanges running at the same time, agrees with \eqref{eq:stoch-tilt}; when
the tilt is faster than the exchanges, each sequence keeps its own $w$, and the decay at $j>n_{\rm tilt}$ is somewhat slower.
The tilt is a motion of another kind than the exchanges. An exchange is triggered, and its rate does not depend on the amplitude of
the noise; the tilt is an ordinary diffusion along a symmetry of the equation, and its rate is proportional to $\epsilon^2$. The
same response moves the turning angle along its line of fixed points, by one unit of $\xi$ in about $n_{\rm tilt}/5$ displacements
at $\xi=2$; in the model the step of the angle is a free parameter (Sec.~\ref{sec:stoch-beta}).

\paragraph{The criterion.} With one displacement in a step of the chain, the ratio of the two times is
\begin{equation}
 \Lambda=\frac{n_{\rm rel}}{n_{\rm tilt}}\simeq\frac a{2\pi^2}\,N^2\epsilon^2=\frac{2a}{\pi^2}\,\frac{L^2}{|C|^2}\,N\,m|C| ,
 \label{eq:stoch-Lambda}
\end{equation}
where the last form is for the thermal noise of the perimeter law $e^{-m|C|}$, with $\epsilon^2=4L^2m/(N|C|)$
(Sec.~\ref{sec:stoch-trigger}). If $\Lambda\ll1$ the plane stands still while the spins relax, and it is quenched; if
$\Lambda\gg1$ it is randomized first, and it is annealed. The border is at a displacement of the vertices of the order of $1/N$,
far below the links, which are of the order of $N^{-1/2}$: at $\xi=2$, $\Lambda\simeq0.33\,N\epsilon^2/\tan^2\frac\delta2$. Both
cases occur. In the limit of a vanishing noise at a fixed $N$, the limit that defines spontaneous stochasticity, $\Lambda\to0$:
the exchanges go on at their rate, which the noise does not set, and the plane is frozen. For a noise in a fixed ratio to the
links $\Lambda$ grows as $N$, and in the statistical limit the plane would be randomized before the spins. In terms of the loop
functional, the plane is quenched if the thermal noise changes the loop functional of a loop of the size $L$ by less than about
$1/N$. That the rotations add up is an assumption about the response of the singular equation to the part of the noise that does
not trigger an exchange. In the model as it stands they do not add up, and the plane is conserved.

\paragraph{What depends on the order.} In the conditional correlation the quenched order is that of
Definition~\ref{post:cond}: the ratio $C=\widehat G/\widehat Z$ is taken at a fixed plane, that is, at a fixed projected
separation $h$, and averaged over the planes with the isotropic weight. The annealed order is the ratio at a fixed separation,
Eq.~\eqref{eq:ord-PA}, in which the planes are weighted by the partition function
$\widehat Z(h)$. We have computed the spectrum in both orders (Sec.~\ref{sec:3d-orders}, Fig.~\ref{fig:ev-spectrum}). For the even ensemble at zero winding they have in common the value $C(0)=\frac18$, the ultraviolet law
$E\propto k^{-7/2}$ (its coefficient is multiplied by $0.8727$ in the annealed order), a positive spectrum at every computed wave
number, and the threshold at $kL=\sqrt2$ with its wave: the oscillating term belongs to the numerator, and the law
\eqref{eq:ev-sqrt2} holds in the annealed order with $a_2$ and $\kappa_2$ multiplied by $2/\pi$
and $\bar\kappa$ replaced by $\bar\kappa-\log2$, as the computed slopes confirm. They differ in the infrared. The vorticity correlation decays as $\log r/r^2$ in the quenched order and as
$1/(r\log r)$ in the annealed one, so the spectrum at small wave numbers is $E\propto k^{-1}\log(1/kL)$ in the first and
$E\propto k^{-2}/\log(1/kL)$ in the second, and the energy at a fixed small wave number decays as $(t+t_0)^{-1}\log(1/kL)$ and as
$(t+t_0)^{-3/2}/\log(1/kL)$. The difference is observable in the finite-size effects (Sec.~\ref{sec:3d-size}): the energy of a
flow of a finite size and its dissipation depend on this size logarithmically in the quenched order and almost linearly in the
annealed one, and the energy decays with an index of $1.1$ to $1.2$ in the first and of $1.44$ to $1.48$ in the second.

The formulas of the following sections are written in the order of Definition~\ref{post:cond}, and the annealed order is given
beside it in Secs.~\ref{sec:3d-orders}, \ref{sec:3d-size} and~\ref{sec:osc-orders}. The quenched order is the order of the model, in which the plane is conserved, and of the limit
of a vanishing noise; and it does not depend on the times at all if the conditioning is read as in Sec.~\ref{sec:cond-justify}:
an Euler state is planar, so the walk of its slow angle around the loop is killed at a rate that its plane fixes. Which order the
thermal noise of a real fluid selects is an open question. The ultraviolet law and the threshold do not depend on the answer; the
infrared law and the finite-size effects do, and measurements can decide it.
% ===== Secs. 6 and 7 (the former Sec. 6, split on 2026-11-14, revision (o)); statements and equations of the former Sec. 6 revised 2026-11-13 (h) =====

\section{The vorticity correlation}
\label{sec:3d-hairpin}

The energy spectrum is measured by the two-point correlation of vorticity. In the loop language this correlation is measured by a
hairpin (Fig.~\ref{fig:hairpin}): two area derivatives of the loop functional are taken at the points $\boldsymbol r_1$ and
$\boldsymbol r_2$, and the loop is then contracted to a thin wire that runs from $\boldsymbol r_1$ to $\boldsymbol r_2$ and
back~\cite{migdal2024quantum,ReviewPaperAM}. We first state how the correlation is defined: as a conditional expectation, by Bayes' formula. The rest of
the section is a computation: the correlation of the Euler ensemble, exactly at finite $N$ and then in the limit $N\to\infty$. The probabilistic meaning of this definition is given in Sec.~\ref{sec:cond-justify}, after the hairpin has been reduced to a Gibbs state of two quantum pendulums.

\subsection{The definition: a conditional expectation}
\label{sec:cond-def}

The hairpin is the instrument that measures the correlation, and it has an amplitude of its own, the loop functional of the same
hairpin without the two vorticities,
\begin{equation}
 \Psi[\hbox{hairpin}]=\bigl\langle e^{\ii\Gamma/\nu}\bigr\rangle ,
 \label{eq:cond-bare-amplitude}
\end{equation}
where $\Gamma$ is the circulation around the hairpin. For a single smooth flow at a fixed viscosity a hairpin of zero width has $\Gamma=0$,
and its amplitude is one. In decaying turbulence it is not. The vorticity is singular on the viscous scale, the limits of zero width and of
zero viscosity do not commute, and the narrow hairpin of the Euler ensemble keeps a phase of order one
(Supplement~\ref{sec:hairpin-not-one}). Our earlier work~\cite{migdal2024quantum} set this amplitude to one. Here we divide by it.

\begin{definition}[The conditional correlation]
\label{post:cond}
The two-point correlation of vorticity is the expectation of the product of the two vorticities conditioned on the hairpin that
measures them,
\begin{equation}
 \boxed{\bigl\langle\boldsymbol\omega(\boldsymbol r_1)\cdot\boldsymbol\omega(\boldsymbol r_2)\bigr\rangle
 =\frac{\bigl\langle\boldsymbol\omega(\boldsymbol r_1)\cdot\boldsymbol\omega(\boldsymbol r_2)\,e^{\ii\Gamma/\nu}\bigr\rangle}
 {\Psi[\hbox{hairpin}]},}
 \label{eq:cond-postulate}
\end{equation}
taken at a fixed separation projected on the plane of the momentum polygon and then averaged over the orientation of the separation with
the isotropic weight. As $N\to\infty$ the hairpin amplitude becomes, in the critical layer $\beta\to\pi$, a positive Gibbs measure of two
quantum pendulums (Sec.~\ref{sec:cond-pendulum}). The normalization $\Psi[\hbox{hairpin}]$ is the partition function $\widehat Z(h)$ of
this measure, and the phases at the two ends of the hairpin belong to the observable.
\end{definition}

This is not a second postulate. The division by the amplitude of the hairpin is the definition of a conditional probability,
Bayes' formula, and the measure that this amplitude defines is not postulated either: the equal-weight sum over the Euler states is
a quantum trace over fermions, which is transformed exactly, in the continuum limit, into the Gibbs trace of two quantum pendulums
(Secs.~\ref{sec:3d-fermion} and~\ref{sec:cond-pendulum}). The only postulate of the statistics remains the equal weight of the Euler
states (Sec.~\ref{sec:3d-ensemble}). The order stated here, the conditioning at a fixed projected separation and then the average
over the orientation, is the quenched one; the annealed order is its alternative (Secs.~\ref{sec:stoch-rotation}
and~\ref{sec:3d-orders}), and which of them a real fluid selects is an open question. The numerator
of~\eqref{eq:cond-postulate} is the second area derivative of Sec.~\ref{sec:contact} with its contact terms, and the ratio is the
conditional average~\eqref{eq:ct-bare}.

\subsection{The vorticity carried by a vertex}

The vorticity of the Euler polygon sits at its vertices. Each vertex carries a null complex vector, and the part of it that
survives in the invariant correlation is normal to the plane of the polygon and imaginary. Let us compute it.
By \eqref{eq:advection-fourier-translation} the vorticity at the jump $k$ is
$\widehat{\boldsymbol\omega}_k=\ii\nu\bar P_k\times\Delta P_k$. With
$P_k=G_k/\sqrt{2\nu\tau}$ from \eqref{eq:3d-complex-euler},
\begin{equation}
 \widehat{\boldsymbol\omega}_k=\frac{\Omega_k}{2\tau},\qquad
 \Omega_k=\ii\,m_k\times\delta_k,\qquad
 m_k=\frac{G_{k+1}+G_k}{2},\quad \delta_k=G_{k+1}-G_k .
 \label{eq:3d-vertex-vorticity-def}
\end{equation}
The midpoint has a real part and the constant imaginary normal,
$m_k=h_k+\ii B\mathbf e_3$. The real midpoint has length $|B|$ and lies along
$\mathbf e_{\gamma_k}=(\cos\gamma_k,\sin\gamma_k,0)$, where
$\gamma_k=\alpha_k+\sigma_k\beta/2$ is the direction of the middle of the edge;
the jump is the unit chord $\delta_k=\sigma_k\mathbf e^\perp_{\gamma_k}$, with
$\mathbf e^\perp_\gamma=(-\sin\gamma,\cos\gamma,0)$. These are the identities
$2R\sin(\beta/2)=1$ and $R\cos(\beta/2)=B$. Therefore
\begin{equation}
 \boxed{\Omega_k=\ii\,h_k\times\delta_k+\ii(\ii B\mathbf e_3)\times\delta_k
 =B\sigma_k\left(\mathbf e_{\gamma_k}+\ii\,\mathbf e_3\right).}
 \label{eq:3d-vertex-vorticity}
\end{equation}
The normal term, $\ii B\sigma_k\mathbf e_3$, is the vorticity of the real representative $f_k$. The planar term, $B\sigma_k\mathbf
e_{\gamma_k}$, comes from the imaginary normal $\ii B\mathbf e_3$, a translation of the momentum loop: it drops out of every closed-loop
phase but not out of the uncentred vorticity (Supplement~\ref{sec:S-cond-translation}). The translation-invariant vorticity of
Sec.~\ref{sec:contact} uses the centred midpoint, $m_k-\langle G\rangle=h_k-\langle f\rangle$, $\langle f\rangle=\frac1N\sum_jf_j$, and
the imaginary normal cancels from it:
\begin{equation}
 \boxed{\Omega^c_k=\ii\,(h_k-\langle f\rangle)\times\delta_k=\ii\,a_k\,\mathbf e_3,\qquad
 a_k=B\sigma_k-\bigl(\langle f\rangle\times\delta_k\bigr)_3\in\mathbb R,\qquad
 \Omega^c_i\cdot\Omega^c_j=-a_ia_j .}
 \label{eq:3d-centred-vorticity}
\end{equation}
The invariant vorticity is normal and imaginary. Its total is $\sum_ka_k=BM$, since $\sum_k\delta_k=0$.

\subsection{The hairpin of the Euler polygon at finite \texorpdfstring{$N$}{N}}
\label{sec:cond-finiteN}

At finite $N$ everything is an exact finite sum: the numerator of the correlation, the amplitude of the hairpin and their ratio.
For the polygon the two area derivatives sit at two vertices $m$ and $n$, the ends of the hairpin. The arc from $m$ to $n$ is mapped onto
the wire on the way out, and the complementary arc on the way back. For endpoints in cyclic order let $\ell=n-m\pmod N$, $1\le\ell<N$, and
define the difference of the two arc averages,
\begin{equation}
 Q_{mn}=\frac1\ell\sum_{j=m}^{n-1}f_j
       -\frac1{N-\ell}\sum_{j=n}^{m+N-1}f_j,
 \qquad
 K_{mn}(\tau)=\frac{|Q_{mn}|}{\sqrt{2\nu\tau}},
 \qquad \tau=t+t_0.
 \label{eq:3d-hairpin-averages}
\end{equation}
The constant imaginary normal cancels from this difference, so $Q_{mn}$ is a real vector in the plane of the polygon, and the phase of the
hairpin is $\exp\bigl(\ii(\boldsymbol r_2-\boldsymbol r_1)\cdot Q_{mn}/\sqrt{2\nu\tau}\bigr)$; the conventions of the endpoints are fixed in
Supplement~\ref{sec:S-hp-closure}. We measure lengths in units of
\begin{equation}
 L=\sqrt{\widetilde\nu\,\tau}=\sqrt{\widetilde\nu\,(t+t_0)},\qquad
 \widetilde\nu=\nu N^2,\qquad K=kL .
 \label{eq:3d-scaling-length}
\end{equation}
Here $\nu$ is the viscosity of the loop equation that the ensemble solves; in the limit $N\to\infty$ at fixed $\widetilde\nu$ it becomes the
vanishing viscosity $\nu_N=\widetilde\nu/N^2$ of an auxiliary flow (Sec.~\ref{sec:3d-renorm}). The phase depends on the separation only
through its projection $\boldsymbol r_\parallel$ on the plane of the polygon, and the average over the rotations in this plane turns it
into $J_0(hN|Q_{mn}|)$, with
\begin{equation}
 h=\frac{|\boldsymbol r_\parallel|}{\sqrt2\,L}=\frac{r\sin\vartheta}{\sqrt2\,L},
 \label{eq:cond-def-h}
\end{equation}
where $\vartheta$ is the angle between $\boldsymbol r_2-\boldsymbol r_1$ and the normal to the plane. The average over the orientation is
then the average over $\vartheta$ with the weight $\sin\vartheta$:
$\operatorname{sinc}(kr)=\int_0^{\pi/2}J_0(kr\sin\vartheta)\sin\vartheta\,\dd\vartheta$.

\paragraph{The numerator.} The two vorticities at the ends with their contact terms give the pair weight $\Phi_{mn}$,
Eq.~\eqref{eq:3d-pair-weight} of Supplement~\ref{sec:S-hp-closure}. At a coincident pair the vanishing arc averages to the midpoint, and
$Q_{kk}=h_k-\langle f\rangle$. The vorticities~\eqref{eq:3d-vertex-vorticity-def} give the factor $1/(4\tau^2)$, and the number of states
cancels. At a fixed projected separation the numerator of~\eqref{eq:cond-postulate} is the source function
\begin{equation}
 W_N(h)=\left\langle\sum_{m<n}\Phi_{mn}\,J_0\!\left(hN|Q_{mn}|\right)-\frac1{2N}\sum_ka_k^2J_0\!\left(hN|Q_{kk}|\right)\right\rangle .
 \label{eq:3d-source-function}
\end{equation}
Averaged over the orientation, it is the transform of a measure $V_N$ on the wave numbers,
\begin{equation}
 \boxed{\begin{gathered}
 \mathcal W_N(r)=\int_0^\infty V_N(K)\operatorname{sinc}\Bigl(\frac{Kr}L\Bigr)\dd K,\\
 V_N(K)=\left\langle\sum_{m<n}\Phi_{mn}\,
 \delta\!\left(K-\frac{N|Q_{mn}|}{\sqrt2}\right)-\frac1{2N}\sum_ka_k^2\,\delta\!\left(K-\frac{N|Q_{kk}|}{\sqrt2}\right)\right\rangle ,
 \end{gathered}}
 \label{eq:3d-energy-spectrum}
\end{equation}
and $W_N(h)=\int_0^\infty V_N(K)J_0(\sqrt2hK)\dd K$; the Hankel inversion, Eq.~\eqref{eq:3d-hankel-inversion}, returns $V_N$. Both forms
are exact at every $N$.

\paragraph{The hairpin amplitude.} The same sum without the two vorticities is the amplitude~\eqref{eq:cond-bare-amplitude} of the
hairpin,
\begin{equation}
 \boxed{D_N(h)=\Bigl\langle\sum_{m<n}J_0\bigl(hN|Q_{mn}|\bigr)\Bigr\rangle,\qquad D_N(0)=\frac{N(N-1)}2.}
 \label{eq:hp-bare}
\end{equation}
At zero separation it is the number of pairs. At $h>0$ it is not: it grows as $N^{3/2}$, since only the fraction $N^{-1/2}$ of the angles
near $\pi$ contributes. The ratio
\begin{equation}
 C_N(h)=\frac{W_N(h)}{D_N(h)}
 \label{eq:cond-def-ratio}
\end{equation}
is exact wherever $D_N(h)>0$. It keeps in its denominator the phases at the two ends of the hairpin, which the definition assigns to the observable; we use it at zero separation, where these phases are one (Sec.~\ref{sec:3d-zero-separation}).

\subsection{The limit \texorpdfstring{$N\to\infty$}{N to infinity}}
\label{sec:cond-limit}

In the limit the correlation is a ratio of two functions of the quantum pendulums, $C(h)=\widehat G(h)/\widehat Z(h)$, in units of
$1/N$. As $N\to\infty$ at a fixed $h>0$, both sums are dominated by the critical layer $\beta\to\pi$, the zig-zag polygons of
Sec.~\ref{sec:3d-fermion}, where the Euler ensemble reduces exactly to two quantum pendulums; the numerical evidence for this dominance
is given in Sec.~\ref{sec:3d-even}. The numerator tends to $\sqrt N\,\widehat G(h)$, Eq.~\eqref{eq:ev-whole}. The hairpin amplitude, with
the phases at its ends removed, tends to $N^{3/2}\widehat Z(h)$, the partition function of the Gibbs state of the two pendulums
(Sec.~\ref{sec:cond-thermal}); its finite-$N$ form is Eq.~\eqref{eq:cond-bulk-mid}. By Definition~\ref{post:cond} the correlation is
$C(h)/N$, with
\begin{equation}
 C(h)=\frac{\widehat G(h)}{\widehat Z(h)},\qquad C(0)=\lim_{h\to0}C(h)=\frac18,
 \label{eq:cond-def-limit}
\end{equation}
the limit $h\to0$ taken after $N\to\infty$. The vorticity correlation and the energy spectrum follow by the same transforms as for the
numerator:
\begin{equation}
 \boxed{\begin{gathered}
 C_\omega(r,t)=\frac1{4\tau^2N}\int_0^{\pi/2}C\Bigl(\frac{r\sin\vartheta}{\sqrt2L}\Bigr)\sin\vartheta\,\dd\vartheta
 =\frac1{4\tau^2N}\int_0^\infty P(K)\operatorname{sinc}\Bigl(\frac{Kr}L\Bigr)\dd K,\\
 E(k,t)=\frac{\widetilde\nu^{\,2}}{8LN}\,\frac{P(kL)}{(kL)^2},\qquad
 P(K)=2K\int_0^\infty h\,J_0(\sqrt2hK)\,C(h)\,\dd h .
 \end{gathered}}
 \label{eq:cond-def-transforms}
\end{equation}
The whole time dependence sits in the prefactors and in the scaling variables $kL$ and $r/L$. The remaining sections compute
$\widehat G$ and $\widehat Z$.

\subsection{Zero separation: the finite-cutoff test}
\label{sec:3d-zero-separation}

The simplest test of an ensemble is the sign of its enstrophy, the correlation at zero separation, and here the test is exact at
every $N$. The even ensemble at zero winding passes it for every state; the odd ensemble fails it at every $N$ computed.

At zero separation the hairpin is a point, its amplitude is the number of pairs, and the correlation is $C_N(0)=W_N(0)/D_N(0)$ at every
$N$. By the identity \eqref{eq:ct-zero}, with \eqref{eq:3d-centred-vorticity}, the numerator is exact for every state:
\begin{equation}
 \boxed{\int_0^\infty V_N(K)\dd K=W_N(0)=\frac12\Bigl\langle2\sum_ka_k^2-\Bigl(\sum_ka_k\Bigr)^2\Bigr\rangle
 =\frac12\Bigl\langle2\sum_ka_k^2-B^2M^2\Bigr\rangle .}
 \label{eq:3d-zero-separation}
\end{equation}
The first term comes from the contact terms, the second is the square of the total vorticity of the polygon.

\paragraph{Even ensemble at zero winding.} Here $M=0$ for every state, the total vorticity vanishes, and
\eqref{eq:3d-zero-separation} gives $W_N(0)=\langle\sum_ka_k^2\rangle>0$. The zero-separation correlation is nonnegative for every state,
not only on average; it vanishes only for the two alternating words of each label, where all $a_k=0$.

\begin{proposition}[The finite-cutoff enstrophy test]
\label{prop:3d-positivity}
With the translation-invariant correlation of Sec.~\ref{sec:contact}, the zero-separation value
\eqref{eq:3d-zero-separation} is nonnegative for every state of the even ensemble at zero winding, and zero only for the two alternating
words of each label. Its average over the odd ensemble is negative at every $N$ computed ($5\le N\le201$, exact).
\end{proposition}

The even ensemble at zero winding passes the test, and it is also the limit of the grand canonical ensemble, which it
dominates~\cite{Zah23} (Sec.~\ref{sec:3d-ensemble}). The odd ensemble fails it, and its limit has no positive spectrum; we do not use it
(Supplement~\ref{sec:S-cond-odd-zero}). In the limit the conditional correlation is finite at zero separation, $C(0)=\frac18$ in units of
$1/N$ (Sec.~\ref{sec:cond-zero}); after the renormalization of Sec.~\ref{sec:3d-renorm} this is
$\langle\boldsymbol\omega\cdot\boldsymbol\omega\rangle=(\nu/\widetilde\nu)^{1/2}/[32(t+t_0)^2]$. It is not the limit of
\eqref{eq:3d-formal-dissipation}, since the limits $h\to0$ and $N\to\infty$ do not commute (Supplement~\ref{sec:S-hp-finiteN} and
Remark~\ref{rem:3d-dissipation}).

\section{The Euler ensemble as a fermionic trace: reduction to the quantum pendulum}
\label{sec:3d-fermion}

This section contains the central derivation of the paper. The equal-weight sum over the Euler states is a quantum trace over
fermions, and we transform this trace, exactly at every $N$ and then by one asymptotic step, into the Euclidean evolution of two
quantum pendulums. So, the Gibbs measure of the pendulums is derived from the equal weights; it is not postulated.

Which Euler polygons produce the spectrum at large $N$, and what is the dynamics
of their slow variables? Let us derive the answer directly from the Quantum
Statistical Principle (Sec.~\ref{sec:qsp}). In Appendix~B of Ref.~\cite{migdal2024quantum} (see also the
Supplementary Material of Ref.~\cite{NTDT2}) the Euler ensemble was written as a
quantum trace over fermions on a ring. We take this trace, add the trace over
the arithmetic states $(p,q,r)$, and follow it to the continuum. No saddle point
is used. The pendulum Hamiltonian, the factors at the endpoints of the hairpin and
the vorticity insertions come out of one exact operator identity and one asymptotic step, the continuum limit of the paired transfer operator. In Ref.~\cite{migdal2024quantum} the same trace was
evaluated by a complex saddle point, an instanton, of the path integral over the density of the fermions, which is a WKB
approximation. The reduction derived here supersedes it: in the local limit the trace is dominated by the zig-zag
polygons near $\beta=\pi$, which the pendulums describe, and the saddle point of the previous work is not the dominant
contribution (Remark~\ref{rem:S-small-angles} in Supplement~\ref{sec:S-hairpin}). We write the formulas for the even ensemble at zero winding; the odd ensemble differs only in the closure,
and its formulas, together with the numerical tests of this section, are in Supplement~\ref{sec:S-fermion}.

\subsection{The trace over all Euler states}

Put a fermion on each step of the polygon, $k=1,\dots,N$, with occupation numbers
$\hat\nu_k=a_k^\dagger a_k$ and spins $\hat\sigma_k=2\hat\nu_k-1$. The occupation
numbers commute with each other, and the Fock space $\mathcal F$, of dimension $2^N$,
has the basis $|\boldsymbol\sigma\rangle$ of spin sequences. The vertex angles
$\hat\alpha_l=\phi+\beta\sum_{k<l}\hat\sigma_k$ become operators; since $\hat\nu_k^2=\hat\nu_k$,
\begin{equation}
 e^{\ii\hat\alpha_l}=e^{\ii\phi}\,e^{-\ii\beta(l-1)}
 \prod_{k<l}\left[1+\left(e^{2\ii\beta}-1\right)\hat\nu_k\right].
 \label{eq:fermion-angle}
\end{equation}
Let us stress that this is a product of arbitrarily many occupation numbers.
The trace below is exact, but it is not the Gaussian trace of free fermions.

The Quantum Statistical Principle gives equal weight to every admissible state
$|p,q,r\rangle\otimes|\boldsymbol\sigma\rangle$ with $\sum_k\sigma_k=qr$. In other words, the
ensemble average is the normalized trace of the projector onto the admissible
states (the normalized projector is the density matrix of \eqref{eq:qsp}),
\begin{equation}
 \langle\mathcal O\rangle_N=\frac{\operatorname{Tr}\bigl[\hat\Pi_N\,\hat{\mathcal O}\bigr]}{\operatorname{Tr}\hat\Pi_N},\qquad
 \hat\Pi_N=\sum_{(p,q,r)}|p,q,r\rangle\langle p,q,r|\otimes\hat P_{qr},\qquad
 \hat P_M=\oint\frac{\dd\omega}{2\pi}\,e^{\ii\omega(\sum_k\hat\sigma_k-M)},
 \label{eq:fermion-qsp-trace}
\end{equation}
with the average over the global orientation understood.

For the even ensemble at zero winding the winding is $r=0$, the projector is $\hat P_0$, and the rational angles run over
$3\le q<N$, so that $\operatorname{Tr}\hat\Pi_N=Z^{(0)}_N$ of \eqref{eq:grand-Z0}; the Markov product (Supplement~\ref{sec:S-fermion-main}) and the transfer operator below
are the same for every sector, and only the closure differs.

Appendix~B of Ref.~\cite{migdal2024quantum} wrote the trace with the Markov
operators of Ref.~\cite{migdal2023exact}, which reveal the spins one by one (Supplement~\ref{sec:S-fermion-main}).
Each sector of the Markov trace is normalized to one; the Quantum Statistical
Principle gives the same weight to every state, so each sector enters with its
number of states $d_N(q,r)=\binom N{(N+qr)/2}$. This is the counting rule of the Euler ensemble
(Supplement~\ref{sec:S-3d-odd}).

\subsection{The transfer operator at a rational angle}

The hairpin phase depends on the spins only through the running spin
$S_l=\sum_{k<l}\sigma_k$, and at $\beta=2\pi p/q$ only through $j=S_l\bmod q$. This cumulative
structure replaces the many-fermion products \eqref{eq:fermion-angle} by a walk on
$\mathbb Z_q$. On an arc where the source is constant, say $b$, a step is the operator
\begin{equation}
 (T_b\psi)_j=\frac{e^{\ii b\cos(\phi+\beta j)}}{2}\bigl(\psi_{j+1}+\psi_{j-1}\bigr),
 \qquad \psi_{j+q}=z\,\psi_j ,
 \label{eq:fermion-transfer}
\end{equation}
where the twist $z$ counts the windings: the coefficient of $z^r$ fixes
$\sum_k\sigma_k=qr$. For the hairpin, $b_1=hNR/n$ on the first arc and $b_2=-hNR/(N-n)$ on
the second. The conditional two-spin characteristic function of one sector, with the steps that carry the vorticity at the two
endpoints, is then exactly a trace over $\mathbb Z_q$ of products of these transfers, Eq.~\eqref{eq:fermion-two-spin} of
Supplement~\ref{sec:S-fermion-transfer}; at zero winding the coefficient of $z^0$ is taken.

\subsection{The zig-zag and its slow angle}

Now let $\beta$ approach $\pi$. Write the turning angle as
\begin{equation}
 \beta=\pi+\delta,\qquad \delta=-\frac{\pi d}{q},\qquad d=q-2p ,
 \label{eq:3d-detuning}
\end{equation}
with $d$ odd and of either sign (for odd $q$; for even $q$ see Supplement~\ref{sec:S-arith}). Since $S_k=\sum_{l<k}\sigma_l$ has the parity of
$k$, we have $e^{\ii\pi S_k}=(-1)^k$, and in complex notation for the plane
\begin{equation}
 f_k=R\,(-1)^k e^{\ii\theta_k},\qquad \theta_k=\phi+\delta S_k .
 \label{eq:3d-zigzag}
\end{equation}
The polygon is an alternating zig-zag whose direction $\theta_k$ turns by
$\pm\delta$ at every step. Over the whole loop the slow angle diffuses by
$\delta\sqrt N$, so the natural variable is
\begin{equation}
 \xi=\sqrt N\,|\delta| ,
 \label{eq:3d-xi}
\end{equation}
and the relevant polygons have $\beta-\pi=O(N^{-1/2})$, that is, in the odd ensemble,
$q=O(\sqrt N)$; at zero winding $q$ ranges up to $N$ (Sec.~\ref{sec:ev-arith}). At zero winding, $S_N=0$, and the slow angle returns to itself, $\theta_N=\theta_0$; in the odd ensemble
the closure is a half-turn (Supplement~\ref{sec:S-fermion}).

\subsection{Two pendulums without a saddle-point approximation}

In the slow coordinate a step multiplies by the phase and shifts $\theta$ by $\pm\delta$.
By \eqref{eq:3d-zigzag}, $\cos(\phi+\beta S_l)=(-1)^l\cos\theta_l$ exactly, so the transfers
\eqref{eq:fermion-transfer} alternate between
\begin{equation}
 D_bC_\delta,\qquad D_{-b}C_\delta,\qquad
 D_b=e^{\ii b\cos\theta},\qquad C_\delta=\tfrac12\left(e^{\delta\partial_\theta}+e^{-\delta\partial_\theta}\right).
 \label{eq:fermion-alternating}
\end{equation}
Multiplying a pair and expanding only in $\delta$,
\begin{equation}
 \boxed{\begin{aligned}
 &D_bC_\delta D_{-b}C_\delta=U_b\bigl(1-2\delta^2H_b\bigr)U_b^{-1}+O(\delta^4),\\
 &U_b=e^{\ii b\cos\theta/2},\qquad H_b=-\frac12\partial_\theta^2+\frac{b^2}{8}\sin^2\theta .
 \end{aligned}}
 \label{eq:fermion-pair}
\end{equation}
The physical meaning is clear. Half of the alternating phase is a pure gauge: it
is removed by $U_b$ and survives only at the ends of the arc. The other half is a
fast noise, a sum of many terms of size $\delta$ with random signs $(-1)^l\sigma_l$. It
points perpendicular to the zig-zag, along $\ii e^{\ii\theta}$; its projection on the
source is proportional to $\sin\theta$, and its average gives the potential
$(b^2/8)\sin^2\theta$. An arc of $LN$ steps contains $LN/2$ pairs, and with
$\delta^2=\xi^2/N$ the product becomes $U_b\,e^{-\xi^2LH_b}\,U_b^{-1}$. The errors, $O(\delta^4)$ per
pair, accumulate to $O(1/N)$ at fixed $\xi$, $b$ and $L$. For the hairpin $R=\frac12+O(\delta^2)$,
so $b_1=h/(2x)$ and $b_2=-h/(2y)$ with $x=n/N$, $y=1-x$, and the two arcs carry the
Euclidean evolutions $K_L=e^{-\xi^2LH_L}$ with
\begin{equation}
 \boxed{H_L=-\frac12\partial_\theta^2+\frac{h^2}{32L^2}\sin^2\theta,}
 \qquad L\in\{x,y\} .
 \label{eq:3d-pendulum}
\end{equation}
The kinetic term is the diffusion of the slow angle; the potential is the price of
the fast noise. This is the Hamiltonian of a quantum pendulum, or the Mathieu
operator.

\paragraph{The oscillations of the trace cancel.} In this asymptotic form the oscillations of the original phase factor of the trace
have cancelled in the bulk of the arcs. The alternating phases $e^{\pm\ii b\cos\theta}$ of the transfers
\eqref{eq:fermion-alternating}, whose products oscillate strongly in the sum over the spins, leave only the gauge factors $U_b$ at
the ends of the arcs and the real potential $\frac{b^2}8\sin^2\theta$. The pendulum Hamiltonian \eqref{eq:3d-pendulum} is a
nonnegative operator, so the evolution along each arc is a contraction. The spectrum of $H_L$ consists of the bands of the pendulum,
labeled by the Bloch quasi-momentum, and the evolution is a contour integral of its resolvent, Eq.~\eqref{eq:fermion-arc-spectral} of
Supplement~\ref{sec:S-fermion-main}. Each level $E$ of the pendulum, a pole of the resolvent, contributes a Gaussian factor $e^{-\xi^2LE}$ in the scaled
angular detuning $\xi$. For the two arcs of a hairpin, with the endpoint operators $\mathcal O_0$ and $\mathcal O_n$, the two evolutions
combine into one, $K_xXK_y=e^{-\xi^2\mathbb L_x}X$, generated by the Sylvester operator $\mathbb L_xX=xH_xX+yXH_y$, $y=1-x$, which acts
on the operators $X$ of the slow angle. The integrals over $\xi$ give its inverse powers,
\begin{equation}
 \int_0^\infty\operatorname{Tr}\bigl[\mathcal O_0K_x\mathcal O_nK_y\bigr]\,2\xi^{2k-1}\dd\xi
 =(k-1)!\,\operatorname{Tr}\bigl[\mathcal O_0\,\mathbb L_x^{-k}(\mathcal O_n)\bigr],
 \label{eq:fermion-two-arc}
\end{equation}
which in the eigenstates of the two pendulums are the inverse powers of the sums $xE+yE'$ of two energies, one of each pendulum: with
$k=2$ the squared energy denominators of the source (Sec.~\ref{sec:3d-even}), with $k=1$ the single ones of the partition function of
the hairpin. The bulk damping factors are positive. Their coefficients, products of the matrix elements of the endpoint
insertions, need not be, and the residual oscillation of the endpoint factors $e^{\pm\ii z\cos\theta}$ produces the threshold at
$K=\sqrt2$ (Sec.~\ref{sec:3d-osc}). With unit endpoint operators, without the vorticity insertions and without the gauge factors
$e^{\pm\ii z\cos\theta}$, the coefficients become the squared overlaps of the Bloch states of the two pendulums, all positive;
$\operatorname{Tr}[K_xK_y]$ is then the bulk part of the bare hairpin, the partition function of the two pendulums at the inverse
temperature $\xi^2$, which is the measure of the conditional correlation (Sec.~\ref{sec:cond-pendulum}, Definition~\ref{post:cond}).

The operator representation thus resolves the oscillation problem of the trace, in a quantum analogue of the resolution of an
oscillating integral into thimbles~\cite{Witten2010PathIntegral,Witten2011ChernSimons}. The oscillating bulk propagation, which the
path integral of Ref.~\cite{migdal2024quantum} evaluated by a complex saddle point, is reorganized into exponentially damped
spectral components. We call these fully resummed spectral contributions \emph{quantum thimbles}. This is our terminology: a literal
identification with the integration cycles of Witten's construction, which arise from the downward flow of the complexified action
from its critical points, would need an additional construction. The quantum pendulum is therefore not a formal solution with
cancellations left to be resolved by analytic continuation: the analytic solution computes the sum over the spin paths directly. How the
quantum thimbles differ from the single instanton of Ref.~\cite{migdal2024quantum} is discussed in Supplement~\ref{sec:S-fermion-main}.

The gauge factors of the two arcs meet at the endpoints. At the endpoint $0$ they combine
into $U_{b_2}^{-1}U_{b_1}=e^{\ii z\cos\theta_0}$, and at the endpoint $n$ into
$e^{-\ii\varepsilon z\cos\theta_n}$, where $\varepsilon=(-1)^n$ and
\begin{equation}
 z=\frac{b_1-b_2}{2}=\frac{h}{4xy} .
 \label{eq:3d-z}
\end{equation}

At zero winding, where $\theta_N=\theta_0$, the trace over the slow angle runs over all quasi-momenta of the Bloch states of
the pendulum (Sec.~\ref{sec:3d-even}); the half-turn closure of the odd ensemble is treated in Supplement~\ref{sec:S-fermion}.

In the framework of the Quantum Statistical Principle this chain of reductions, from the covariant-derivative operator
$\hat D=\partial+\ii v/\nu$ that implements the Navier--Stokes flow (Sec.~\ref{sec:qsp-identity}), to the fermionic occupation
operators of the Euler states, and finally to a Schr\"odinger operator on $S^1$, is nothing but a change of representation inside the
same trace. Let us stress the status of each step. The operator identity, the Fock trace and the walk on $\mathbb Z_q$ are exact at
finite $N$; only the last step, to the Schr\"odinger operator, is asymptotic, and it is checked against the exact transfer matrix, with
differences that decrease as $1/N$. Without such an operator-level representation of the classical flow, this sequence of
transformations would be difficult to motivate or justify: a classical sum over polygons does not suggest, by itself, that its
continuum limit is a Schr\"odinger operator. Here it is one because the sum was a quantum trace from the start.
% ===== Sec. 8.1-8.3 rewritten 2026-10-31 =====
\section{The conditional state as a Gibbs state of two quantum pendulums}
\label{sec:cond-pendulum}

The vorticity correlation is defined as a conditional expectation on the hairpin (Definition~\ref{post:cond}). In this section we
derive the measure of this expectation: in the critical layer $\beta\to\pi$, the zig-zag layer of Sec.~\ref{sec:3d-fermion}, the hairpin
amplitude without the vorticities becomes a positive Gibbs measure of two quantum pendulums. Let us stress that this measure is derived from the
equal weights of the Euler states, not postulated. We then show why its partition function is the normalization of the correlation:
it is Bayes' formula.

\subsection{Counting the states and conditioning on the hairpin}
\label{sec:cond-two-steps}

The first step is the one used so far, and it is the only hypothesis. The decaying trajectories $F_k(s)=\sqrt{2\nu\tau}\,P_k(t)$ of~\eqref{eq:rescaled-vertices} are
carried to the fixed points of~\eqref{eq:rescaled-full-mle}, the Euler states, and every admissible state, a rational angle with a spin
sequence, counts once, with the Haar measure for the orientation $\Omega\in O(3)$: the microcanonical density matrix
$\hat\Pi_N/\operatorname{Tr}\hat\Pi_N$ of~\eqref{eq:fermion-qsp-trace}. It tells how to count the states, not how to condition on a loop.

The second step, Definition~\ref{post:cond}, conditions on the hairpin, and it adds no hypothesis. The equal-weight sum of the hairpin is transformed, exactly at
every $N$ (Sec.~\ref{sec:3d-fermion}) and then by the continuum limit near $\beta=\pi$, into the trace of the Euclidean evolution of two
pendulums, ordered along the loop, with the phases $e^{\pm\ii z\cos\theta}$ at its two marks. Without these phases it is a partition
function with a positive weight, a path measure of the slow angle (Sec.~\ref{sec:cond-state}), and this partition function is the normalization of the conditional expectation. Section~\ref{sec:cond-justify} shows that the weight is the survival probability of a random walk, and that the conditioning on the hairpin is the conditioning on survival, by Bayes' formula.

\subsection{The thermostat}
\label{sec:cond-state}

The basic observation is that the zig-zag layer has fast and slow degrees of freedom. Of the four spin configurations of a pair of steps,
the two with equal spins turn the slow angle $\theta_k=\phi+\delta S_k$ of~\eqref{eq:3d-zigzag} by $\pm2\delta$, and the two with
opposite spins leave it in place and carry a fast noise, the fluctuating part of the alternating phase. The equal weights of the spin
sequences, a microcanonical ensemble at zero total spin, make the slow angle a random walk, which wanders by $\xi$ around the loop, and
the average of the noise gives it a potential: together, a Boltzmann weight, the Feynman--Kac weight~\eqref{eq:cond-survival} of
its paths. The fast spins are the thermostat of the slow angle, which
inherits from their microcanonical ensemble a canonical distribution, as a subsystem does from its reservoir, with one difference: no
energy is exchanged, and the potential is the damping of a random phase. The transfer matrices, ordered along the loop,
accumulate into one exponential on each arc, in the order of the arcs on the circle: a Gibbs factor $e^{-\xi^2(xH_x+yH_y)}$ in the
ordered sense, the product $e^{-\xi^2xH_x}e^{-\xi^2yH_y}$, with $H_x$ on the fraction $x$ of the Euclidean time and $H_y$ on the rest.

In the bulk the weight of a slow path is positive. The zero total spin returns the slow angle to its starting point on the line, and the
uniform initial angle closes the trace (Supplement~\ref{sec:S-even-steps}, Steps~4 and~5). With the trace per cell on the covering line,
$\operatorname{Tr}\mathcal A=\int_0^\pi\mathcal A(\theta,\theta)\dd\theta$, the return contributes $1/Z_0(\xi)$, the inverse of the free
trace $Z_0=\operatorname{Tr}e^{-\xi^2H_0}=\sqrt{\pi/2}\,/\xi$, $H_0=-\partial_\theta^2/2$, and the Feynman--Kac formula~\cite{Kac1949}
gives
\begin{equation}
 \frac{\operatorname{Tr}\bigl[K_xK_y\bigr]}{Z_0(\xi)}
 =\mathbb E\Bigl[\exp\Bigl(-\int_0^{x\xi^2}V_x(\theta_\tau)\dd\tau-\int_{x\xi^2}^{\xi^2}V_y(\theta_\tau)\dd\tau\Bigr)\Bigr],\qquad
 K_L=e^{-\xi^2LH_L},
 \label{eq:cond-survival}
\end{equation}
the expectation over the Brownian bridge $\theta_\tau$ with the generator $\partial_\theta^2/2$, from a uniform $\theta_0$ back to
$\theta_0$ in the time $\xi^2$, in the potentials $V_L=(h^2/32L^2)\sin^2\theta\ge0$: the Boltzmann weight
$\exp[-\int(\dot\theta^2/2+V)\dd\tau]$ of the slow path. The phases at the marks do not weigh the path; they sit where the vorticities are
inserted. The gauge is not chosen: $U_b$ is the only one in which both arcs evolve with real pendulum Hamiltonians.

No energy of a reservoir fixes the temperature. The rational angles are uniform in $\xi$ (Supplement~\ref{sec:S-even-steps}, Step~7), and
the return adds $1/Z_0=\sqrt{2/\pi}\,\xi$, so that $\dd\xi/Z_0=\dd\xi^2/\sqrt{2\pi}$: with the unnormalized Gibbs factors the measure is
flat in $\xi^2$ and in $x$. The uniform density of the angles near $\pi$, all that remains of their arithmetic, and the return of the
slow angle set the distribution of the temperature.

\paragraph{The Gibbs measure.} Normalized by its trace, the bulk weight of the hairpin is the positive path measure of the slow angle
\begin{equation}
 \dd\mu_h=\frac{Z_0(\xi)\,e^{-S_{h,x,\xi}[\theta]}}{(2\pi)^{3/2}\,\widehat Z(h)}\;\dd x\,\dd\xi^2\,\dd\mathbb P_\xi(\theta),\qquad
 S_{h,x,\xi}[\theta]=\int_0^{x\xi^2}\!V_x(\theta_\tau)\dd\tau+\int_{x\xi^2}^{\xi^2}\!V_y(\theta_\tau)\dd\tau ,
 \label{eq:cond-state}
\end{equation}
with the Brownian bridge $\mathbb P_\xi$ of~\eqref{eq:cond-survival} and the flat measure $\dd x\,\dd\xi^2$, normalized to one by the
partition function~\eqref{eq:cond-C}. The marked insertion $\mathcal O_h$ of the Euler ensemble consists of the vorticities with the
phases $e^{\pm\ii z\cos\theta}$ at the two marks, at the Euclidean times $0$ and $x\xi^2$, and of the response to a uniform source. By
Definition~\ref{post:cond} the correlation is its expectation in this measure,
\begin{equation}
 C(h)=\int\mathcal O_h\,\dd\mu_h=\frac{\widehat G(h)}{\widehat Z(h)} .
 \label{eq:cond-expectation}
\end{equation}

The measure~\eqref{eq:cond-state} is an inhomogeneous Euclidean Gibbs measure on the thermal circle, with $H_x$ on the interval $x\xi^2$
and $H_y$ on the rest: the Gibbs state of no single Hamiltonian. The ordered product
$e^{-\xi^2xH_x}e^{-\xi^2yH_y}=e^{-\xi^2\mathbb L_x}(\mathbb I)$, the Gibbs factor of the Sylvester generator $\mathbb L_xX=xH_xX+yXH_y$
of Sec.~\ref{sec:3d-fermion} acting on the identity, represents this path integral in the ordered trace, by~\eqref{eq:cond-survival}; it
is in general not Hermitian and not a density matrix: divided by its trace, it can give complex expectation values to positive operators.
At $h=0$ the weight~\eqref{eq:cond-survival} is one, and the measure reduces to the count of states: a hairpin of zero size is a point
(Supplement~\ref{sec:S-cond-checks}); the integral over $\xi^2$ then diverges (Sec.~\ref{sec:cond-thermal}), and $C(0)=\frac18$ is the
limit $h\to0^+$. Here the measure defines and tests a two-point function; higher correlators, the evolution and the responses of a
passive scalar need a common marked loop functional, still to be constructed.

\subsection{Why the hairpin is normalized by its Gibbs trace}
\label{sec:cond-justify}

The loop equation determines the amplitudes but not how to normalize them as probabilities. The bulk weight of the hairpin,
however, has a direct probabilistic meaning, and with it Definition~\ref{post:cond} is the standard rule of conditional
probability, Bayes' formula.

\paragraph{The weight is a survival probability.} Write the measure~\eqref{eq:cond-state} as $\dd\mu_h=W_h\,\dd m/\widehat Z(h)$, with
the reference measure $\dd m=(2\pi)^{-3/2}Z_0(\xi)\,\dd x\,\dd\xi^2\,\dd\mathbb P_\xi(\theta)$ and the weight
$W_h=e^{-S_{h,x,\xi}[\theta]}\le1$, so that $\widehat Z(h)=\int W_h\,\dd m$. Let the slow angle walk on its circle as the Brownian bridge
of~\eqref{eq:cond-survival}, and let it be killed at the rate $V_x(\theta_\tau)$ on the first arc and $V_y(\theta_\tau)$ on the second.
Given the path, the walk survives the thermal circle with the probability $W_h$, the probability that an independent exponential time of
unit mean exceeds the accumulated rate $S_{h,x,\xi}[\theta]$. The partition function $\widehat Z(h)$ is the total probability of survival,
summed with the flat measure over the durations $\xi^2$ and the positions $x$ of the marks.

\paragraph{What kills the walk.} The killing rate is the phase of the hairpin averaged over the fast spins. On each pair of steps the fast
spins add a random increment to the phase of the hairpin, of a size set by the projected separation $h$ and by the slow angle; averaged,
the random phase damps the amplitude by $e^{-V\dd\tau}$, with $V\propto h^2\sin^2\theta$. A path that accumulates a large random phase
decoheres, and the hairpin measures nothing along it (Sec.~\ref{sec:3d-fermion}, Eq.~\eqref{eq:3d-pendulum}). Survival means that the phase of the hairpin has stayed coherent around the whole
loop.

\paragraph{Conditioning on survival.} The two vorticities are inserted on the same paths, with the same weight:
$\int\mathcal O_hW_h\,\dd m=\widehat G(h)$. Hence, exactly,
\begin{equation}
 \boxed{C(h)=\int\mathcal O_h\,\dd\mu_h=\frac{\int\mathcal O_hW_h\,\dd m}{\int W_h\,\dd m}
 =\mathbb E_m\bigl[\mathcal O_h\,\big|\,\hbox{survival}\bigr],}
 \label{eq:cond-survival-conditioning}
\end{equation}
the expectation of the product of the vorticities over the walks that survive the hairpin. To divide instead by the number of states
would be to count every killed walk as a zero product of vorticities. A killed walk does not carry a zero correlation; it carries none
that this hairpin can measure. The fraction of survivors depends on the instrument, through the separation that sets the killing rate,
and not on the vorticities at its ends. Dividing by it corrects for the efficiency of the instrument, as the Bayes formula does for a
conditional probability. At $h=0$ nothing is killed, and the conditioning reduces to the count of states, the normalization of the
zero-separation test of Sec.~\ref{sec:3d-zero-separation}.

\paragraph{The phases at the marks and the orientation.} The factor $e^{\ii\Gamma/\nu}$ with its phases at the marks is not a probability.
These phases sit at the insertions, not along the path, so they belong to the observable $\mathcal O_h$. Kept in the denominator, they make
the normalization oscillate, and the spectrum fails the positivity test at a fixed separation (Supplement~\ref{sec:S-cond-literal}). The
orientation of the hairpin is not a variable of the walk either: through $h$ it sets the killing rate, a parameter of the process, and a
survival probability is defined at a fixed rate. The correlation is therefore conditioned at a fixed projected separation, and the
orientation is averaged afterwards with the isotropic weight $\sin\vartheta$, as Definition~\ref{post:cond} states. Only the
relative orientation of the hairpin and of the plane of the polygon enters, and this argument takes the plane as given: the state
is then a mixture of ensembles with fixed planes, each conditioned separately, and the plane is a quenched variable. If the plane is
itself a variable of one state, in equilibrium with the spins, it is weighted by the survival as the path is and is averaged in
both traces: this is the annealed order, whose spectrum is computed in Sec.~\ref{sec:3d-orders}. Section~\ref{sec:stoch-rotation}
estimates how fast the thermal noise could randomize the plane, and Supplement~\ref{sec:S-cond-orders} compares the other orders of
the averages.

\paragraph{What is derived, what is defined and what is open.} Given the pendulum limit, each step above is exact. Since $m$ has infinite mass, the conditioning is done
at $\xi\le\Xi$; at $h>0$, where $\widehat Z(h)$ is finite, the conditioned laws tend to $\mu_h$ in total variation as $\Xi\to\infty$.
The measure is derived from the equal weights of the Euler states. The conditional
expectation~\eqref{eq:cond-survival-conditioning}, the product of the vorticities averaged over the walks that survive the hairpin,
is the vorticity correlation by definition: Bayes' formula for the hairpin that measures it. What stays open is the order of the
conditioning and of the average over the orientation, quenched or annealed (Sec.~\ref{sec:3d-orders}), which the comparison with
the data should settle (Sec.~\ref{sec:comparison}).

\subsection{The thermal circle}
\label{sec:cond-thermal}\label{sec:cond-partition}

The closed loop is the thermal circle of the slow angle. Its circumference in diffusive units, $\xi^2=N(\beta-\pi)^2$, the variance of the
walk of the slow angle around the loop, is the Euclidean time and, in the Gibbs state, the inverse temperature; at the critical angle
$\beta=\pi$ the temperature is infinite. This is the temperature of the slow angle on the loop, not of the fluid of Supplement~\ref{sec:qsp-gibbs}.
The two marks cut the circle into the intervals $x\xi^2$ and $y\xi^2$, with different Hamiltonians, and the vorticity correlation is a
thermal two-point function of the pendulums, with the insertions at the imaginary times $0$ and $x\xi^2$, together with its response to a
uniform source.
Its normalization, the denominator of $C=\widehat G/\widehat Z$ in Eq.~\eqref{eq:cond-def-limit}, is the partition function
\begin{equation}
 \boxed{\widehat Z(h)=\frac1{(2\pi)^{3/2}}\int_0^1\dd x\int_0^\infty\dd\xi^2\;\operatorname{Tr}\bigl[e^{-\xi^2xH_x}e^{-\xi^2yH_y}\bigr],}
 \label{eq:cond-C}
\end{equation}
and the numerator $\widehat G$ is the same trace with the vorticities inserted at the two marks, together
with its response to a uniform source:
\begin{equation}
 \widehat G(h)=\frac1{(2\pi)^{3/2}}\int_0^1\dd x\int_0^\infty\dd\xi^2\;\Bigl\{\xi^2\operatorname{Tr}\bigl[\mathcal O_0K_x\mathcal O_nK_y\bigr]
 +\sum_u\partial_u\operatorname{Tr}\bigl[\mathcal O^u_0K_x\mathcal O^u_nK_y\bigr]\Big|_{\epsilon=0}\Bigr\}.
 \label{eq:cond-G}
\end{equation}
The two terms of $\widehat G$, the source~\eqref{eq:ev-G0}, are the contact terms with the normal vorticity at the endpoints $0$ and $n$
and those with the mean $\langle f\rangle$, the responses $\partial_u=\partial/\partial\epsilon$ to a uniform source $\epsilon\mathbf u$ along each of the two in-plane axes $u$, added to the hairpin source in the potentials of both pendulums and taken at $\epsilon=0$; the endpoint operators $\mathcal O_0$, $\mathcal O_n$ and
$\mathcal O^u_{0,n}$ are written out in Supplement~\ref{sec:S-cond-checks}, after Eq.~\eqref{eq:cond-resolvent}. The insertions carry
$\delta^2=\xi^2/N$, the inverse temperature divided by $N$: this is the explicit $1/N$ of the correlation (Sec.~\ref{sec:3d-ensemble}).

There is no free temperature: the flat measure integrates it out. A pair of levels, one of each pendulum, is a level $\Lambda=xE+yE'$ of
$\mathbb L_x$. Besides the squared overlap $|\langle m,x|m',y\rangle|^2$ of the two states, it then has the weight $\int_0^\infty e^{-\xi^2\Lambda}\dd\xi^2=1/\Lambda$, the zero-energy resolvent of
the pair: a power law rather than a Boltzmann factor, so that the statistics has no energy scale of its own. This is
why the higher levels matter. The insertions, with their extra $\xi^2$, turn $1/\Lambda$ into $1/\Lambda^2$. At small $h$ only the bottom
of the ground band counts (Sec.~\ref{sec:cond-zero}). At large $h$ the levels of the deep wells are equidistant, every level below the
barrier contributes with its weight $1/\Lambda$, and their harmonic sum gives a logarithm. Thus (Supplement~\ref{sec:S-cond-checks})
\begin{equation}
 h\widehat Z(h)\simeq\frac{\sqrt\pi}4\quad(h\to0),\qquad
 h\widehat Z(h)\simeq\frac{\log h+\bar\kappa}{4\sqrt{2\pi}}\quad(h\to\infty),
 \label{eq:cond-Z-asym}
\end{equation}
with $\bar\kappa=\log64+\gamma_E-1$, and in between $h\widehat Z$ increases monotonically, without the oscillations of the literal
denominator (Supplement~\ref{sec:S-cond-literal}). At $h=0$ the free pendulum has zero energy at the bottom of its band, and the integral over the inverse
temperature diverges: it counts all states. This is the origin of the singularity $1/h$ of the source divided by the number of states and
of the $1/r$ singularity of the earlier correlation; in the ratio the divergences of the numerator and of the partition function cancel.
The Gibbs form is a property of the critical layer only: away from $\beta=\pi$ there is no slow angle, and no temperature.

\subsection{Resolvents}
\label{sec:cond-resolvents}

The integral over the inverse temperature is a Laplace transform. By~\eqref{eq:fermion-two-arc} it gives the inverse powers of
$\mathbb L_x$, the generator of the evolution around the thermal circle:
$\widehat Z(h)=(2\pi)^{-3/2}\int_0^1\dd x\operatorname{Tr}[\mathbb L_x^{-1}(\mathbb I)]$, and $\mathbb L_x^{-2}$ for the contact terms
of $\widehat G$ (Supplement~\ref{sec:S-cond-main}). The inverse of the Sylvester operator is an integral over an imaginary energy of the resolvents $R_L(\zeta)=(LH_L-\zeta)^{-1}$ of the
pendulums,
\begin{equation}
 \mathbb L_x^{-k}(X)=\int_{-\infty}^\infty\frac{\dd\omega}{2\pi}\;R_x(\ii\omega)\,X\,R_y(-\ii\omega)^k,\qquad k=1,2,
 \label{eq:cond-split}
\end{equation}
the frequency form of the correlator integrated over the temperature: with the flat measure the thermal circle is integrated out, and
$\omega$ is an imaginary-energy convolution variable; reading it as a continuum of Matsubara frequencies would need a separate
derivation. This is the operator form of Eq.~\eqref{eq:ev-split}. Hence $\widehat Z$ and
$\widehat G$ are the integrals $(2\pi)^{-3/2}\int_0^1\dd x\int\dd\omega/2\pi$ of $\mathcal Z(h,x,\omega)=\operatorname{Tr}[R_x(\ii\omega)R_y(-\ii\omega)]$
and of $\mathcal T(h,x,\omega)/16$, the expression~\eqref{eq:pf-T} with the traces per cell, as in the Floquet form~\eqref{eq:ev-floquet}
of the source. The
conditional correlation is therefore
\begin{equation}
 \boxed{C(h)=\frac1{16}\,\frac{\displaystyle\int_0^1\dd x\int_{-\infty}^\infty\dd\omega\;\mathcal T(h,x,\omega)}
 {\displaystyle\int_0^1\dd x\int_{-\infty}^\infty\dd\omega\;\mathcal Z(h,x,\omega)},}
 \label{eq:cond-floquet}
\end{equation}
a ratio of two quadratures, over the arc fraction, the imaginary energy and one cell, of the Floquet solutions and their derivatives; no
sum over the levels and no matrix element between the states of the two pendulums remains.

\paragraph{An integrable system.} The Mathieu operator is related, by analytic continuation and with its own Floquet conditions, to the
rank-two oper of the two-particle closed Toda chain (the real Toda chain, with its $\cosh$ potential, is a different self-adjoint
Hamiltonian). Its Floquet solutions at complex
energy follow from Baxter's equation of this chain~\cite{BaermanRavazziniTeschner2026}, whose oper equation at rank two is the Fourier
recursion~\eqref{eq:3d-fourier-matrix}; exact WKB and the thermodynamic Bethe
ansatz~\cite{DunneUnsal2016Mathieu,Sueishi2021ExactWKB,Imaizumi2020Mathieu} give the Floquet exponent below the barrier and need, at
complex energy, the continuation of the quantum periods and of their Stokes data (Remark~\ref{rem:3d-tba}). The spectrum of decaying
turbulence is thus a set of quadratures of the Floquet solutions of an integrable system, the quantum pendulum, in a thermal state on the
loop; the integrability concerns this pendulum, not the Navier--Stokes evolution.

\subsection{Small separation}
\label{sec:cond-zero}

As $h\to0$ at fixed $x$ the hairpin opens only a small gap, the mean potential energy $E_0=h^2/(64xy)$ of the two arcs, and the thermal
circle grows long, $\langle\xi^2\rangle=1/(2E_0)=32xy/h^2$. The temperature, of the order of the gap and far below the higher bands, is
effectively zero: the pendulums are at the bottom of their ground band, the long waves of the slow angle. The conditional correlation at fixed
$x$ is $(1-2x)^2/(32xy)$ (Supplement~\ref{sec:S-cond-main}), half of what a single level at $E_0$ would give, since the
long waves fill the bottom of the band; the weight of $x$ is the semicircle $\sqrt{xy}$:
\begin{equation}
 \boxed{C(0)=\frac1{32}\,\frac{\int_0^1(1-2x)^2\dd x/\sqrt{x(1-x)}}{\int_0^1\sqrt{x(1-x)}\dd x}
 =\frac14\int_0^1\frac{(1-2x)^2\dd x}{\pi\sqrt{x(1-x)}}=\frac18 ,}
 \label{eq:cond-zero}
\end{equation}
a quarter of the arcsine average of $(1-2x)^2$, in units of $1/N$, with $h\to0$ taken after $N\to\infty$ (Sec.~\ref{sec:cond-def}): a
number fixed by the ground band, the thermal average over its long waves. It is also the ratio of the asymptotes $\widehat G\simeq\sqrt\pi/(32h)$,
Eq.~\eqref{eq:ev-small}, and $\widehat Z\simeq\sqrt\pi/(4h)$. Since $h\widehat Z-\sqrt\pi/4=O(h^{3/2})$,
the first correction to $C$, of order $\sqrt h$, comes from the numerator and fixes the ultraviolet law of the spectrum
(Sec.~\ref{sec:3d-final}).

\paragraph{What is exact, what is asymptotic, what is hypothesized.} The operator identities of Sec.~\ref{sec:3d-fermion} are exact at
every $N$. The pendulum limit is asymptotic, with errors of order $1/N$ at fixed $\xi$, $x$ and $h$, not uniform in $\xi$, whose relevant
values grow as $1/h$; hence $h\to0$ is taken after $N\to\infty$. Given the limit, the Gibbs form of the bulk trace and its Feynman--Kac, resolvent and
Floquet forms are exact. In the critical layer Monte Carlo confirms the numerator up to $N=10^5$ and the ratio up to $N=32000$
(Sec.~\ref{sec:3d-even}), and the survival~\eqref{eq:cond-survival} at $N=\QWALKN$ (Supplement~\ref{sec:S-cond-checks}); over the
whole even ensemble at zero winding the ratio agrees with $C$ within $1.6$ standard deviations, with errors of $0.45$ to $1.0$ percent,
at $h=\frac14$, $\frac12$ and $1$ (Supplement~\ref{sec:S-finiteN}). Hypothesized is only the equal weight of the Euler states. The correlation is the conditional
expectation of Definition~\ref{post:cond}, whose probabilistic meaning is given in Sec.~\ref{sec:cond-justify}.

\section{The even ensemble at zero winding: the conditional source}
\label{sec:3d-even}

Now we solve the ensemble that dominates the grand ensemble (Sec.~\ref{sec:3d-ensemble}) and passes the
positivity test (Proposition~\ref{prop:3d-positivity}): even $N$, total spin $M=\sum_k\sigma_k=0$, angles
$\beta=2\pi p/q$ with $(p,q)=1$ and $3\le q<N$, all states with equal weights, $Z^{(0)}_N=\binom N{N/2}\Phi_N$. The
correlation itself is the conditional source $C=\widehat G/\widehat Z$ of Eq.~\eqref{eq:cond-def-limit}, in units of $1/N$, the
expectation value of the vorticities at the ends of the hairpin in its Gibbs state, Definition~\ref{post:cond}, which the transforms
\eqref{eq:cond-def-transforms} carry to the energy spectrum. Its numerator $\widehat G$ is the limit of the source function of this
ensemble; its denominator $\widehat Z$ is the partition function of the two pendulums of Sec.~\ref{sec:cond-pendulum}.
This section states the results; the computations, the values, the fits
and the tests are in Supplement~\ref{sec:S-even-steps}, \ref{sec:S-cond} and~\ref{sec:S-even-details}.

\subsection{The source function}

What we compute is the source function $W_N(h)$ of Eq.~\eqref{eq:3d-source-function}, whose Hankel transform is the measure $V_N$ of the numerator, in
the limit of large $N$. Only the polygons with turning angles near $\pi$ contribute: there the polygon goes back and forth, a zig-zag
whose direction turns slowly as the spins dictate, and the two arcs of the hairpin become the two quantum pendulums of
Sec.~\ref{sec:3d-fermion}. The small angles and, by the symmetry $\beta\to2\pi-\beta$, the angles near $2\pi$ contribute at fixed
$h<\pi/2$ only terms exponentially small in $\sqrt N$ (Supplement~\ref{sec:S-small-angles}); the bound rests on one lattice spectral
inequality that we verify numerically, and for larger $h$ the decay persists in our computations with a smaller rate, which we have not
bounded. Summing over the rational angles, the positions of the endpoints and the spin sequences of zero total spin gives
\begin{equation}
 \boxed{W_N(h)=\sqrt N\,\widehat G(h)+o\bigl(\sqrt N\bigr),}
 \label{eq:ev-G0}
\end{equation}
where $\widehat G$ is a double quadrature over the fraction $x$ of the polygon on one arc of the hairpin and an imaginary energy $\omega$
of the pendulums, Eq.~\eqref{eq:ev-floquet}. The factor $\sqrt N$ counts the pairs of endpoints, the
fraction of the angles near $\pi$ and the order of the insertions; the partition function, without the insertions, grows as $N^{3/2}$
(Sec.~\ref{sec:cond-pendulum}).

\subsection{The arithmetic at zero winding}
\label{sec:ev-arith}

Near $\beta=\pi$ the rational angles are counted by a sum of Euler totients, cut off at $q<N$. In the limit their density becomes uniform:
of the arithmetic only the factor $\zeta(2)$ is left, and it cancels against the number of angles in the normalization. In the conditional source even this cancellation is not needed: the
same uniform density weighs the partition function (Sec.~\ref{sec:cond-pendulum}), and the ratio does not depend on it at all. At
finite $N$ the zeros of the Riemann zeta function are present, but they disappear in the limit (Remark~\ref{rem:ev-zeros}). The counting is
in Supplement~\ref{sec:S-even-arith-steps} and Supplement~\ref{sec:S-arith}.

\subsection{The Floquet form of the source and of the partition function}
\label{sec:ev-floquet}

The sums over the Bloch states and the integral over the Bloch angle can be carried out exactly (Supplement~\ref{sec:S-floquet-steps}),
and the source becomes an integral of solutions of the Mathieu equation at complex energies,
\begin{equation}
 \boxed{\widehat G(h)=\frac{\sqrt{2\pi}}{64\pi^2}\int_0^1\dd x\int_{-\infty}^\infty\frac{\dd\omega}{2\pi}\;\mathcal T(h,x,\omega),}
 \label{eq:ev-floquet}
\end{equation}
where $\mathcal T$ is an integral over one period of products of the Floquet solutions of the two pendulums at the energies $\ii\omega/x$ and
$-\ii\omega/y$, acted on by the endpoint operators. The partition function is the same double integral without the endpoint operators,
$\widehat Z(h)=(2\pi)^{-3/2}\int_0^1\dd x\int_{-\infty}^\infty\mathcal Z(h,x,\omega)\,\dd\omega/2\pi$, with the cell trace
$\mathcal Z=\operatorname{Tr}[R_x(\ii\omega)R_y(-\ii\omega)]$ of the resolvents $R_L(\zeta)=(LH_L-\zeta)^{-1}$ of the two pendulums and a
prefactor sixteen times that of \eqref{eq:ev-floquet}; the ratio of the two quadratures is \eqref{eq:cond-floquet}
(Sec.~\ref{sec:cond-resolvents}). In other words, the two arcs of the hairpin are two resolvents at opposite imaginary energies, and the
vorticities at its ends are operators between them: this is all that distinguishes the numerator from the denominator.

\subsection{The conditional source: values and asymptotes}
\label{sec:ev-source}

The results are best stated for three functions that stay finite as $h\to0$: the numerator $g=h\widehat G$, the partition function
$\Pi=h\widehat Z$ and their ratio, the conditional source $C=g/\Pi$. The partition function is positive and increases monotonically, from
$\sqrt\pi/4$ at $h\to0$ to a logarithmic growth, so the zeros of $C$ are those of the numerator. The conditional source is positive
except in the interval $\QCDIPLO<h<\QCDIPHI$, where it dips to $\QCMIN$ at $h\simeq\QCMINH$, a fraction $\QCMINREL$ of $C(0)$, and it
stays positive at all its later minima (Supplement, Fig.~\ref{fig:ev-G0}). Does the negative dip of $C$ contradict positivity? No. Let us stress that $C(h)$ is a source, not a spectrum. It is the correlation at a
fixed planar separation, before the average over the orientation of the hairpin, and nothing requires it to be positive. Positivity is a
property of the spectrum, the Hankel transform $P(K)$ of Eq.~\eqref{eq:cond-def-transforms}, and $P$ is positive at every computed wave
number (Sec.~\ref{sec:3d-final}).

\paragraph{Small $h$.} For a weak potential only the lowest band of each pendulum counts in the bulk of the hairpin, and it gives the
leading terms exactly; the corrections come from the end layers $x\sim h$, where the short arc is not a weak pendulum:
\begin{equation}
 \boxed{\begin{gathered}
 h\widehat G(h)=\frac{\sqrt\pi}{32}-a_{1/2}\,h^{1/2}+O\bigl(h^{3/2}\bigr),\qquad
 h\widehat Z(h)=\frac{\sqrt\pi}4+c_{3/2}\,h^{3/2}+O\bigl(h^2\bigr),\\
 C(h)=\frac18-\frac{4a_{1/2}}{\sqrt\pi}\,h^{1/2}+O\bigl(h^{3/2}\bigr).
 \end{gathered}}
 \label{eq:ev-small}
\end{equation}
The constant terms come from the ground band, and their ratio is $C(0)=\frac18$ (Sec.~\ref{sec:cond-zero}). Neither the numerator nor the
partition function has a term linear in $h$, and the partition function has no term in $h^{1/2}$: the weight $\sqrt{xy}$ of its bulk
vanishes at the ends. In the end layers, $x=h/(4w)$, the short arc is a pendulum of frequency $w$ with all its levels and the long arc
is a free rotor with the kinetic energy $\mu$; the two coefficients are quadratures of the resolvent $R_\mu=(H_w+\mu)^{-1}$ of this one
pendulum, $a_{1/2}=\QGA$ and $c_{3/2}=\QPIC$, Eq.~\eqref{eq:ev-small-a}. Fits of the computed values reproduce $a_{1/2}$ and $c_{3/2}$ to a
relative $\QSMALLAGREE$.

\paragraph{Large $h$.} In the semiclassical limit of deep potential wells the leading term of order $h^{-2}$ cancels identically, and what
remains is an oscillating term and a logarithm (Supplement~\ref{sec:S-asym-steps}). For $\QGFITLO\le h\le\QGHMAX$ the computed values of the numerator, and for $\QZFITLO\le h\le\QZHMAX$ those of the partition function,
follow, up to terms of higher order in $1/h$,
\begin{equation}
 \begin{gathered}
 h\widehat G(h)={\rm Re}\Bigl[e^{2\ii h}\Bigl(D+\frac{D_1}h\Bigr)\Bigr]h^{-5/2}+\frac{E'\log h+F}{h^2},\\
 h\widehat Z(h)=\frac{\log h+\bar\kappa}{4\sqrt{2\pi}}+\frac{c_0}h+O\Bigl(\frac{\log h}{h^2}\Bigr),\qquad \bar\kappa=\log64+\gamma_E-1,
 \end{gathered}
 \label{eq:ev-large}
\end{equation}
with $D$ computed in the semiclassical limit in Sec.~\ref{sec:3d-osc}, the derived constant $\bar\kappa$, and with $E'=\QEPRIME$,
$F=\QFTAIL$, $|D_1|=\QDONE$ and $c_0=\QPICZERO$ fitted, with the next orders included, to a relative $\QGTAILRMS$ (rms). The fitted
$E'$, $c_0$ and $D$ agree with the closed forms $E'=1/(1024\sqrt\pi)$, $c_0=7\pi^{3/2}/(512\sqrt2)$ and $D=(I_2+\ii I_1)/(128\pi)$
derived in Ref.~\cite{JiangMigdalLean2026} to better than $10^{-6}$ (Supplement~\ref{sec:S-tail-closed}). Hence the conditional source at large $h$,
\begin{equation}
 h^2C(h)=4\sqrt{2\pi}\Bigl[E'+\frac{F-\bar\kappa E'}{\log h+\bar\kappa}\Bigr]
 +\frac{4\sqrt{2\pi}\,{\rm Re}\bigl(De^{2\ii h}\bigr)}{h^{1/2}\,(\log h+\bar\kappa)}+O\Bigl(\frac1h\Bigr),
 \label{eq:ev-C-large}
\end{equation}
with $A_c=4\sqrt{2\pi}E'=1/(128\sqrt2)=\QAC$ and $B_c=4\sqrt{2\pi}(F-\bar\kappa E')=\QBC$, numerical through the fitted $F$; the form of
the law is derived. The conditional source falls as $h^{-2}$, but its
coefficient approaches the limit $A_c$ only logarithmically, and from below; the oscillation is divided by the same logarithm. The
ratio of the two laws reproduces the computed $C$ to a relative $\QCLAWB$ at $h\ge400$. We use \eqref{eq:ev-large} beyond the computed values; the leading infrared law of the spectrum depends on $E'$
and $F$ (Sec.~\ref{sec:3d-final}).

\subsection{The whole ensemble}

The exact sums over the finite ensemble, computed with the transfer formula up to $N=120$, follow the continuum angle by
angle where $\delta=\xi/\sqrt N$ is small, and approach it as $N$ grows where it is not (Supplement~\ref{sec:ev-whole}).
Together with the tests at fixed $x$, which converge as $1/N$, this supports
\begin{equation}
 W_N(h)=\sqrt N\,\widehat G(h)\,\bigl(1+o(1)\bigr),\qquad h>0\ \hbox{fixed},
 \label{eq:ev-whole}
\end{equation}
and that no other region of angles contributes at this order. This is numerical evidence over the whole ensemble, not
a theorem, and the convergence is not uniform in $h$. A direct Monte Carlo simulation reproduces $\sqrt N\,\widehat G(h)$
in the zig-zag layer within \MCDEVFOUR{} at $N=10^4$ for $0.2\le h\le0.5$, and within the statistical errors, \MCERRFIVE{}, at
$N=10^5$ (Supplement~\ref{sec:S-mc}). The partition function is checked at fixed
$(x,\xi,h)$. In the critical layer, with the numerator and the denominator~\eqref{eq:cond-bulk-mid} computed on the
same samples, the ratio $NW_N/D^{\rm bulk}_N$, extrapolated from $N\le32000$, agrees with the pendulum ratio of the layer within one
percent at $h\le1$, with statistical errors of $0.2$ to $0.7$ percent (Supplement~\ref{sec:S-finiteN}); over the whole even ensemble at zero winding it agrees with $C$ within $1.6$ standard deviations at $h=\frac14$, $\frac12$ and $1$,
with errors of $0.45$ to $1.0$ percent (Supplement~\ref{sec:S-whole-evidence}). The
conditional source thus rests on Definition~\ref{post:cond}, which defines it, and on the pendulum limits
of its numerator and of its partition function.

\section{The energy spectrum}
\label{sec:3d-final}

This section gives the main result of the paper, the energy spectrum of decaying turbulence in closed form,
Eq.~\eqref{eq:ev-spectrum}. It follows from the conditional source $C=\widehat G/\widehat Z$ of the even states at zero winding, a
ratio of two quadratures of the resolvents of the pendulums, in the Gibbs state of the hairpin, which is derived from the equal
weights of the Euler states (Remark~\ref{rem:ev-open}). What is left is the limit; the details are in
Supplement~\ref{sec:S-spectrum-details}. Up to
Sec.~\ref{sec:3d-orders} the plane of the polygon is a quenched variable, as in Definition~\ref{post:cond}; Sec.~\ref{sec:3d-orders}
gives the spectrum of the annealed order, and Sec.~\ref{sec:3d-size} the finite-size effects, which distinguish the two.

\subsection{Renormalization and the turbulent viscosity}
\label{sec:3d-renorm}

At fixed turbulent viscosity $\widetilde\nu=\nu_NN^2$ and fixed $kL$ the spectrum decreases as $N^{-1}$: the numerator grows as
$N^{1/2}$, the partition function of the hairpin as $N^{3/2}$, and what remains is the order $\xi^2/N$ of the two vorticity insertions.
One cannot simply set $N=\infty$: the limit would be zero. We met this problem before, in Sec.~VII\,B of
Ref.~\cite{migdal2024quantum}, and solved it there by a renormalization (Supplement~\ref{sec:S-renorm}). The dimensionless scaling functions depend on the circulation and the viscosity only
through $\Gamma/\nu$, and we compute them in an auxiliary flow, whose viscosity, the bare viscosity, vanishes as the cutoff is removed,
\begin{equation}
 \nu_N=\frac{\widetilde\nu}{N^2},\qquad N\to\infty\quad\hbox{at fixed}\quad\widetilde\nu ;
 \label{eq:3d-aux}
\end{equation}%
other powers of $N$ are the subject of Sec.~\ref{sec:3d-limits}. A power of the viscosity in front of a scaling function is not a
function of $\Gamma/\nu$, and it must keep its physical value. If an observable grows as $N^a=(\widetilde\nu/\nu_N)^{a/2}$, the
dependence on the cutoff and the dependence on the viscosity coincide, and the calculation alone cannot separate them. The
renormalization condition assigns the whole power to the viscosity: the
product $\nu^{a/2}E$ has the same value in every flow at large circulation, so that in the physical flow the power of the viscosity
in front takes its physical value. The homogeneity in $\Gamma/\nu$ motivates this condition but does not prove it. In the sense of
Remark~\ref{rem:intro-status} it is part of the definition of the theory, as a renormalization condition is in quantum field theory
(Table~\ref{tab:status}); experiments that vary
the viscosity alone, at a fixed geometry, can test it; the effective indices and the normalized spectra cannot, since the amplitude
cancels from them. Thus
\begin{equation}
 \boxed{E(k,t)=\Bigl(\frac{\widetilde\nu}{\nu}\Bigr)^{a/2}\frac{\widetilde\nu^{3/2}}{\sqrt{t+t_0}}\,
 f\bigl(k\sqrt{\widetilde\nu\,(t+t_0)}\bigr),}
 \label{eq:3d-renorm-result}
\end{equation}
with the scaling function $f$ of the auxiliary flow, the physical viscosity $\nu$, and the turbulent viscosity $\widetilde\nu$, a free
parameter of the solution, the single scale of circulation, as the QCD scale is the single mass scale of chromodynamics. In a given
flow $\widetilde\nu$ is measured, together with $t_0$.

For the spectrum $a=-1$. Normalized by the number of states alone, as in the
previous versions of this paper, the spectrum had $a=\frac12$. This is the multiplicative renormalization of one observable, the two-point function. It is evidence for, but not a proof of, the
renormalizability of the whole statistical theory, which needs the same for every correlation function and a limit that does not
depend on the regularization (Remark~\ref{rem:intro-status}). The factor $(\nu/\widetilde\nu)^{1/2}$ does not depend on time and
changes no index; at fixed $\widetilde\nu$ it vanishes as $\nu^{1/2}$. Let us stress that the ratio $\widetilde\nu/\nu$ is not the
Reynolds number. The Reynolds number is, as usual, the ratio $\Gamma/\nu$ of a typical circulation to the viscosity, and it
depends on the size of the loop. The turbulent viscosity $\widetilde\nu$ is a free parameter of the theory, produced by the formal limit
of vanishing viscosity, renormalized by the cutoff, as the physical scale of the turbulence; it is found by matching the scaling
variable $k\sqrt{\widetilde\nu\,(t+t_0)}$ to the data, and nothing requires $\widetilde\nu/\nu$ to be large. The renormalized
spectrum has no dissipation range; a physical flow follows it up to its own, which we estimate to begin at
\begin{equation}
 k_\nu=\frac{1}{\sqrt{\nu\,(t+t_0)}},\qquad k_\nu L=\sqrt{\frac{\widetilde\nu}{\nu}} .
 \label{eq:3d-knu}
\end{equation}

\subsection{The renormalized spectrum}
\label{sec:3d-renormalized}

By \eqref{eq:3d-renorm-result} with $a=-1$ and $f(K)=P(K)/(8K^2)$ the renormalized spectrum is
\begin{equation}
 \boxed{E(k,t)=\Bigl(\frac{\nu}{\widetilde\nu}\Bigr)^{1/2}\frac{\widetilde\nu^{\,2}}{8L}\,\frac{P(kL)}{(kL)^2},\qquad
 P(K)=2K\int_0^\infty h\,J_0(\sqrt2hK)\,C(h)\dd h,}
 \label{eq:ev-spectrum}
\end{equation}
with $L=\sqrt{\widetilde\nu(t+t_0)}$ and the conditional source $C=\widehat G/\widehat Z$ of \eqref{eq:cond-def-limit}. This is our
result for the energy spectrum of decaying turbulence. In what sense is it a limit? At every $N$ the function $V_N(K)$ of the
numerator is a finite sum of point masses, some of them negative (Remark~\ref{rem:intro-positivity}) and some at $K=0$. The limit is
meant in the weak sense on $K>0$, and it needs dominations that we assume without proof (Supplement~\ref{sec:S-weak-limit}). The
result supersedes the spectrum of the previous versions of this paper, $E\propto1/k$ in the infrared and $E\propto k^{-2}$ in the
ultraviolet, which was normalized by the number of states and thus set the hairpin factor to one
(Supplement~\ref{sec:hairpin-not-one}). Values are tabulated in Supplement~\ref{sec:S-spectrum}.

Every link of the chain \eqref{eq:intro-main} is a quadrature, and every integrand is explicit; the chain with the resolvents written
through the Floquet solutions is Eq.~\eqref{eq:ev-chain}. No sum over the levels is left anywhere. The energy spectrum of decaying turbulence is
thus a set of multiple quadratures of the explicit resolvents of an integrable system, the quantum pendulum, in the Gibbs state of
Sec.~\ref{sec:cond-pendulum}.

\begin{figure}[t]
 \centering
 \includegraphics[width=\textwidth]{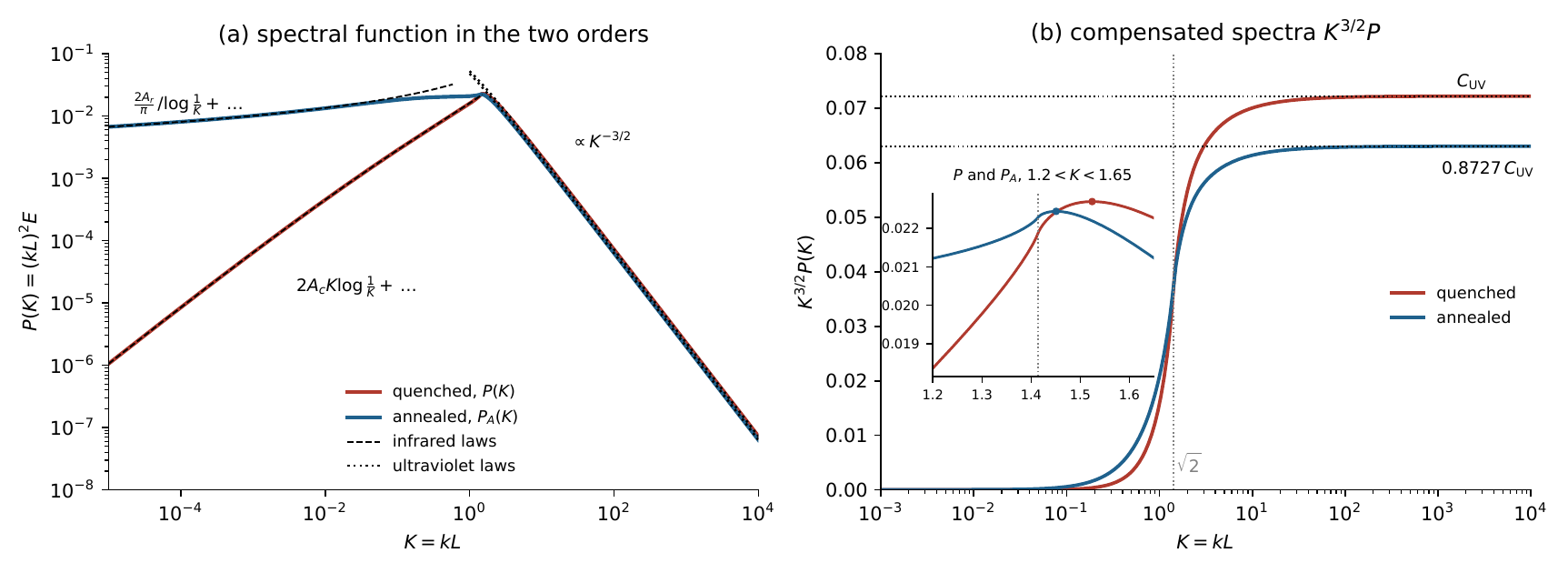}
 \caption{The two solutions. (a) The spectral functions $(kL)^2E$, in units $(\nu/\widetilde\nu)^{1/2}\,\widetilde\nu^{\,2}/(8L)$,
 on logarithmic axes: $P(K)$ of the quenched order, Eq.~\eqref{eq:ev-spectrum} (red, the lower curve at $K<1$), and $P_A(K)$ of the
 annealed order, Eq.~\eqref{eq:ord-PA} (blue), with their infrared laws \eqref{eq:ev-IR} and \eqref{eq:ord-IR} (dashed) and their
 ultraviolet laws \eqref{eq:ev-UV} and \eqref{eq:ord-UV} (dotted). The energy spectrum $E\propto P/K^2$ falls as $\log(1/kL)/k$ in
 the quenched order and as $1/(k^2\log(1/kL))$ in the annealed one, and then as $k^{-7/2}$ in both.
 (b) The compensated spectra $K^{3/2}P$ and $K^{3/2}P_A$, which tend to $C_{\rm UV}$ and to $0.8727\,C_{\rm UV}$ (horizontal dotted
 lines); vertical dotted line: $K=\sqrt2$. Inset: $P$ and $P_A$ on linear axes for $1.2<K<1.65$: the rise to the threshold, where both
 slopes diverge as a double logarithm, Eqs.~\eqref{eq:ev-sqrt2} and \eqref{eq:ord-sqrt2}, and the maxima at $K\simeq\QSPKMAX$ and
 $K\simeq1.451$.}
 \label{fig:ev-spectrum}
\end{figure}

\paragraph{Positivity.} The function $P$ is positive at every computed $K$, $\QSPPKLO\le K\le\QSPPKHI$, and beyond, where the infrared
and ultraviolet laws \eqref{eq:ev-IR} and \eqref{eq:ev-UV} hold; it rises to a maximum, $P=\QSPPMAX$ at $K\simeq\QSPKMAX$, just above
the threshold $K=\sqrt2$, and then falls (Fig.~\ref{fig:ev-spectrum}). Positivity is computed, not proved: a ratio of two linear
functionals of the ensemble is not itself a linear functional.

\paragraph{Ultraviolet.} A term $C_kh^{k/2}$ of the source gives the term $U_kK^{-1-k/2}$ of the spectrum, with
$U_k=2^{1+k/4}\,\Gamma(1+\frac k4)\,C_k/\Gamma(-\frac k4)$, which vanishes for $k=0,4,8,\dots$. The value $C(0)=\frac18$ and the even
powers of $h$ therefore do not reach the spectrum, the source has no term $h^1$, and the end-layer term $h^{1/2}$ of
\eqref{eq:ev-small} gives
\begin{equation}
 \boxed{P(K)=C_{\rm UV}K^{-3/2}+U_3K^{-5/2}+O(K^{-7/2}),\qquad
 C_{\rm UV}=\frac{2^{1/4}\,\Gamma(\frac14)}{2\sqrt\pi\,\Gamma(\frac34)}\,a_{1/2}=\QSPCUV ,}
 \label{eq:ev-UV}
\end{equation}
a closed form times the end-layer quadrature; $U_3=\QSPUthree$ comes from the $h^{3/2}$ terms, fitted for the numerator. The spectrum falls as $k^{-7/2}$ for $kL\gg1$, up to the dissipation range \eqref{eq:3d-knu}. Its integral
converges to the value of the correlation at zero separation,
\begin{equation}
 \int_0^\infty P(K)\,\dd K=C(0)=\frac18 ;
 \label{eq:ev-sum-rule}
\end{equation}
the computed spectrum satisfies this sum rule to a relative $\QSPsumacc$, an internal check, not physical accuracy: the source has
$C(0)=\frac18$ built in.

\paragraph{Infrared.} At large $h$ the smooth part of the conditional source decays as $hC\simeq A_c/h+B_c/[h(\log h+\bar\kappa)]$,
Eq.~\eqref{eq:ev-C-large}, and its first moment diverges logarithmically. The Bessel kernel acts on such a slowly varying tail as a sharp
cutoff at $h=\sqrt2\,e^{-\gamma_E}/K$, with a small correction. Hence
\begin{equation}
 \boxed{\begin{gathered}
 \frac{P(K)}K=2A_c\Bigl(\log\frac{\sqrt2}K-\gamma_E\Bigr)+2B_c\log L_K+2M-\frac{\zeta(3)\,B_c}{3L_K^3}+\dots,\\
 E\simeq\Bigl(\frac{\nu}{\widetilde\nu}\Bigr)^{1/2}\frac{\widetilde\nu\,A_c}{4k(t+t_0)}\,\log\frac1{kL}\qquad(kL\ll1),
 \end{gathered}}
 \label{eq:ev-IR}
\end{equation}
with $L_K=\log(\sqrt2/K)-\gamma_E+\bar\kappa$ and the regularized first moment $M=\QCOM$ of the source, Eq.~\eqref{eq:ev-IR-next};
the $\zeta(3)$ comes from the Bessel kernel, not from the arithmetic. The spectrum computed from the Floquet values of $C$ follows this
law to a relative $\QSPIRacc$ for $K\le10^{-2}$. The index tends to one from above, $n\simeq1+1/\log(1/K)$; the double
logarithm, of opposite sign, is comparable with the leading term at every computed $K$; the leading law depends on the tail
of the numerator through $E'$, whose closed form gives $A_c=1/(128\sqrt2)$ (Supplement~\ref{sec:S-tail-closed}). The energy diverges as $\log^2$ in the infrared, and a finite flow cuts it at its lowest wave number $k_0$
(Sec.~\ref{sec:3d-size}, Remark~\ref{rem:3d-energy-normalization}); the enstrophy is finite, Eq.~\eqref{eq:ev-sum-rule}.

\paragraph{The singularity at $K=\sqrt2$.} The hairpins with two equal arcs make the spectrum singular at $K=\sqrt2$.
At $x=\frac12$ the endpoint phase $z=h/(4xy)$ is stationary, and the source oscillates as $\cos2h$; at $\sqrt2K=2$ this oscillation
resonates with the Bessel kernel. The contact kernel $\frac12-x$ vanishes at the stationary point and weakens the oscillation of the
numerator by a factor $1/h$, but in the conditional source it is divided by the partition function, which grows only as $\log h$:
\begin{equation}
 h\widehat G\supset{\rm Re}\bigl(De^{2\ii h}\bigr)\,h^{-5/2},\qquad
 hC\supset\frac{{\rm Re}\bigl(De^{2\ii h}\bigr)}{h^{3/2}\,\Pi(h)},\qquad
 \Pi(h)=h\widehat Z(h)\simeq\frac{\log h+\bar\kappa}{4\sqrt{2\pi}} .
 \label{eq:ev-D}
\end{equation}
The normalization thus gives back the power of $h$ that the contact kernel took away, up to the logarithm of $\Pi$. With
$\epsilon=2-\sqrt2K$, the cosine and the sine act on the slowly varying $1/\Pi$ as a sharp cutoff at $h=e^{-\gamma_E}/|\epsilon|$, as
the Bessel kernel does in the infrared:
\begin{equation}
 \boxed{\begin{aligned}
 \frac{\dd P}{\dd K}={}&a_2\,\log\Bigl(\log\frac1{\sqrt2\,|K-\sqrt2|}-\gamma_E+\bar\kappa\Bigr)
 -\frac{\kappa_2\;{\rm sgn}(K-\sqrt2)}{\log\frac1{\sqrt2\,|K-\sqrt2|}-\gamma_E+\bar\kappa}\\
 &+\hbox{regular}+O\bigl(\log^{-2}|K-\sqrt2|\bigr),
 \end{aligned}}
 \label{eq:ev-sqrt2}
\end{equation}
with
\begin{equation}
 a_2=8\sqrt2\;{\rm Im}\bigl(De^{\ii\pi/4}\bigr)=\QCOATWO,\qquad \kappa_2=4\sqrt2\,\pi\;{\rm Re}\bigl(De^{\ii\pi/4}\bigr)=\QCOKAPPA .
 \label{eq:ev-a2}
\end{equation}
The spectrum is continuous at $K=\sqrt2$; its slope diverges to $+\infty$ on
both sides, as $a_2\log\log$, and the slope drops across the threshold by $2\kappa_2/\log(1/|K-\sqrt2|)$, which vanishes slowly. The local index therefore goes to
$-\infty$ at $K=\sqrt2$, but it is negative only within about $\QSPnegwin$ of it. The computed
slopes follow the double logarithm down to $|K-\sqrt2|=\QSPTHDMIN$, which checks the transform and the form of the law, not the
amplitude. The singularity is still weak,
but stronger than before: it is a nonanalyticity of the limiting statistical spectrum at fixed turbulent viscosity, at which the spectrum
stays finite and continuous and its slope diverges. In the earlier spectrum the slope was continuous there, and only the
second derivative jumped and diverged logarithmically.

\paragraph{The Mellin amplitude.} The Mellin amplitude of $P$ is the Mellin transform of the conditional source, a ratio of pendulum
quantities with no zeta function in it, times the Mellin transform of the Bessel kernel; the singularities of the derived expansions
are real, with no global classification claimed, and the double pole
at $z=-1$ gives the logarithm of the infrared law. The threshold at $K=\sqrt2$ is not a pole: it shows in the slow decay of the
amplitude along vertical lines (Supplement~\ref{sec:ev-mellin}).

\subsection{Time dependence and the finite-time singularity}

\begin{figure}[t]
 \centering
 \includegraphics[width=\textwidth]{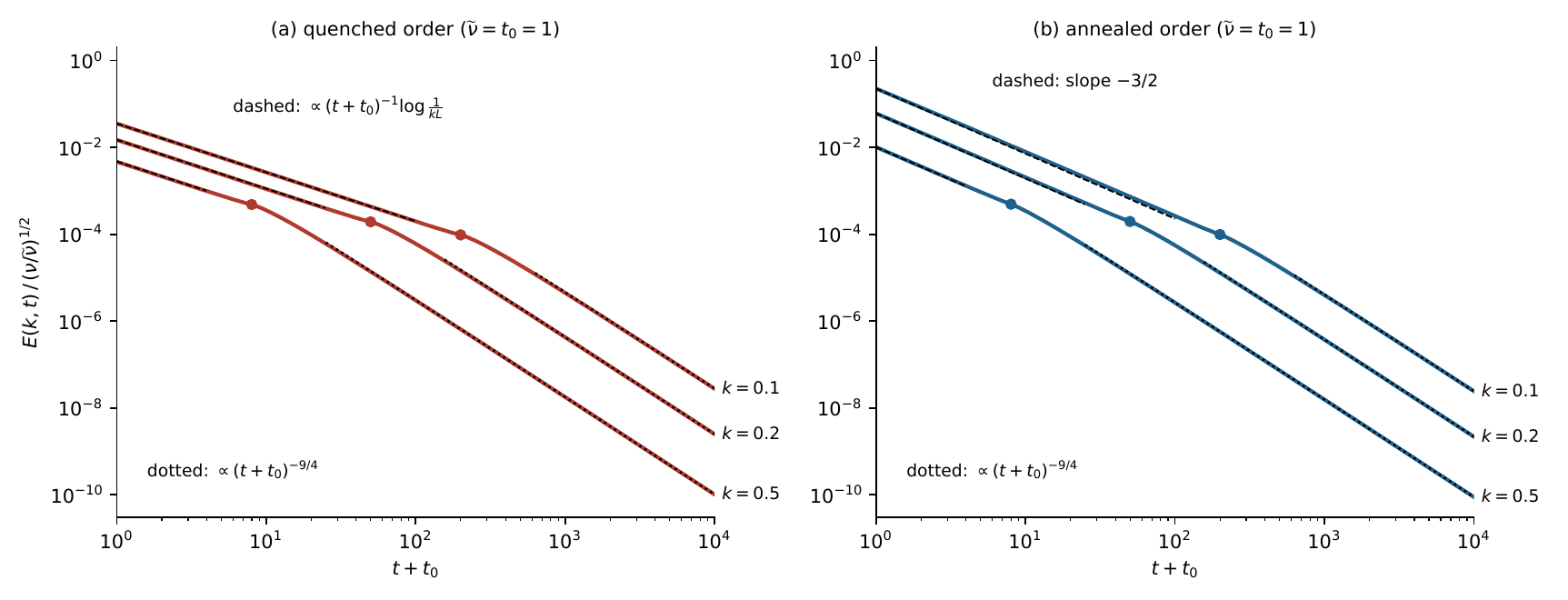}
 \caption{The spectrum, divided by $(\nu/\widetilde\nu)^{1/2}$, as a function of time at the fixed wave numbers
 $k=0.1$, $0.2$ and $0.5$ (solid lines), in units $\widetilde\nu=t_0=1$, on logarithmic axes of $t+t_0$ and $E$: (a) the quenched
 order, Eq.~\eqref{eq:ev-spectrum}; (b) the annealed order, Eq.~\eqref{eq:ord-PA}. At early times, $kL\ll1$, the quenched spectrum
 follows the infrared law \eqref{eq:ev-IR}, $\propto(t+t_0)^{-1}\log(1/kL)$ (dashed in (a)), and the annealed spectrum falls nearly
 as $(t+t_0)^{-3/2}$ (dashed in (b): the slope $-\frac32$; its infrared law \eqref{eq:ord-IR} is reached only at $kL\lesssim10^{-2}$).
 At late times, $kL\gg1$, both follow their ultraviolet laws, $\propto(t+t_0)^{-9/4}$ (dotted). The dots mark the singular times
 \eqref{eq:ev-tstar}, $t_*=199$, $49$ and $7$, the same in the two orders, at which $kL=\sqrt2$ and the time derivative of $E$ diverges
 to $+\infty$.}
 \label{fig:ev-time}
\end{figure}

Time enters only through $L=\sqrt{\widetilde\nu(t+t_0)}$, and $t=0$ is the physical initial time with $t_0>0$ kept everywhere. At a
fixed wave number the spectrum decays as $(t+t_0)^{-1}\log(1/kL)$
while $kL\ll1$ and as $(t+t_0)^{-9/4}$ once $kL\gg1$; in the annealed order of Sec.~\ref{sec:3d-orders} it decays as
$(t+t_0)^{-3/2}/\log(1/kL)$ and then as $(t+t_0)^{-9/4}$. The threshold \eqref{eq:ev-sqrt2}, common to the two orders, is a singularity in time. At a fixed wave number $k$
the spectrum is singular at the moment when
\begin{equation}
 \boxed{k\sqrt{\widetilde\nu\,(t+t_0)}=\sqrt2,\qquad t_*(k)=\frac{2}{\widetilde\nu k^2}-t_0,}
 \label{eq:ev-tstar}
\end{equation}
which is a finite positive time for $k<\sqrt{2/(\widetilde\nu t_0)}$. Since $\dd K/\dd t>0$, $E(k,t)$ is continuous at $t=t_*(k)$,
while, by $\partial\log E/\partial\log(t+t_0)=\frac12(KP'/P-3)$, its first time derivative diverges to $+\infty$ as
$\log\log(1/|t-t_*|)$: at this wave number the spectrum rises locally through a continuous cusp (Fig.~\ref{fig:ev-time}), which is
neither a divergent rate of decay nor a blow-up of the energy. The asymmetry $1/\log(1/|t-t_*|)$ of the two sides belongs to this
derivative; in $E$ it carries the extra factor $|t-t_*|$. The energy above a fixed $k_0$ decays with an index that drifts from
$1+1/\log(1/k_0L)$ to $\frac94$, not as a power law, and that inherits the threshold: it stays finite and continuous, and only its time
derivative diverges (Supplement~\ref{sec:S-spectrum-details}). In the annealed order the index starts near $\frac32$ instead of one
(Sec.~\ref{sec:3d-size}).

\paragraph{The energy balance.} The balance equation of the energy spectrum, $\partial_tE=T-2\nu k^2E$, follows from the
Navier--Stokes equation, which the loop equation encodes, but our solution cannot verify it. Its transfer is a three-point function,
which we do not compute (Sec.~\ref{sec:cond-state}), and it contains the product of two velocities at one point, in general a new
operator, sensitive to the scale of the viscosity, which the continuum theory, valid at finite separations, does not determine: as for
the operator product expansion in quantum field theory and in critical phenomena, the balance defines $T(k,t)$ and is not a relation to
be checked against the triple correlator at separated points. The potential part of the velocity, which enters the energy budget through
the boundaries, drops from the circulation. The triple correlator linear in the separation, a $\nabla\delta(\boldsymbol k)$ term with no
support at $k>0$ (Refs.~\cite{migdal2024quantum,ReviewPaperAM}), describes, taken at all separations, the inertial range of stationary forced
turbulence, not the decay. This answers the question of the energy balance (Supplement~\ref{sec:S-energy-balance}; for the global balance,
Remark~\ref{rem:3d-dissipation}).

\subsection{The annealed orientation: the second spectrum}
\label{sec:3d-orders}

The spectrum \eqref{eq:ev-spectrum} treats the plane of the polygon as a quenched variable: the ratio $C=\widehat G/\widehat Z$ is
formed at a fixed projected separation $h$ and then averaged over the orientations (Definition~\ref{post:cond}). If the plane is in
equilibrium with the spins, an annealed variable (Sec.~\ref{sec:stoch-rotation}), the two averages come first and the ratio is taken
at a fixed separation $r$. The renormalized prefactors are the same, and so is the relation
$c(R)=\int_0^\infty P(K)\,\mathrm{sinc}(KR)\dd K$ between the correlation and the spectrum:
\begin{equation}
 \boxed{\begin{gathered}
 C^A_\omega(r,t)=\Bigl(\frac{\nu}{\widetilde\nu}\Bigr)^{1/2}\frac{c_A(r/L)}{4(t+t_0)^2},\qquad
 c_A(R)=\frac{\int_0^{\pi/2}\widehat G(h_\vartheta)\sin\vartheta\,\dd\vartheta}{\int_0^{\pi/2}\widehat Z(h_\vartheta)\sin\vartheta\,\dd\vartheta},\qquad
 h_\vartheta=\frac{R\sin\vartheta}{\sqrt2},\\
 E_A(k,t)=\Bigl(\frac{\nu}{\widetilde\nu}\Bigr)^{1/2}\frac{\widetilde\nu^{\,2}}{8L}\,\frac{P_A(kL)}{(kL)^2},\qquad
 P_A(K)=\frac{2K}{\pi}\int_0^\infty R\,c_A(R)\,\sin(KR)\dd R .
 \end{gathered}}
 \label{eq:ord-PA}
\end{equation}
The planes are now weighted by the partition function of the hairpin instead of the isotropic weight
(Supplement~\ref{sec:S-cond-orders}); nothing else changes, and $P_A$ is again a chain of quadratures of the two functions
$\widehat G$ and $\widehat Z$. The last integral converges in the sense of Abel; we compute it after one integration by parts, from
the same Floquet values as $P$ (Supplement~\ref{sec:S-annealed}). Both spectra are shown in Fig.~\ref{fig:ev-spectrum}.

\paragraph{What the two orders share.} At zero separation the orientation does not matter: $c_A(0)=C(0)=\frac18$, so $P_A$ obeys the
sum rule \eqref{eq:ev-sum-rule}, and the computed $P_A$ satisfies it to a relative $4\times10^{-11}$. The hairpins with one short arc
give the same cusp of the correlation, and the ultraviolet law has the same form with other coefficients,
\begin{equation}
 \begin{gathered}
 P_A(K)=C^A_{\rm UV}K^{-3/2}+U^A_3K^{-5/2}+O(K^{-7/2}),\\
 C^A_{\rm UV}=\frac2\pi\,\frac{I_{1/2}}{I_{3/2}}\,C_{\rm UV}=0.8727\,C_{\rm UV}=0.0630,\qquad U^A_3=-0.0153,
 \end{gathered}
 \label{eq:ord-UV}
\end{equation}
with $I_s=\int_0^{\pi/2}\sin^s\vartheta\,\dd\vartheta$. The oscillating term of the source belongs to the numerator, and the
partition function averaged over the orientations grows as slowly as $\Pi$ itself:
$\int_0^{\pi/2}\widehat Z(h_\vartheta)\sin\vartheta\,\dd\vartheta\simeq(\pi/2h)\,[\Pi(h)-\log2/(4\sqrt{2\pi})]$ at $h=R/\sqrt2$.
The threshold law \eqref{eq:ev-sqrt2} therefore holds for $P_A$ with $a_2$ and $\kappa_2$ multiplied by $2/\pi$ and $\bar\kappa$
replaced by $\bar\kappa-\log2$:
\begin{equation}
 \begin{gathered}
 \frac{\dd P_A}{\dd K}=\frac{2a_2}\pi\,\log\Lambda_2-\frac{2\kappa_2}\pi\,\frac{{\rm sgn}(K-\sqrt2)}{\Lambda_2}
 +\hbox{regular}+O\bigl(\Lambda_2^{-2}\bigr),\\
 \Lambda_2=\log\frac1{\sqrt2\,|K-\sqrt2|}-\gamma_E+\bar\kappa-\log2 .
 \end{gathered}
 \label{eq:ord-sqrt2}
\end{equation}
Fits of the computed slopes at $10^{-12}\le|K-\sqrt2|\le10^{-4}$ return $2a_2/\pi=0.0229$ to $3\times10^{-4}$ and
$2\kappa_2/\pi=0.0113$ to $7\times10^{-4}$, as well as in the quenched order; with $\bar\kappa$ in place of $\bar\kappa-\log2$ the
second would be off by $4\%$. The singular time \eqref{eq:ev-tstar} and the period of the wave, Eq.~\eqref{eq:osc-period}, are the
same in the two orders. The function $P_A$ is positive at every computed wave number, $10^{-9}\le K\le10^4$. It has its maximum,
$0.02245$, at $K\simeq1.451$, just above the threshold, and it crosses $P$ at $K=1.453$: it is larger than $P$ below this point and
smaller above, by the factor $0.8727$ at large $K$.

\paragraph{Where they differ: the infrared.} At a large separation the numerator of $c_A$ comes from the planes nearly normal to
$\boldsymbol r$, within an angle of the order of $L/r$, where the projected separation is of the order of $L$: it is $2M_1/R^2$,
with the first moment $M_1=\int_0^\infty h\,\widehat G(h)\dd h=0.015931$ of the numerator. It is divided by the partition function
averaged over all the planes, which decays only as $\log R/R$. In the quenched order each plane carries its own ratio $C(h)$,
which decays as $A_c/h^2$, and all the angles between $L/r$ and one contribute equally, which gives $c\simeq2A_c\log R/R^2$,
Eq.~\eqref{eq:osc-c-large}. Thus
\begin{equation}
 \boxed{c_A(R)\simeq\frac{A_r}{R\,(\log R+C'')},\quad A_r=\frac{16M_1}{\sqrt\pi}=0.1438,\quad
 C''=\bar\kappa-\frac32\log2=2.696 ,}
 \label{eq:ord-c-large}
\end{equation}
up to the wave and to terms of the relative order $1/R$; the computed $c_A$ follows this law to $0.7\%$ at $R=100$ and to
$7\times10^{-5}$ at $R=10^4$. The sine transform of this tail is a series in the inverse logarithm,
\begin{equation}
 \boxed{\begin{gathered}
 P_A(K)=\frac{2A_r}{\pi}\Bigl[\frac1{\Lambda}+\frac{\gamma_E}{\Lambda^2}+\frac{\gamma_E^2-\pi^2/12}{\Lambda^3}+\dots\Bigr]+O(K),
 \qquad\Lambda=\log\frac1K+C'',\\
 E_A\simeq\Bigl(\frac{\nu}{\widetilde\nu}\Bigr)^{1/2}\frac{A_r\,\widetilde\nu^{1/2}}{4\pi\,k^2\,(t+t_0)^{3/2}\,\log(1/kL)}\qquad(kL\ll1),
 \end{gathered}}
 \label{eq:ord-IR}
\end{equation}
whose $n$th coefficient is $n!$ times that of $s^n$ in $\Gamma(1-s)\cos(\pi s/2)$. The series is asymptotic; its first five terms
reproduce the computed $P_A$ to $3\times10^{-7}$ at $K=10^{-8}$ and to $10^{-4}$ at $K=10^{-4}$, and at $K=10^{-2}$ the terms of the
order of $K$ are $1\%$. The spectrum of the annealed order is therefore close to $k^{-2}$ on the whole range $kL<1$. On the computed
range $10^{-8}\le kL\le1$ its local index $n=-\partial\log E/\partial\log k$ lies between $1.86$ and $1.98$ and tends to two from
below, as $2-1/\Lambda$, while that of the quenched order lies between $1.06$ and $1.27$ and tends to one. At $kL=0.01$ the
annealed spectrum is $29$ times the quenched one, and at $kL=0.1$ six times. Unlike the amplitude $A_c$ of the quenched order,
$A_r$ is not a closed form: it is a quadrature of the numerator.

\subsection{Finite size: the energy and its dissipation in the two orders}
\label{sec:3d-size}

The two infrared laws are not a detail of the largest scales: they decide how the energy of the flow and its dissipation depend on
the size of the system, and this is an observable difference between the two orders. A flow of a finite size has a lowest wave
number $k_0$, of the order of its inverse size, and the energy above it is
\begin{equation}
 \boxed{U_{k_0}(t)=\int_{k_0}^\infty E(k,t)\dd k=\Bigl(\frac{\nu}{\widetilde\nu}\Bigr)^{1/2}\frac{\widetilde\nu}{8(t+t_0)}\,I(k_0L),
 \qquad I(x)=\int_x^\infty\frac{P(K)}{K^2}\dd K ,}
 \label{eq:size-energy}
\end{equation}
with $P_A$ and $I_A$ in the annealed order. Since $x=k_0L$ grows as $\sqrt{t+t_0}$, the rate at which this energy is dissipated and
its decay index are
\begin{equation}
 \boxed{\varepsilon=-\frac{\dd U_{k_0}}{\dd t}=\Bigl(\frac{\nu}{\widetilde\nu}\Bigr)^{1/2}\frac{\widetilde\nu}{8(t+t_0)^2}
 \Bigl[I(x)+\frac{P(x)}{2x}\Bigr],\quad
 \alpha_E=-\frac{\dd\log U_{k_0}}{\dd\log(t+t_0)}=1+\frac{P(x)}{2x\,I(x)} .}
 \label{eq:size-diss}
\end{equation}
This $\varepsilon$ is the loss of energy that the balance of energy requires; the viscous range, in which it takes place, lies
outside our solution (Remark~\ref{rem:3d-dissipation}). By the infrared laws \eqref{eq:ev-IR} and \eqref{eq:ord-IR}, for $x\ll1$,
\begin{equation}
 \boxed{\begin{aligned}
 \hbox{quenched:}\quad&I(x)\simeq A_c\log^2\frac1x,&\quad\alpha_E&\simeq1+\frac1{\log(1/x)},&\quad
 U_{k_0}&\propto\frac{\log^2(1/k_0L)}{t+t_0},\\
 \hbox{annealed:}\quad&I_A(x)\simeq\frac{2A_r}{\pi}\,\frac1{x\,\lambda},&\quad\alpha_E&\simeq\frac32-\frac1{2\lambda},&\quad
 U_{k_0}&\propto\frac{1}{k_0\,\lambda\,(t+t_0)^{3/2}},
 \end{aligned}}
 \label{eq:size-laws}
\end{equation}
with $\lambda=\log(1/x)+C''-\gamma_E$. In the quenched order every octave of the scales
between $L$ and the size of the system carries a comparable energy, which grows only as the logarithm of the scale, and the total
depends on this size as the square of a logarithm. In the annealed order the energy
sits at the largest scales that the system admits: per unit mass it is proportional to the size of the system, up to a logarithm,
and so is its dissipation.

\begin{table}[htp]
\centering\small
\caption{Finite-size effects in the two orders of the orientation average: the function $I(x)$ of \eqref{eq:size-energy}, which
gives the energy above the lowest wave number $k_0$, and the decay index $\alpha_E$ of \eqref{eq:size-diss}, against $x=k_0L$.}
\label{tab:size}
\setlength{\tabcolsep}{4.5pt}
\begin{tabular}{lrrrrrrrrr}
\toprule
$k_0L$ & $10^{-4}$ & $10^{-3}$ & $10^{-2}$ & $0.1$ & $0.3$ & $1$ & $\sqrt2$ & $3$ & $10$\\
\midrule
$I$, quenched & $0.444$ & $0.273$ & $0.148$ & $0.0636$ & $0.0363$ & $0.0142$ & $0.00885$ & $0.00170$ & $8.94\times10^{-5}$\\
$I_A$, annealed & $89.0$ & $11.4$ & $1.52$ & $0.188$ & $0.0618$ & $0.0144$ & $0.00821$ & $0.00150$ & $7.83\times10^{-5}$\\
$\alpha_E$, quenched & $1.095$ & $1.117$ & $1.153$ & $1.223$ & $1.296$ & $1.556$ & $1.874$ & $2.191$ & $2.238$\\
$\alpha_E$, annealed & $1.452$ & $1.441$ & $1.438$ & $1.482$ & $1.539$ & $1.721$ & $1.960$ & $2.201$ & $2.240$\\
\bottomrule
\end{tabular}
\end{table}

\begin{figure}[t]
 \centering
 \includegraphics[width=\textwidth]{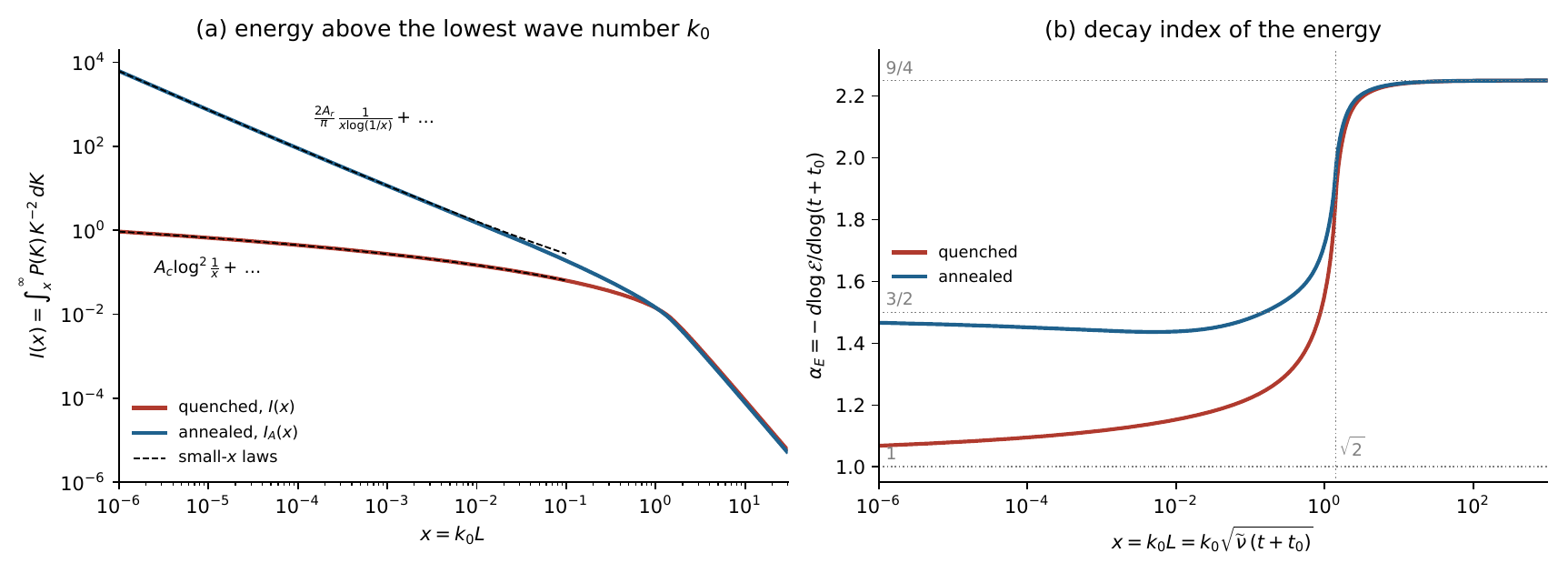}
 \caption{Finite-size effects in the two orders of the orientation average, against $x=k_0L$, with $k_0$ the lowest wave number of
 the flow. (a) The function $I(x)$ of \eqref{eq:size-energy}, to which the energy above $k_0$ is proportional at a given time, in
 the quenched order (red) and $I_A(x)$ in the annealed order (blue), with their small-$x$ laws (dashed,
 Supplement~\ref{sec:S-annealed}): the first grows as $\log^2(1/x)$, the second as $1/(x\log(1/x))$. (b) The decay index $\alpha_E$
 of the energy, Eq.~\eqref{eq:size-diss}: at small $x$ it is near one in the quenched order and near $\frac32$ in the annealed one,
 and it tends to $\frac94$ in both; vertical dotted line: $x=\sqrt2$.}
 \label{fig:ev-size}
\end{figure}

Both laws are reached slowly, and Table~\ref{tab:size} and Fig.~\ref{fig:ev-size} give the computed values. While the scale $L$ is
much smaller than the system, $10^{-4}\le k_0L\le0.1$, the energy decays with an index between $1.10$ and $1.22$ in the quenched
order and between $1.44$ and $1.48$ in the annealed one, where the index has a shallow minimum, $1.436$ at $k_0L=0.005$, and passes
$\frac32$ at $k_0L=0.15$. In both orders it rises to $\frac94$ when $L$ reaches the size of the system. The response to the size of
the system at a given time is different as well. If the size is doubled, $k_0\to k_0/2$, at $k_0L=0.01$, the energy grows by the
factor $1.22$ and its dissipation by $1.21$ in the quenched order, and by $1.83$ and $1.83$ in the annealed one; at $k_0L=0.1$ the
factors are $1.33$ and $1.30$ against $1.92$ and $1.89$. The decay of the energy in two flows that differ only in their size, or the
decay index of one flow, can therefore decide between the two orders.

\subsection{What remains to be settled}

Besides exact identities and asymptotic limits, the spectrum rests on the equal weight of the Euler states, the only physical
hypothesis; on the conditional correlation (Definition~\ref{post:cond}), with its order of the orientation average
(Sec.~\ref{sec:3d-orders}); on the renormalization condition and the assumed dominations; on the ratio at finite $N$, tested at five values of $h$ only; and on the matching
to physical flows (Remark~\ref{rem:ev-open}). Four things should not be confused: the accuracy of the quadratures and interpolations;
the uncertainty of the source, propagated to $P$ (Supplement~\ref{sec:S-spectrum}); the fitted constants $F$ and $D_1$ of the
large-$h$ tail, which enter $B_c$ and $M$, while the forms of the laws and the constants $E'$, $A_c$ and $D$ are derived (Supplement~\ref{sec:S-tail-closed}); and the physical hypothesis with the order of the conditioning, which no accuracy of the computation can test. The dissipation, the normalization by energy and the convergence of the limit are discussed in
Remarks~\ref{rem:3d-dissipation}, \ref{rem:3d-energy-normalization} and~\ref{rem:3d-matching} (Supplement~\ref{sec:S-open}).

\section{Oscillations in coordinate space}
\label{sec:3d-osc}

The singularity of the spectrum at $K=\sqrt2$ sits at a finite wave number, and this has a
direct consequence in coordinate space. A spectrum that is smooth except at a wave number
$k_*$ makes the correlation functions oscillate: at large separation they acquire a term
$\cos(k_*r+\psi)$ times a decreasing function of $r$, with an amplitude and a phase fixed by the strength of
the singularity~\cite{Lighthill1958}. Here $k_*=\sqrt2/L$, and the period of the oscillation is
\begin{equation}
 \boxed{\lambda_*=\frac{2\pi}{k_*}=\sqrt2\,\pi L=\sqrt2\,\pi\sqrt{\widetilde\nu\,(t+t_0)}\simeq\OSCPER\,L .}
 \label{eq:osc-period}
\end{equation}
It is not a free parameter: like the singular time \eqref{eq:ev-tstar}, it is fixed by the scale
$L(t)$ of the similarity solution, and it grows with it. We write $R=r/L$
throughout; the derivations are in Supplement~\ref{sec:S-osc-derivation}.

\subsection{Where the wave comes from}

The wave comes from the hairpins with two equal arcs, whose weight accumulates at the threshold. Here is how.
In the semiclassical limit the endpoint factor takes the values $1$ and $\cos2z$, with $2z=h/(2xy)$, and since $h$ is conjugate to
$\sqrt2K$, the factor $\cos2z$ carries the weight of the hairpins with arcs $xN$ and $(1-x)N$ to
\begin{equation}
 \sqrt2K=\frac1{2x(1-x)}\ge2 .
 \label{eq:osc-edge}
\end{equation}
The lower bound is reached by the hairpins with two equal arcs, $x=\frac12$, where $\dd K/\dd x=0$: their weight accumulates at the
threshold, as the density of states of a crystal is singular where the group velocity vanishes. The oscillating amplitude of the contact
density vanishes there quadratically, $A(x)=A_2(x-\frac12)^2+\dots$, and the stationary point gives
\begin{equation}
 \boxed{h\widehat G\supset{\rm Re}\bigl(De^{2\ii h}\bigr)h^{-5/2},\qquad D=\frac{e^{3\ii\pi/4}}{32\sqrt{2\pi}}\,A_2,\qquad
 A_2=\ATWORE\ATWOIM\,\ii ,}
 \label{eq:osc-source}
\end{equation}
the form \eqref{eq:ev-D}, with $|D|=\ABSD$ and $\arg D=\ARGD$. The coefficient $A_2$ is computed from the full pendulums, with an accuracy of about $\ATWOACC$
(Supplement~\ref{sec:S-osc}), and agrees with its closed form $e^{-3\ii\pi/4}(I_2+\ii I_1)/(2\sqrt{2\pi})$~\cite{JiangMigdalLean2026} (Supplement~\ref{sec:S-tail-closed}). The source has no other oscillating term, and the partition function none, so the threshold is the only
singularity of $P$ at finite $K$. The large-$R$ expansion of every correlation function of the
limiting spectrum therefore consists of non-oscillating terms, which come from $K\to0$, and
of the wave, which comes from $K=\sqrt2$.

\subsection{The vorticity correlation}

\begin{figure}[t]
 \centering
 \includegraphics[width=\textwidth]{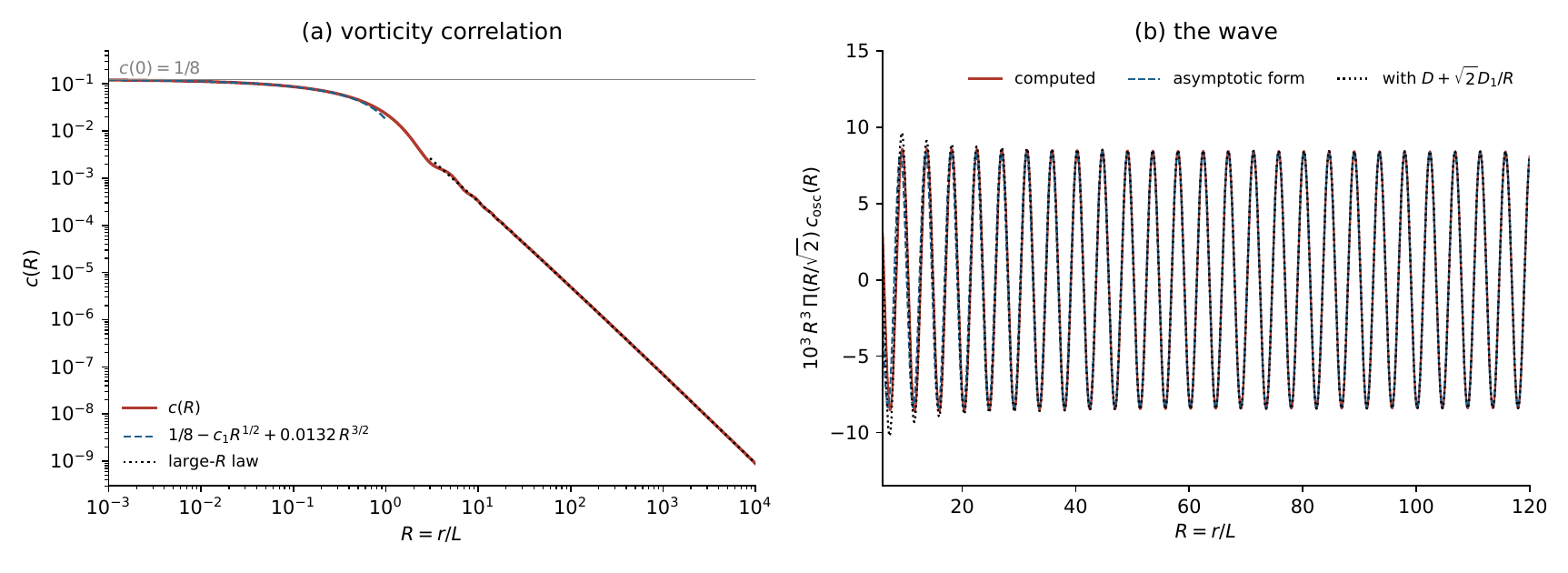}
 \caption{The vorticity correlation \eqref{eq:ev-corr} of the conditional state. (a) The scaling function $c(R)$, in units
 $(\nu/\widetilde\nu)^{1/2}/[4(t+t_0)^2]$, against $R=r/L$ for $10^{-3}\le R\le10^4$ on logarithmic axes (line), with the small-$R$ law
 \eqref{eq:osc-c-small} up to the term $R^{3/2}$ (dashed), the large-$R$ law \eqref{eq:osc-c-large} without the wave (dotted), and the
 value $c(0)=\frac18$ (thin horizontal line). (b) The wave, on linear axes, $6\le R\le120$, in units of $10^{-3}$: the difference between the correlations computed with the full source and with
 its oscillating term removed (for $h>2$, fully from $h=4$), multiplied by $R^3\,\Pi(R/\sqrt2)$, with $\Pi=h\widehat Z$, minus its
 average over one period (line; the subtraction of the non-oscillating part computed from the moments of the removed term, as in
 Supplement~\ref{sec:S-osc}, gives the same curve to about $2\%$ of $|D|$); the asymptotic form
 $\sqrt{2\pi}\,|D|\cos(\sqrt2R-\frac\pi4+\arg D)$ of \eqref{eq:osc-c} (dashed); and the same with $D+\sqrt2D_1/R$ in place of $D$, the
 next oscillating order of the source \eqref{eq:ev-large} read at the edge (dotted). The period is $2\pi/\sqrt2$ in $R$: the Abel
 average reads the source at $h=R/\sqrt2$.}
 \label{fig:ev-corr}
\end{figure}

The vorticity correlation needs no spectrum: it is an Abel transform of the conditional source itself. Indeed, with
$C_\omega(r)=2\int_0^\infty k^2E\,\mathrm{sinc}(kr)\dd k$ and the identity
$\int_0^\infty J_0(aK)\sin(RK)\dd K=(R^2-a^2)^{-1/2}\theta(R-a)$, the limiting correlation is the first of the
transforms~\eqref{eq:cond-def-transforms} with the renormalized prefactor:
\begin{equation}
 \boxed{C_\omega(r,t)=\Bigl(\frac{\nu}{\widetilde\nu}\Bigr)^{1/2}\frac{c(r/L)}{4(t+t_0)^2},\qquad
 c(R)=\int_0^{\pi/2}C\!\left(\frac{R\sin\vartheta}{\sqrt2}\right)\sin\vartheta\,\dd\vartheta .}
 \label{eq:ev-corr}
\end{equation}
Here $c$ is the average of the source over the orientation of the separation with respect to the plane of the
hairpin, and every law of $c$ is a law of $C$ seen through this average. The correlation is positive at every computed separation, $\QCORMIN\le r/L\le\QCORMAX$ (Fig.~\ref{fig:ev-corr}a): the
Abel average over $[0,R/\sqrt2]$ always includes the large positive values of $C$ at small $h$.

\paragraph{Small separation.} At $R=0$ the average is trivial: $c(0)=C(0)=\frac18$, Eq.~\eqref{eq:cond-zero}, with $R\to0$ taken
after $N\to\infty$; the $1/R$ of the earlier correlation, from the numerator alone, is removed by the partition function. Every power of
the source, the constant included, becomes the same power of $R$, and \eqref{eq:ev-small} gives
\begin{equation}
 \boxed{c(R)=\frac18-c_1R^{1/2}+\QCOcthree\,R^{3/2}+O(R^2),\qquad c_1=2^{3/4}\,\frac{\Gamma(\frac54)}{\Gamma(\frac74)}\,a_{1/2}=\QCOcone .}
 \label{eq:osc-c-small}
\end{equation}
The cusp $R^{1/2}$ of the correlation and the ultraviolet tail $K^{-3/2}$ of $P$ come from the same hairpins, those with one short arc.

\paragraph{Large separation.} Here the Abel average reads the tail \eqref{eq:ev-C-large} of the source and acts on it like a sharp cut
at $h=\sqrt2R$. With the regularized first moment $M$ of the source,
\begin{equation}
 \boxed{R^2c(R)\simeq2A_c\log(\sqrt2R)+2B_c\log L_R+2M+\frac{\pi^2}{12}\,\frac{B_c}{L_R^2}+\zeta(3)\,\frac{B_c}{L_R^3},}
 \label{eq:osc-c-large}
\end{equation}
with $L_R=\log(\sqrt2R)+\bar\kappa$, up to the wave and to terms of order $\log R/R$. With the wave \eqref{eq:osc-c} added, the law
reproduces the computed correlation to $\QCOLAWACC$ for $R\ge\QCOLAWFROM$. The earlier correlation fell as $1/R^2$, without a
logarithm.

\paragraph{The wave.} The edge of the Abel average, $\vartheta=\frac\pi2$, probes the source at $h=R/\sqrt2$, where the phase
$2h=\sqrt2R\sin\vartheta$ of its oscillating term is stationary, and the stationary phase gives the wave:
\begin{equation}
 \boxed{\begin{aligned}
 c(R)&=c_{\rm reg}(R)+c_{\rm osc}(R),\\
 c_{\rm osc}(R)&=\frac{\sqrt{2\pi}\,|D|}{\Pi(R/\sqrt2)\,R^3}\cos\Bigl(\sqrt2R-\frac\pi4+\arg D\Bigr)
 \simeq\frac{8\pi\,|D|\cos\bigl(\sqrt2R-\frac\pi4+\arg D\bigr)}{R^3\,[\log(R/\sqrt2)+\bar\kappa]}.
 \end{aligned}}
 \label{eq:osc-c}
\end{equation}
A fit of the computed correlation on $50\le R\le120$ returns $|D|$ to $\QCOWAVEFITD$ and $\arg D$ to $\QCOWAVEFITARG$
(Fig.~\ref{fig:ev-corr}b). The smooth part falls as $c_{\rm reg}\simeq2A_c\log R/R^2$, so the relative size of the wave
decreases as $1/(R\log^2R)$, and a logarithmic derivative multiplies it by $\sqrt2R$. The effective index of the vorticity correlation,
$n_\omega=-\partial\log C_\omega/\partial\log r$, therefore carries a wave that decays only as the inverse square of a logarithm:
\begin{equation}
 \boxed{\begin{aligned}
 n_\omega&=n_\omega^{\rm reg}+n_\omega^{\rm osc},\qquad
 n_\omega^{\rm osc}\simeq-R\,\partial_R\frac{c_{\rm osc}}{c_{\rm reg}},\\
 n_\omega^{\rm osc}&\simeq\frac{2\sqrt\pi\,|D|\,\sin\bigl(\sqrt2R-\frac\pi4+\arg D\bigr)}{\Pi(R/\sqrt2)\,R^2c_{\rm reg}(R)}
 \simeq\frac{\sqrt\pi\,|D|}{E'\log^2R}\,\sin\Bigl(\sqrt2R-\frac\pi4+\arg D\Bigr).
 \end{aligned}}
 \label{eq:osc-n}
\end{equation}
The last form is reached at very large $R$ only; with the computed $c_{\rm reg}$ the oscillation of the index is about
$\pm\QCONOMSIX$ near $r=6L$, $\pm\QCONOMTEN$ at $r=10L$ and $\pm\QCONOMFIFTY$ at $r=50L$.

\subsection{The velocity structure function}

The structure function $D_v(r,t)=\langle(\mathbf v(\mathbf r)-\mathbf v(0))^2\rangle=4\int_0^\infty E\,[1-\mathrm{sinc}(kr)]\dd k$ is
finite at every separation, although the energy diverges. With \eqref{eq:ev-spectrum} and a Bessel integral we find
\begin{equation}
 \boxed{\begin{gathered}
 D_v(r,t)=\Bigl(\frac{\nu}{\widetilde\nu}\Bigr)^{1/2}\frac{\widetilde\nu}{t+t_0}\,\mathcal S\Bigl(\frac rL\Bigr),\\
 \mathcal S(R)=\int_0^{R/\sqrt2}hC(h)\Bigl[\operatorname{arccosh}\frac R{\sqrt2h}-\sqrt{1-\frac{2h^2}{R^2}}\,\Bigr]\dd h ,
 \end{gathered}}
 \label{eq:osc-D}
\end{equation}
with the effective index
\begin{equation}
 \zeta_2=\frac{\partial\log D_v}{\partial\log r}=\frac1{\mathcal S(R)}\int_0^{R/\sqrt2}hC(h)\sqrt{1-\frac{2h^2}{R^2}}\,\dd h .
 \label{eq:osc-zeta}
\end{equation}
At short distance $\zeta_2\to2$: the integral of $P=K^2E$ converges, to $c(0)=\frac18$, $\mathcal S\simeq R^2/96$, and below the scale
$L$ the velocity is smooth in the mean square; the cusp of $c$ gives $\zeta_2\simeq2-\QCOZSMALL\,R^{1/2}$. At large distance the
energy of the large scales takes over. Integrating $c=2R^{-2}\partial_R(R^2\partial_R\mathcal S)$ with \eqref{eq:osc-c-large} gives
\begin{equation}
 R\,\partial_R\mathcal S\simeq A_c\,\ell+B_c\log(\ell+\bar\kappa)+M,\qquad
 \mathcal S\simeq\frac{A_c}2\log^2R,\qquad \zeta_2\simeq\frac2{\log R},
 \label{eq:osc-S-large}
\end{equation}
with $\ell=\log(\sqrt2R/e)$: the image of the divergence of the energy. The index decreases only logarithmically, from below: it is
$\QCOZTEN$ at $r=10L$, $\QCOZHUNDRED$ at $r=100L$ and $\QCOZTENFOUR$ at $r=10^4L$. There is no inertial plateau. The index passes
the Kolmogorov value $2/3$ at $r=\QCOZTWOTHIRDS\,L$, where nothing special happens. The wave changes this index by only about
$\QCOZOSCSIX$ near $r\simeq6L$.

\subsection{What the wave is and what it is not}

The oscillations are periodic in $r$, not log-periodic: the derived expansions give the Mellin amplitude only real singularities. At a
fixed separation the
indices oscillate in time: this is the threshold \eqref{eq:ev-tstar} seen at a fixed $r$ instead of a fixed $k$. Velocity gradients are
the place to look: the vorticity correlation $C_\omega=\frac12\nabla^2D_v$ carries the same wave on a background that decays. Let us
stress the status of these numbers. Given the
conditional source, the period \eqref{eq:osc-period} is exact, the forms \eqref{eq:osc-c}, \eqref{eq:osc-n} and
\eqref{eq:osc-zeta-osc} are asymptotically exact, and their amplitudes are computed; the source itself rests on the conditional correlation, Definition~\ref{post:cond}. These amplitudes are those of the quenched order; in the annealed order the period and
the phase are the same and the amplitudes are smaller (Sec.~\ref{sec:osc-orders}). The comparison with simulations and experiment is left to Ref.~\cite{MigdalWang2026QPTF} (Sec.~\ref{sec:comparison}).

\subsection{The two orders in coordinate space}
\label{sec:osc-orders}

\begin{figure}[t]
 \centering
 \includegraphics[width=\textwidth]{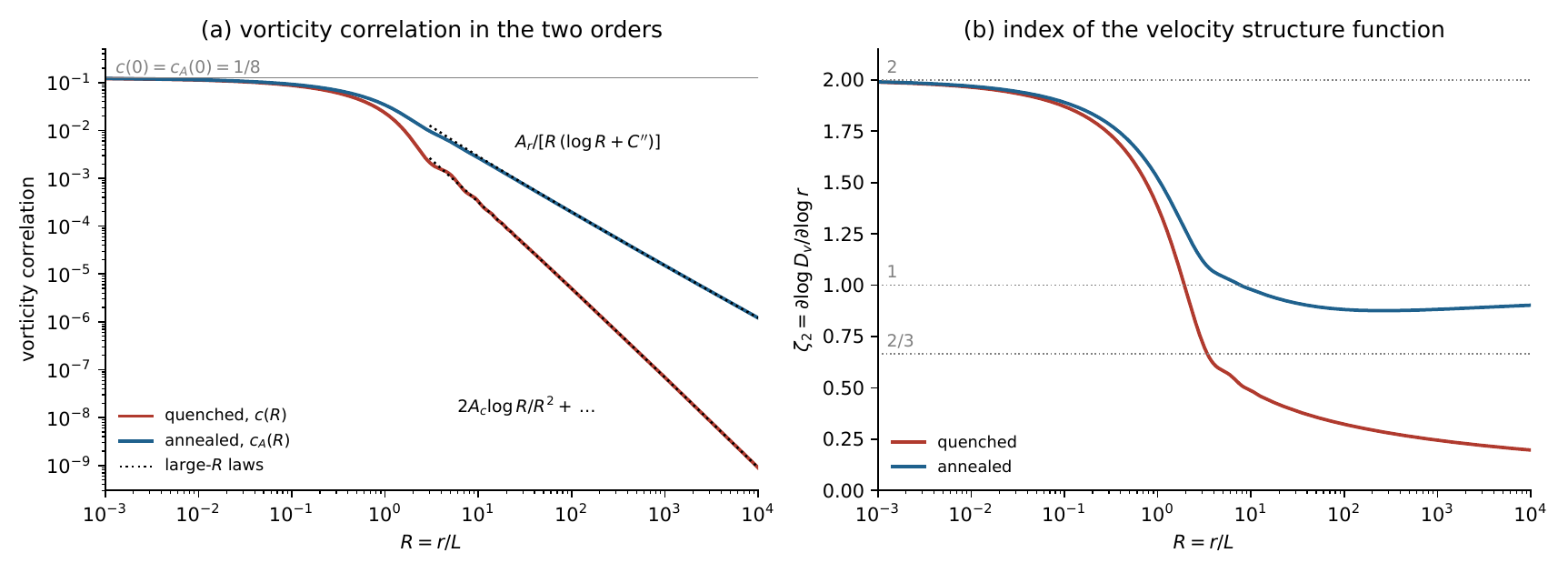}
 \caption{The two orders in coordinate space. (a) The scaling functions of the vorticity correlation, $c(R)$ of the quenched order,
 Eq.~\eqref{eq:ev-corr} (red), and $c_A(R)$ of the annealed order, Eq.~\eqref{eq:ord-PA} (blue), in units
 $(\nu/\widetilde\nu)^{1/2}/[4(t+t_0)^2]$, against $R=r/L$ on logarithmic axes, with their large-$R$ laws \eqref{eq:osc-c-large} and
 \eqref{eq:ord-c-large} without the wave (dotted) and the common value $\frac18$ at $R=0$ (thin horizontal line). (b) The index
 $\zeta_2$ of the velocity structure function, Eq.~\eqref{eq:osc-zeta}: it tends to zero as $2/\log R$ in the quenched order and to
 one, from below, in the annealed order; horizontal dotted lines: $2$, $1$ and the Kolmogorov value $2/3$.}
 \label{fig:ev-orders-coord}
\end{figure}

The correlation of the annealed order needs no spectrum either: it is the ratio \eqref{eq:ord-PA} of two Abel averages
(Fig.~\ref{fig:ev-orders-coord}a). At small separation the two orders nearly coincide,
\begin{equation}
 c_A(R)=\frac18-c_1^AR^{1/2}+0.0102\,R^{3/2}+O(R^2),\qquad c_1^A=\frac2\pi\,\frac{I_{1/2}}{I_{3/2}}\,c_1=0.1053 ,
 \label{eq:ord-c-small}
\end{equation}
the cusp of \eqref{eq:osc-c-small} with the factor of \eqref{eq:ord-UV}. At large separation the annealed correlation is
\eqref{eq:ord-c-large}: it decays as $1/(R\log R)$ instead of $\log R/R^2$, its index tends to one, as $n_\omega\simeq1+1/\log R$,
instead of two, and at $r=100L$ it is $40$ times the quenched correlation. The wave has the same period and the same phase; its
amplitude is divided by the averaged partition function,
\begin{equation}
 c^A_{\rm osc}(R)\simeq\frac{16\,|D|\cos\bigl(\sqrt2R-\frac\pi4+\arg D\bigr)}{R^3\,(\log R+C'')} ,
 \label{eq:ord-c-osc}
\end{equation}
that is, the last form of \eqref{eq:osc-c} with $8\pi$ replaced by $16$ and $\bar\kappa$ by $\bar\kappa-\log2$; a fit of the computed
correlation on $40\le R<120$ returns $|D|$ to $0.5\%$ and $\arg D$ to $10^{-3}$. Since the smooth part is now larger, the relative
size of the wave decreases as $1/R^2$, and the oscillation of the index, $n_\omega^{\rm osc}\simeq[16\sqrt2\,|D|/(A_rR)]
\sin(\sqrt2R-\frac\pi4+\arg D)$, with $16\sqrt2\,|D|/A_r=0.52$, decreases as $1/R$ and not as the inverse square of a logarithm: it
is about $\pm0.07$ near $r=10L$ and $\pm0.01$ near $r=50L$, against $\pm\QCONOMTEN$ and $\pm\QCONOMFIFTY$ in the quenched order. The wave is a
much weaker feature of the annealed order.

For the structure function, $c_A=2R^{-2}\partial_R(R^2\partial_R\mathcal S_A)$ and \eqref{eq:ord-c-large} give
\begin{equation}
 \mathcal S_A(R)\simeq\frac{A_r}{4}\,\frac{R}{\log R+C''},\qquad \zeta_2^A\simeq1-\frac1{\log R+C''}\qquad(R\gg1),
 \label{eq:ord-S-large}
\end{equation}
up to terms of the relative order $1/\log R$: the structure function grows almost linearly, where that of the quenched order grows
as $\log^2R$, Eq.~\eqref{eq:osc-S-large}. This is the image of the linear divergence of the energy. The index $\zeta_2^A$ falls from
two to $0.98$ at $r=10L$, to $0.88$ at $r=100L$ and to a minimum, $0.876$ at $r\simeq270L$, and then rises slowly toward one: it is
$0.90$ at $r=10^4L$ (Fig.~\ref{fig:ev-orders-coord}b). It never reaches the Kolmogorov value $2/3$, which the quenched index passes at
$r=\QCOZTWOTHIRDS\,L$ on its way to zero. In a flow of a finite size both indices are cut at separations of the order of the size of
the system.

\section{The scaling limits of the even ensemble}
\label{sec:3d-limits}

The limit of Sec.~\ref{sec:3d-renorm} removes the cutoff at fixed $\widetilde\nu=\nu_NN^2$. Why this power? The loop equation does
not fix it. Here we classify the limits
\begin{equation}
 N\to\infty\quad\hbox{at fixed}\quad\widetilde\nu=\nu_NN^{\chi},\qquad K=kL,\qquad L=\sqrt{\widetilde\nu\,(t+t_0)},
 \label{eq:lim-chi}
\end{equation}
with an exponent $\chi>0$, for which the bare viscosity vanishes. Only two of them have structure. The zig-zag layer,
$|\beta-\pi|\sim N^{-1/2}$, gives $\chi=2$, and the generic angles, with Gaussian arc averages, give $\chi=1$; any other exponent
gives a single power of $k$, without a scale, if the layers separate, which is not proven. The two limits with structure are limits
of the same ensemble, and the loop equation admits both: the choice is the law by which the bare viscosity vanishes with the cutoff.

Two properties of the ensemble, neither of which refers to data, single out $\chi=2$. The first is its criticality: a correlation
length along the loop of the order of the whole loop, which makes the limit an interacting continuum theory, integrable, that keeps
the arithmetic of the rational angles. The second is its crossover scale, which only $\chi=2$ keeps at a finite $kL$. The selection
of $\chi=2$ is therefore an explicit part of the definition of Remark~\ref{rem:intro-status}: it selects the universality class. The
limit $\chi=1$ was computed with the earlier normalization only (Supplement~\ref{sec:S-limits-full} and~\ref{sec:S-limits}).

\section{Predictions for simulations and experiment}
\label{sec:comparison}

The theory makes predictions that the existing data can test, and here we list them. We do not make the comparison in this paper.
It needs a detailed analysis of the data, with both solutions of Sec.~\ref{sec:3d-orders}, and it is the subject of a forthcoming
paper with Wei Wang~\cite{MigdalWang2026QPTF}, on the direct numerical simulations of decaying turbulence of Rodhiya and
Sreenivasan~\cite{RodhiyaSreenivasan2026} and on the measurements of K\"uchler, Bewley and Bodenschatz in the Max Planck Variable
Density Turbulence Tunnel~\cite{KuchlerBewleyBodenschatz2023}.

What is there to adjust? Almost nothing. The spectrum has a fixed shape and a single scale $L$. Its amplitude, with the power of
the viscosity in it, drops out of the effective indices and of the normalized spectra, so the theory can only be shifted in scale
and, in a bounded flow, cut at the lowest wave number $k_0$. This is where the theory meets the data.

\paragraph{Two solutions.} The simplest test between the two solutions is the decay of the energy. Its index drifts slowly in
both: from $1.15$ to $1.30$ in the quenched solution \eqref{eq:ev-spectrum} and from $1.44$ to $1.54$ in the annealed solution
\eqref{eq:ord-PA} as $k_0L$ grows from $0.01$ to $0.3$ (Table~\ref{tab:size}). The two solutions have the same ultraviolet law, up
to the factor $0.8727$, and the same threshold; they differ at $kL<1$ and in the finite-size effects of Sec.~\ref{sec:3d-size}. The
other tests are the dependence of the energy and of its dissipation on the size of the flow, logarithmic in the first solution and
almost linear in the second, and the index $\zeta_2$ of the structure function at $r\gg L$, which tends to zero in the first and to
one in the second (Sec.~\ref{sec:osc-orders}). Which of the two solutions a given flow follows, what selects it, and what plays the
role of $k_0$ in a simulation whose box is much larger than the scales of its initial data, are questions for this comparison.

\paragraph{The fine features.} The oscillations of the vorticity correlation (Sec.~\ref{sec:3d-osc}) and the slope of the spectrum
near $kL=\sqrt2$ (Sec.~\ref{sec:3d-final}) need flows with $\widetilde\nu/\nu\gg1$: simulations in boxes much larger than $L$, and a
wider wind tunnel with several stations.

\section{Conclusion}
\label{sec:conclusion}

We have reduced the statistics of decaying turbulence to the quantum mechanics of two pendulums, and its energy spectrum to a
chain of explicit quadratures. The theory is an operator theory of classical flows. Let us recall its logic, a chain of five steps,
and then say what it rests on.

First, the statistics of a flow is quantum mechanics in loop space. The loop functional, the average exponential of the
circulation, is the wave function of this quantum mechanics and evolves by the linear loop equation. By an exact operator identity
it is the loop-ordered exponential of the covariant derivative $\partial+\ii v/\nu$, with the loop parameter as time and the
viscosity in the role of Planck's constant (Secs.~\ref{sec:mle} and~\ref{sec:qsp-identity}).

Second, the turbulent state is the state of equal weights, and the equation alone does not produce it. Statistical states are
density matrices in loop space, the Quantum Statistical Principle (Sec.~\ref{sec:qsp}). The principle itself assumes nothing; the
physics enters through the choice of the state. For decaying turbulence it is the state at the end of the decay, in which all
solutions on the fixed decaying trajectory have equal weight: a quantum form of the ergodic hypothesis, and the only physical hypothesis of the theory. One should not look for
these weights by integrating the deterministic equation. In the even ensemble at zero winding its time derivative term is singular
at every solution, the velocity near the solutions is undetermined, and perturbed trajectories grow instead of relaxing
(Sec.~\ref{sec:stoch-singular}). We interpret this instability as the source of spontaneous stochasticity: the thermal noise that
is present in every fluid only triggers the random motion, a diffusion of the momentum loop over the Euler states, and neither its
amplitude nor the temperature of the fluid enters the result (Secs.~\ref{sec:stoch-picture} and~\ref{sec:stoch-trigger}). In a
simple model of this diffusion, random exchanges of neighbouring spins and symmetric reflected shifts of the turning angle, the
equal weights are proved to be the attractor: the entropy grows monotonically, and every initial distribution converges to the
uniform one, for any doubly stochastic probabilities of the moves (Sec.~\ref{sec:stoch}).

The model has two consequences for what follows. It explains the thermal circle of the pendulum: in the continuum limit the
exchanges are a diffusion of the slow angle of the polygon, and its stationary measure is the path measure of the free rotor on
this circle (Sec.~\ref{sec:stoch-attractor}), whose temperature is a variable of the loop, the inverse of the variance of the walk
of the slow angle around it, and not the temperature of the fluid. And it leaves one question open: is the global orientation of
the polygon quenched or annealed in the conditional correlation? We give both solutions. The ultraviolet law and the threshold do
not depend on the order; the infrared law and the finite-size effects do (Secs.~\ref{sec:stoch-rotation} and~\ref{sec:3d-size}).
The model itself is a Markov chain in the discrete logarithmic time, and its exact solution shows that the discretization is valid
for the distribution of the states in the statistical limit, with a finite relaxation time if the step decreases as $N^{-3}$.

Third, turbulence at infinite Reynolds number is defined, as in quantum field theory, by a regularization and a renormalization
(Remark~\ref{rem:intro-status}). The regularization is the discretized loop equation on a polygon with $N$ sides. The renormalized
limit leaves the turbulent viscosity $\widetilde\nu$ as the single physical scale, much as the QCD scale survives the removal of
the lattice cutoff (Sec.~\ref{sec:3d-renorm}).

Fourth, the exact solution of the discretized loop equation has the statistics of two quantum pendulums. The equation has an exact
manifold of decaying solutions, the Euler ensemble, built from rational angles and spin sequences
(Sec.~\ref{sec:three-dimensional}). Written as a trace over fermions, its partition function and its conditional vorticity
correlations become, near the critical angle $\beta=\pi$, Gibbs averages of two quantum pendulums with the Mathieu Hamiltonian: the
fast spins are a thermostat for the slow direction of the polygon, and the closed loop is its thermal circle
(Secs.~\ref{sec:3d-fermion} and~\ref{sec:cond-pendulum}). The Gibbs measure is thus derived, not postulated. Without an operator representation of the flow this chain would be hard
to motivate: a classical sum over polygons does not suggest by itself that its continuum limit is a Schr\"odinger operator. Here it
is one because the sum was a quantum trace from the start.

Fifth, the integrable pendulums give the observables in closed form. The energy spectrum \eqref{eq:intro-main} is a chain of
quadratures of explicit resolvents; it is positive, its enstrophy is finite, and it falls as $k^{-7/2}$ at large wave numbers. At
small wave numbers it falls as $k^{-1}\log(1/kL)$ in the quenched order of the orientation average and as $k^{-2}/\log(1/kL)$ in the
annealed one (Secs.~\ref{sec:3d-final} and~\ref{sec:3d-orders}). The two solutions differ in an observable way: the energy and its
dissipation depend on the size of the flow logarithmically in the first and almost linearly in the second, and the energy decays
with an index of $1.1$ to $1.2$ and of about $1.45$ (Sec.~\ref{sec:3d-size}).

The most unexpected result is the threshold. In both solutions the hairpins with two equal arcs produce a threshold at
$k\sqrt{\widetilde\nu(t+t_0)}=\sqrt2$, where the slope of the spectrum diverges as a double logarithm. In time this is a
singularity of the decay at a fixed wave number, reached at a finite time; in space it is a wave, an oscillation of the vorticity
correlation with the period $\sqrt2\,\pi L$ and a slowly decreasing amplitude (Sec.~\ref{sec:3d-osc}). This is a prediction for the
statistics of velocity gradients in decaying turbulence.

One earlier claim is withdrawn. The number theory of the Euler ensemble, which earlier suggested a role for the zeros of the
Riemann zeta function, survives the local limit only as a flat density of rational angles. The zeta singularities disappear, and
their place is taken by a singularity that is not arithmetic but geometric (Remark~\ref{rem:ev-zeros}); the zeta zeros do not
govern the decay at long times.

What does all this rest on? The mathematics is a theorem, and the physics is one hypothesis; Table~\ref{tab:status} gives the status
of each statement. The reduction of the Euler ensemble to the quantum pendulum is proved and formally verified in Lean~4 in the
forthcoming paper by Y.~Jiang, A.~Migdal et al.~\cite{JiangMigdalLean2026} (Sec.~\ref{sec:lean}): the finite-$N$ statements are
exact, and the convergence of the ensemble to the pendulum formulas, outlined in Supplement~\ref{sec:S-finiteN-proof} and supported
by exact sums and Monte Carlo simulations, holds at every nonzero separation, together with the grand canonical limit. What is not
proved is the physics, and it is a single hypothesis: the equal probability of all the states of the Euler ensemble. We do not
derive it, but there is now a mechanism for this ergodicity (Secs.~\ref{sec:stoch-mechanism} and~\ref{sec:stoch}): the thermal
fluctuations of the vertices of the polygon trigger a diffusion over the Euler states, by infinitesimal moves between close states,
and given this noise-triggered singular diffusion the equal probabilities are a theorem, an H-theorem. That the instability of the
equation acts as this diffusion is not derived. Nothing else is postulated: the Gibbs measure of the pendulums is derived from the
equal weights, and the vorticity correlation is the expectation conditioned on the hairpin, by Bayes' formula. The theory is
therefore not rigorous: as statistical mechanics rests on the ergodic hypothesis, it rests on a physical principle, and it is tested
by its consequences.

What remains is concrete. The central mathematical question is whether the renormalized limit exists and is universal, independent
of the regularization. The central physical questions are whether the random
moves of the stochastic model can be derived from the singular evolution near the Euler states (Sec.~\ref{sec:stoch-attractor}), and
whether the orientation of the polygon is quenched or annealed in the conditional correlation (Sec.~\ref{sec:stoch-rotation}).
The next steps are the theory of the correlation at the cutoff scale, which carries the dissipation, the completion in the
infrared, and the passive scalar, whose correlations must be recomputed with the new measure. The comparison with simulations and
the wind tunnel is not made in this paper (Sec.~\ref{sec:comparison}): it needs both solutions and a detailed analysis of the data,
and will be given with Wei Wang~\cite{MigdalWang2026QPTF}; the fine features need higher Reynolds numbers. The full discussion of
the results, of their status and of the open problems is in Supplement~\ref{sec:S-discussion}.

\subsection{The surprising finite-time singularity}

The phenomenology of turbulence expects scaling laws: a power-law spectrum and power-law decay
rates. A singularity of the limiting spectrum at $k=\sqrt2/L(t)$, which moves with time and separates
two regimes, each with its own drifting index, therefore comes as a surprise. It originates in the
number-theoretic Euler ensemble and in the analytic solution of a quantum pendulum, which emerges
from the operator representation of the statistical evolution of the classical Navier--Stokes flow
by quantum mechanics in loop space, together with the hypothesis of equal probabilities of the Euler states. The discrete levels of this quantum system appear as a
singularity of the classical decaying energy spectrum.

\subsection{The formal Lean proof}
\label{sec:lean}

The mathematics of this paper is formalized in the proof assistant Lean~4, with its library Mathlib, in the forthcoming joint paper
by Y.~Jiang, A.~Migdal et al., \emph{The Quantum Pendulum of Decaying Turbulence: Exact Euler-Polygon Rigidity, Twisted
Bloch Transfer Operators, and the Unconditional Grand-Canonical Continuum Limit}~\cite{JiangMigdalLean2026}, carried out with colleagues at
Google Research. The formalization
separates the mathematics from the physics. The physical hypothesis, the equal weight of the Euler states, and the
definitions of Table~\ref{tab:status}, the conditional correlation and the renormalization with its matching of the amplitude, are
not part of the formal statements: they say which finite sums represent the
physical observables, and the theorems are statements about these sums.

The exact finite-$N$ statements of this paper are proved in Lean, with no unproved steps and no axioms beyond those of the Lean kernel:
the operator identity of Proposition~\ref{prop:qsp-identity} and the calculus of jumps; the Euler solutions of the discretized loop
equation, for every $N$ and every winding; the vorticity of a vertex, the contact terms and the invariance of the correlation; the
zero-separation values and their signs; the counts of the sectors of the grand ensemble; and the fermionic trace with its exact
transfer-matrix form. Several asymptotic results are proved as well: the expansion of the paired transfer operator that yields the
pendulum Hamiltonian, the ground-band value $\frac18$ of Eq.~\eqref{eq:cond-zero}, and closed forms of the constants of the
large-$h$ tail, $E'=1/(1024\sqrt\pi)$, $A_c=1/(128\sqrt2)$ and $c_0=7\pi^{3/2}/(512\sqrt2)$, which agree with the values fitted in
Sec.~\ref{sec:ev-source} to every printed digit. That the correlation tends to this value at zero separation,
$C(0)=\lim_{h\to0}C(h)=\frac18$, is not yet proved, in Lean or on paper: the proof of the limits covers only $h>0$. The route to a
proof is clear and the numbers agree (Sec.~\ref{sec:ev-source}), but one key estimate is still missing.

The continuum limit itself is proved as well: the theorem that, as $N\to\infty$, the source and the partition function of the even
ensemble at zero winding, divided by $N^{1/2}$ and $N^{3/2}$, converge to the pendulum traces $\widehat G$ and $\widehat Z$ of
Eqs.~\eqref{eq:cond-G} and~\eqref{eq:cond-C}, so that their ratio converges to $C(h)$. This is the convergence that the present paper
supports by exact sums and Monte Carlo simulations (Remark~\ref{rem:3d-matching}). Its reduction to a set of estimates on the critical
layer $\beta\to\pi$ and on the rest of the ensemble was first proved by conventional means: our own tools, Claude Opus and GPT
Astra, working in collaboration, produced that proof, with uniform rates of convergence (Supplement~\ref{sec:S-finiteN-proof}); the
rates are not formalized. In Ref.~\cite{JiangMigdalLean2026} the
reduction of the Euler ensemble to the quantum pendulum is proved unconditionally and formally verified in Lean~4, in
twenty-eight main theorems: for every fixed scaled separation $h>0$, the critical layer converges to the continuum Mathieu traces
$\widehat G$ and $\widehat Z$, with $\widehat Z\ge\sqrt\pi/(4h)>0$; the angles away from the critical layer, the sectors with nonzero
winding and the odd sectors are suppressed; and the grand canonical ratio converges, as $\mu\to0^+$, to $C(h)=\widehat G/\widehat Z$.
The reduction of the Euler ensemble to the two pendulums is therefore a formally verified theorem, and the physical hypothesis
concerns only the choice of the statistical state and its identification with decaying turbulence, not the computation of its
observables; Ref.~\cite{JiangMigdalLean2026} states the physical input separately from the theorems. The Lean code will be made public with that paper.

\section*{Data availability}
The programs behind every number and figure, with the data files they read, form an archive described in
Sec.~\ref{sec:S-code} of the Supplementary Material. Its principal files are provided with this paper as ancillary files
(the archive \texttt{anc.zip} of the supplementary data), listed in the file \texttt{README.txt} among them: the Floquet values of the conditional source of Sec.~\ref{sec:3d-even} with their
interpolation; the programs that compute from them the spectrum and the correlation function in the two orders of the orientation
average (Secs.~\ref{sec:3d-final} and~\ref{sec:3d-osc}), with the tables of the quenched order and the recorded output for the
annealed one; three programs of the stochastic model of Secs.~\ref{sec:stoch-mechanism} and~\ref{sec:stoch}, with their outputs;
and the note \texttt{ergodicityNotes}. The complete archive, with the Floquet engine for the resolvents of the two pendulums and
the programs of the other results, is available at
\url{https://drive.google.com/file/d/17V9Fr7S5OTTX0IZQFjvJDBx-aT2bqPU8/view?usp=sharing};
the large raw outputs of the runs are available from the author. This paper uses no simulation or experimental data: the
comparison with the data of Refs.~\cite{RodhiyaSreenivasan2026,KuchlerBewleyBodenschatz2023} is left to
Ref.~\cite{MigdalWang2026QPTF}.

\section*{Acknowledgments}
I am grateful to K.~R.~Sreenivasan and E.~Witten for inspiring discussions. I thank A.~Rodhiya for providing the DNS data, and
E.~Bodenschatz and his group for the wind-tunnel data.
I am grateful to Michael Brenner, Yanchen (Jeff) Jiang and their colleagues at Google Research for the formalization in Lean of the
finite-$N$ results of this paper, for the closed forms of the large-$h$ constants, and for the formal proof of the reduction to the quantum pendulum and of the
continuum limit~\cite{JiangMigdalLean2026}.

\section*{Declaration of generative AI and AI-assisted technologies in the manuscript preparation process}
The loop-space formulation and the earlier Euler-ensemble papers are the author's work. During the preparation of this work the
author used, under his direction, OpenAI ChatGPT/Codex and GPT-6 (``Astra''), Anthropic's Claude (Opus and Fable~5.2) and
Google's Gemini. The division of labour was the same throughout. The author posed the questions, made the guesses, supplied the
physical intuition and rejected the conventional solutions. The AI systems explored the algebra, suggested alternatives and
solved the problems that were posed; they also audited the derivations, wrote and ran the programs of the archive, and wrote the
text in the author's style. In particular, Claude Opus and GPT-6 produced, in collaboration, the proof of the limits of the
pendulum reduction (Supplement~\ref{sec:S-finiteN-proof}), and the stochastic model of ergodicity
(Secs.~\ref{sec:stoch-mechanism} and~\ref{sec:stoch}) was formulated in discussions with Claude Fable~5.2, after the author's
initial idea of stochastic perturbations from thermal noise along the zero mode of the momentum-loop equation. The Lean formalization of the
forthcoming paper~\cite{JiangMigdalLean2026} was carried out by the Google Research team with their own AI systems and is not
part of this paper. After using these tools, the author reviewed and edited the content as needed and takes full responsibility
for the content of the publication.
\bibliographystyle{elsarticle-num}
\bibliography{references}
\end{document}